# The NANOGrav 15 yr Data Set: Targeted Searches for Supermassive Black Hole Binaries

Nikita Agarwal,[1,2] Gabriella Agazie,[3] Akash Anumarlapudi,[4] Anne M. Archibald,[5] Zaven Arzoumanian,[6] Jeremy G. Baier,[7] Paul T. Baker,[8] Bence Bécsy,[9] Laura Blecha,[10] Adam Brazier,[11,12] Paul R. Brook,[9] Sarah Burke-Spolaor,[1,2] Rand Burnette,[7] Robin Case,[7] J. Andrew Casey-Clyde,[13] Yu-Ting Chang,[14] Maria Charisi,[15,16] Shami Chatterjee,[11] Tyler Cohen,[17] Paolo Coppi,[14] James M. Cordes,[11] Neil J. Cornish,[18] Fronefield Crawford,[19] H. Thankful Cromartie,[20] Kathryn Crowter,[21] Megan E. DeCesar,[22] Paul B. Demorest,[23] Heling Deng,[24,25] Lankeswar Dey,[1,2] Timothy Dolch,[26,27] Daniel J. D'Orazio,[28,29,30] Ellis Eisenberg,[14] Elizabeth C. Ferrara,[31,32,33] Graham Doskoch,[1,2] William Fiore,[21] Emmanuel Fonseca,[1,2] Gabriel E. Freedman,[33] Emiko C. Gardiner,[34] Nate Garver-Daniels,[1,2] Peter A. Gentile,[1,2] Kyle A. Gersbach,[35] Joseph Glaser,[1,2] Matthew J. Graham,[36] Deborah C. Good,[37] Kayhan Gültekin,[38] C. J. Harris,[38] Jeffrey S. Hazboun,[7] Forrest Hutchison,[39] Ross J. Jennings,[1,2,*] Aaron D. Johnson,[3,40] Megan L. Jones,[3] David L. Kaplan,[3] Luke Zoltan Kelley,[41] Matthew Kerr,[42] Joey S. Key,[43] Nima Laal,[35] Michael T. Lam,[44,45,46] William G. Lamb,[35] Bjorn Larsen,[39] T. Joseph W. Lazio,[47] Natalia Lewandowska,[48] Tingting Liu,[49] Duncan R. Lorimer,[1,2] Jing Luo,[50,†] Ryan S. Lynch,[51] Chung-Pei Ma,[34,52] Dustin R. Madison,[53] Cayenne Matt,[38] Alexander McEwen,[3] James W. McKee,[54] Maura A. McLaughlin,[1,2] Natasha McMann,[35] Bradley W. Meyers,[55,56] Patrick M. Meyers,[40] Chiara M. F. Mingarelli,[39] Andrea Mitridate,[57] Priyamvada Natarajan,[58,39,59] Cherry Ng,[60] David J. Nice,[61] Shania Nichols,[44] Stella Koch Ocker,[40,62] Ken D. Olum,[63] Timothy T. Pennucci,[64] Benetge B. P. Perera,[65] Polina Petrov,[35] Nihan S. Pol,[66] Henri A. Radovan,[67] Scott M. Ransom,[68] Paul S. Ray,[42] Joseph D. Romano,[66] Jessie C. Runnoe,[35] Alexander Saffer,[68,*] Shashwat C. Sardesai,[3] Ann Schmiedekamp,[69] Carl Schmiedekamp,[69] Kai Schmitz,[70] Federico Semenzato,[71,72] Brent J. Shapiro-Albert,[1,2,73] Rohan Shivakumar,[14,74] Xavier Siemens,[7,3] Joseph Simon,[7] Sophia V. Sosa Fiscella,[45,46] Ingrid H. Stairs,[21] Daniel R. Stinebring,[75] Kevin Stovall,[23] Abhimanyu Susobhanan,[76] Joseph K. Swiggum,[61] Jacob A. Taylor,[7] Stephen R. Taylor,[35] Mercedes S. Thompson,[21] Jacob E. Turner,[51] Michele Vallisneri,[47,40] Rutger van Haasteren,[76] Sarah J. Vigeland,[3] Haley M. Wahl,[1,2] London Willson,[13] Kevin P. Wilson,[1,2] Caitlin A. Witt,[77] David Wright,[7] Olivia Young[45,46] And Qinyuan Zheng[39]

The NANOGrav Collaboration

[1] Department of Physics and Astronomy, West Virginia University, P.O. Box 6315, Morgantown, WV 26506, USA
[2] Center for Gravitational Waves and Cosmology, West Virginia University, Chestnut Ridge Research Building, Morgantown, WV 26505, USA
[3] Center for Gravitation, Cosmology and Astrophysics, Department of Physics and Astronomy, University of Wisconsin-Milwaukee, P.O. Box 413, Milwaukee, WI 53201, USA
[4] Department of Physics and Astronomy, University of North Carolina, Chapel Hill, NC 27599, USA
[5] Newcastle University, NE1 7RU, UK
[6] X-Ray Astrophysics Laboratory, NASA Goddard Space Flight Center, Code 662, Greenbelt, MD 20771, USA
[7] Department of Physics, Oregon State University, Corvallis, OR 97331, USA
[8] Department of Physics and Astronomy, Widener University, One University Place, Chester, PA 19013, USA
[9] Institute for Gravitational Wave Astronomy and School of Physics and Astronomy, University of Birmingham, Edgbaston, Birmingham B15 2TT, UK
[10] Physics Department, University of Florida, Gainesville, FL 32611, USA
[11] Cornell Center for Astrophysics and Planetary Science and Department of Astronomy, Cornell University, Ithaca, NY 14853, USA
[12] Cornell Center for Advanced Computing, Cornell University, Ithaca, NY 14853, USA
[13] Department of Physics, University of Connecticut, 196 Auditorium Road, U-3046, Storrs, CT 06269-3046, USA
[14] Department of Astronomy, Yale University, New Haven, CT 06520, USA
[15] Department of Physics and Astronomy, Washington State University, Pullman, WA 99163, USA
[16] Institute of Astrophysics, FORTH, GR-71110, Heraklion, Greece
[17] Department of Physics, New Mexico Institute of Mining and Technology, 801 Leroy Place, Socorro, NM 87801, USA
[18] Department of Physics, Montana State University, Bozeman, MT 59717, USA
[19] Department of Physics and Astronomy, Franklin & Marshall College, P.O. Box 3003, Lancaster, PA 17604, USA

Corresponding author: Chiara M. F. Mingarelli
chiara.mingarelli@yale.edu

[20]National Research Council Research Associate, National Academy of Sciences, Washington, DC 20001, USA resident at Naval Research Laboratory, Washington, DC 20375, USA
[21]Department of Physics and Astronomy, University of British Columbia, 6224 Agricultural Road, Vancouver, BC V6T 1Z1, Canada
[22]Department of Physics and Astronomy, George Mason University, Fairfax, VA 22030, resident at the U.S. Naval Research Laboratory, Washington, DC 20375, USA
[23]National Radio Astronomy Observatory, 1003 Lopezville Rd., Socorro, NM 87801, USA
[24]Columbia Astrophysics Laboratory, Columbia University, 538 West 120th Street, New York, NY 10027, USA
[25]Department of Physics, Columbia University, New York, NY 10027, USA
[26]Department of Physics, Hillsdale College, 33 E. College Street, Hillsdale, MI 49242, USA
[27]Eureka Scientific, 2452 Delmer Street, Suite 100, Oakland, CA 94602-3017, USA
[28]Space Telescope Science Institute, 3700 San Martin Drive, Baltimore, MD 21218, USA
[29]Department of Physics and Astronomy, Johns Hopkins University, 3400 North Charles Street, Baltimore, Maryland 21218, USA
[30]Niels Bohr International Academy, Niels Bohr Institute, Blegdamsvej 17, 2100 Copenhagen, Denmark
[31]Department of Astronomy, University of Maryland, College Park, MD 20742, USA
[32]Center for Research and Exploration in Space Science and Technology, NASA/GSFC, Greenbelt, MD 20771
[33]NASA Goddard Space Flight Center, Greenbelt, MD 20771, USA
[34]Department of Astronomy, University of California, Berkeley, 501 Campbell Hall #3411, Berkeley, CA 94720, USA
[35]Department of Physics and Astronomy, Vanderbilt University, 2301 Vanderbilt Place, Nashville, TN 37235, USA
[36]Cahill Center for Astrophysics, California Institute of Technology, MC 249-17, 1200 E California Boulevard, Pasadena, CA 91125, USA
[37]Department of Physics and Astronomy, University of Montana, 32 Campus Drive, Missoula, MT 59812
[38]Department of Astronomy and Astrophysics, University of Michigan, Ann Arbor, MI 48109, USA
[39]Department of Physics, Yale University, New Haven, CT 06520, USA
[40]Division of Physics, Mathematics, and Astronomy, California Institute of Technology, Pasadena, CA 91125, USA
[41]Astrophysics Working Group, NANOGrav Collaboration, Berkeley, CA, USA
[42]Space Science Division, Naval Research Laboratory, Washington, DC 20375-5352, USA
[43]University of Washington Bothell, 18115 Campus Way NE, Bothell, WA 98011, USA
[44]SETI Institute, 339 N Bernardo Ave Suite 200, Mountain View, CA 94043, USA
[45]School of Physics and Astronomy, Rochester Institute of Technology, Rochester, NY 14623, USA
[46]Laboratory for Multiwavelength Astrophysics, Rochester Institute of Technology, Rochester, NY 14623, USA
[47]Jet Propulsion Laboratory, California Institute of Technology, 4800 Oak Grove Drive, Pasadena, CA 91109, USA
[48]Department of Physics and Astronomy, State University of New York at Oswego, Oswego, NY 13126, USA
[49]Department of Physics and Astronomy, Georgia State University, 25 Park Place, Suite 605, Atlanta, GA 30303, USA
[50]Department of Astronomy & Astrophysics, University of Toronto, 50 Saint George Street, Toronto, ON M5S 3H4, Canada
[51]Green Bank Observatory, P.O. Box 2, Green Bank, WV 24944, USA
[52]Department of Physics, University of California, Berkeley, CA 94720, USA
[53]Department of Physics, Occidental College, 1600 Campus Road, Los Angeles, CA 90041, USA
[54]Department of Physics and Astronomy, Union College, Schenectady, NY 12308, USA
[55]Australian SKA Regional Centre (AusSRC), Curtin University, Bentley, WA 6102, Australia
[56]International Centre for Radio Astronomy Research (ICRAR), Curtin University, Bentley, WA 6102, Australia
[57]Deutsches Elektronen-Synchrotron DESY, Notkestr. 85, 22607 Hamburg, Germany
[58]Department of Astronomy, Yale University, 266 Whitney Avenue, New Haven, CT 06511, USA
[59]Black Hole Initiative, Harvard University, 20 Garden Street, Cambridge, MA 02138, USA
[60]Dunlap Institute for Astronomy and Astrophysics, University of Toronto, 50 St. George St., Toronto, ON M5S 3H4, Canada
[61]Department of Physics, Lafayette College, Easton, PA 18042, USA
[62]The Observatories of the Carnegie Institution for Science, Pasadena, CA 91101, USA
[63]Institute of Cosmology, Department of Physics and Astronomy, Tufts University, Medford, MA 02155, USA
[64]Institute of Physics and Astronomy, Eötvös Loránd University, Pázmány P. s. 1/A, 1117 Budapest, Hungary
[65]Arecibo Observatory, HC3 Box 53995, Arecibo, PR 00612, USA
[66]Department of Physics, Texas Tech University, Box 41051, Lubbock, TX 79409, USA
[67]Department of Physics, University of Puerto Rico, Mayagüez, PR 00681, USA
[68]National Radio Astronomy Observatory, 520 Edgemont Road, Charlottesville, VA 22903, USA
[69]Department of Physics, Penn State Abington, Abington, PA 19001, USA
[70]Institute for Theoretical Physics, University of Münster, 48149 Münster, Germany
[71]Dipartimento di Fisica Galileo Galilei, Università di Padova, I-35131 Padova, Italy
[72]INFN Sezione di Padova, I-35131 Padova, Italy
[73]Giant Army, 915A 17th Ave, Seattle WA 98122

[74] *Department of History, Yale University, New Haven, CT 06520, USA*
[75] *Department of Physics and Astronomy, Oberlin College, Oberlin, OH 44074, USA*
[76] *Max-Planck-Institut für Gravitationsphysik (Albert-Einstein-Institut), Callinstraße 38, D-30167 Hannover, Germany*
*Leibniz Universität Hannover, D-30167 Hannover, Germany*
[77] *Department of Physics, Wake Forest University, 1834 Wake Forest Road, Winston-Salem, NC 27109*

## ABSTRACT

We present the first targeted searches for continuous gravitational waves (CWs) from 114 active galactic nuclei (AGN) that may host supermassive black hole binaries, using the NANOGrav 15 yr data set. By incorporating electromagnetic priors on sky location, distance, redshift, and CW frequency, our strain and chirp mass upper limits are typically improved by a factor of $\sim 2$ (median 2.2) relative to all-sky limits at the same frequency. Bayesian comparisons against a model including only a Hellings–Downs correlated background disfavors a CW signal for all targets, with a mean Bayes factor of $0.73 \pm 0.32$. Two targets have Bayes factors slightly above unity, but coherence tests, random targeting experiments, and a conservative accounting of the 114-target trials factor all indicate that they are consistent with noise. We use these two candidates as worked examples to illustrate an end-to-end targeted CW search analysis and a suite of follow up tests that future promising candidates would need to pass. We find that the electromagnetic interpretations of both candidates are ambiguous, and we update the constraints on a putative binary in 3C 66B, ruling out part of its previously allowed parameter space. Ultimately, our results demonstrate the current sensitivity of targeted pulsar timing array searches for CWs and define a roadmap for future multimessenger CW detections.

## 1. INTRODUCTION

Detecting continuous gravitational waves (CWs) from individual supermassive black hole binaries (SMBHBs) is a primary goal of pulsar timing array (PTA) experiments. A CW detection would provide unambiguous evidence that SMBHBs form, harden, and persist into the gravitational wave (GW) emitting regime, directly probing the long-standing "final parsec problem" (Begelman et al. 1980; Milosavljević & Merritt 2003; Burke-Spolaor et al. 2019). Such a detection would give access to key binary parameters, including the strain amplitude, chirp mass, GW phase, and more (Jenet et al. 2004). Furthermore, it would enable the tracking of binary orbital evolution on decade-long timescales (Corbin & Cornish 2010; Lee et al. 2011) to thousands of light-years when pulsar distances are well constrained, testing the post-Newtonian expansion in the very strong gravitational regime (Mingarelli et al. 2012).

A small sample of CW detections could also constrain SMBHB demographics, help calibrate galaxy merger rates across mass and redshift, and anchor predictions for the space-based LISA mission (Amaro-Seoane et al. 2023; Steinle et al. 2023; Barausse et al. 2023). Moreover, such systems would become high-priority targets for multi-wavelength electromagnetic follow-up, enabling studies of circumbinary accretion and evolution, relativistic jet properties, and host galaxy dynamics in the galactic nucleus; see D'Orazio & Charisi (2023) for a recent review. Indeed, if the host galaxy is identified, the system also becomes a nHz "standard siren," allowing for an independent measurement of the Hubble constant (Schutz 1986; Wang et al. 2025). Strongly lensed SMBHBs may also be detectable through their CW signals (Khusid et al. 2023). Detecting these lensed binaries would yield another independent Hubble constant measurement and allow exploration of SMBHB orbital dynamics and evolution not accessible by any other means.

As evidence for the existence of the gravitational wave background (GWB) strengthens (Agazie et al. 2023a; EPTA Collaboration et al. 2023; Reardon et al. 2023; Xu et al. 2023; Miles et al. 2025), the frontier is now the detection of individual SMBHBs. All-sky CW searches, which set uniform priors on the location of the CW source and scan the whole sky, face difficult localization limits due to the low spatial resolution of PTAs: at present, all-sky searches cannot localize SMBHBs to better than 29 deg$^2$ (Goldstein et al. 2019; Petrov et al. 2024), making the host galaxy difficult to identify. Targeted CW searches offer an alternative route, by consid-

* NANOGrav Physics Frontiers Center Postdoctoral Fellow
† Deceased

ering the binary hypothesis in a particular host galaxy and constraining the binary candidate's properties.

Strategies for identifying promising host galaxies for CW searches have evolved steadily. Simon et al. (2014) introduced a top-down $\mathcal{M}_c/D_L$ ranking of nearby galaxies. Rosado & Sesana (2014) modeled the SMBHB population in SDSS to estimate where resolvable systems are statistically likely to occur, and Mingarelli et al. (2017) combined galaxy catalogs, cosmological merger rates, and PTA noise properties to identify nearby galaxies that could plausibly host individually detectable binaries. Subsequent studies refined these approaches (Xin et al. 2021; Bardati et al. 2024a,b; Horlaville et al. 2025).

The first targeted PTA analysis of a specific galaxy came with 3C 66B (Arzoumanian et al. 2020), demonstrating that electromagnetic priors on sky location and GW frequency can tighten chirp-mass limits relative to all-sky searches (Aggarwal et al. 2019). Targeted PTA searches now build on this foundation by using electromagnetic information to restrict sky position, distance, and GW frequency (Liu & Vigeland 2021). Here we present a unified methodology that integrates these priors with PTA data and a suite of statistical and coherence tests.

Independent studies and consistency checks with the recently measured GWB amplitude indicate that between 5 and 8 CRTS systems could, statistically, host true binaries, though it is not known which ones (Sesana et al. 2018; Kelley et al. 2019; Casey-Clyde et al. 2025). These remain candidates rather than confirmed binaries, and most admit non-binary interpretations. A targeted PTA search therefore asks a conditional question: if a given galaxy hosted a binary with catalog-informed properties, what do the PTA data say about that hypothesis?

Despite this astrophysical motivation, our analysis does not yield statistically significant evidence for CWs. The goal of this work is instead to lay out a roadmap for future targeted searches: a coherent framework for combining electromagnetic priors with PTA analyses, testing low significance outliers, and interpreting host specific limits as PTA sensitivity improves.

With more than a hundred targets, occasional Bayes-factor fluctuations of order a few are expected and are not compelling evidence for a CW (background expectations and trials factors are summarized in Table 2). Targeted searches also require care in interpretation. PTA sky localization remains broad (Goldstein et al. 2019; Petrov et al. 2024), and electromagnetic indicators often constrain only coarse ranges of binary parameters. The narrow priors used here should therefore be viewed as practical, catalog-informed hypotheses rather than precise constraints. In this setting, targeted analyses sharpen upper limits and test whether electromagnetic scenarios predict strain strong enough to be detectable.

This paper is laid out as follows. Section 2 summarizes the electromagnetic context for all targets, describes the CRTS and OVRO catalogs, and reviews the NANOGrav 15 yr data. Section 3 presents the CW signal model, the priors used in the targeted searches, and the Bayesian framework for model comparison. Section 5 reports the host-specific upper limits, Bayes factors, and background estimates for all 114 targets; the full numerical results are collected in Appendix B. Table 2 provides an at-a-glance summary of all robustness and coherence tests used throughout the paper. In Section 6 we apply these tests to two case-study systems, SDSS J153636.22+044127.0 (J1536+0441) and SDSS J072908.71+400836.6 (J0729+4008), to illustrate the methodology required for any future CW candidate. Section 7 summarizes the implications for targeted searches and multimessenger follow-up.

## 2. SMBHB CANDIDATES AND NANOGRAV DATA

The targeted analyses in this work use the NANOGrav 15-year data set (Agazie et al. 2023b), which provides high-precision timing of 68 millisecond pulsars with baselines approaching 16 years. The release includes narrowband and wideband time-of-arrival measurements, updated timing models with refined astrometry and binary parameters, and characterizations of both white and red noise. These data form the basis for all continuous-wave searches carried out here.

Our target list consists of 114 AGN that have been proposed as potential SMBHB hosts based on periodic or quasi-periodic electromagnetic variability. Most of these systems originate from the Catalina Real-Time Transient Survey (CRTS), where optical light-curve periodicity has been interpreted as a possible binary signature. We supplement this list with two long-baseline periodic blazars identified in the Owens Valley Radio Observatory (OVRO) monitoring program and with the well-studied radio galaxy 3C 66B.

For several CRTS sources, additional photometric coverage is available from the Zwicky Transient Facility (ZTF; Masci et al. 2019), the Asteroid Terrestrial-impact Last Alert System (ATLAS; Tonry et al. 2018), and the Wide-field Infrared Survey Explorer (WISE; Wright et al. 2010), which we use later when examining individual candidates. These 114 objects constitute the complete catalog analyzed in this work. Their electromagnetic periods provide the GW frequency hypotheses tested in our targeted searches. Section 2.1 describe the selection of these sources in more detail.

### 2.1. *Target List*

Here we focus on placing limits on sources using a single-valued GW frequency derived from electromagnetic information. We therefore target AGN with periodic or quasi-periodic variability that can serve as proxies for the binary orbital period and hence the GW frequency.

Several mechanisms may generate such periodic signals: circumbinary accretion modulations (Farris et al. 2014); relativistic Doppler boosting and gravitational lensing; and hydrodynamics-driven variability (D'Orazio et al. 2015; Charisi et al. 2018; D'Orazio & Di Stefano 2018; Kelley et al. 2019). The mapping between variability and orbital period depends on system properties. For circular, equal-mass binaries, accretion-driven modulation may exceed the orbital period by up to a factor of six, while eccentric systems tend toward a 1:1 mapping (Westernacher-Schneider et al. 2022; D'Orazio et al. 2024). At mass ratios $q \lesssim 0.2$ variability often tracks the orbital period, whereas for $q \lesssim 0.05$ it may become stochastic (D'Orazio et al. 2016; Dittmann & Ryan 2024).

Other proposed binary signatures include double-peaked or velocity-offset broad lines, precessing jets, and S/X-shaped radio lobes (Heckman et al. 1981; Shen & Loeb 2010; Merritt & Ekers 2002); see Bogdanović et al. (2022); Charisi et al. (2022); D'Orazio & Charisi (2023); Mingarelli et al. (2025) for reviews.

Apparent periodicity can also arise from stochastic quasar variability, as shown for CRTS sources (Vaughan et al. 2016); see Zhu & Thrane (2020); Davis et al. (2024) and El-Badry et al. (2025) for recent assessments. We return to these issues in Section 6.3.1.

Given these uncertainties, we adopt a simple 1:1 correspondence between the observed period and the orbital period, taking the GW frequency to be twice the orbital frequency. This provides a clean starting point for this first analysis.

#### 2.1.1. *Catalina Real-time Transient Survey Binary Candidates*

The CRTS candidates were identified by Graham et al. (2015) as AGN exhibiting periodic or semi-periodic variability. CRTS data were collected between 2005 May and 2016 April and calibrated to PS1 $r$-band magnitudes, while ZTF (2018–present) and ATLAS (2015–present) provide continued coverage on the same photometric system. WISE contributes mid-IR light curves in $W1$ and $W2$ from 2010–2011 and 2014–2024 with a half-year cadence.

While these objects were initially plausible SMBHB candidates, independent analyses show that many may be false positives. Short baselines covering only a few apparent periods are particularly susceptible to time-correlated red noise (Vaughan et al. 2016; Witt et al. 2022; Davis et al. 2024). Moreover, Sesana et al. (2018) demonstrated that if all CRTS sources were genuine binaries, the implied GWB would exceed PTA limits; Casey-Clyde et al. (2025) find that at most eight CRTS objects can be true binaries. This motivates treating each CRTS object as a candidate host rather than a confirmed binary, consistent with our conditional interpretation of targeted searches.

#### 2.1.2. *3C 66B and OVRO Candidates*

We augment the CRTS catalog with radio-variable sources. The Owens Valley Radio Observatory (OVRO) 40 m Telescope Monitoring Program surveyed 1,158 $\gamma$-ray bright blazars and identified two SMBHB candidates. PKS 2131−021 exhibits coherent sinusoidal variations from radio to optical wavelengths that have persisted for more than 45 years, consistent in both phase and period across independent windows (O'Neill et al. 2022; Kiehlmann et al. 2025). PKS J0805−0111 shows similar long-term behavior across multiple radio bands (de la Parra et al. 2025; Hincks et al. 2025).

3C 66B was among the earliest suggested PTA targets (Sudou et al. 2003), based on astrometric detections of elliptical motion. Mass constraints were first placed by Jenet et al. (2004) and later revised downward by Iguchi et al. (2010). The observed variability in radio-emitting systems is often attributed to jet precession, so the mapping between observed and orbital periods may differ by factors of several (Katz et al. 1982; Sudou et al. 2003; Britzen et al. 2023). In some circumstances the observed period may closely trace the orbital period, as in the "kinematic orbital model" of Kiehlmann et al. (2025). Here we assume a circular binary and adopt $f_{\mathrm{GW}} = 2/P_{\mathrm{obs}}$ for all candidates.

Together, the CRTS optical sample, the OVRO blazars, and 3C 66B constitute the 114-host catalog used in this targeted CW search.

### 2.2. *The NANOGrav 15 yr Data Set*

The NANOGrav 15 yr data release (Agazie et al. 2023b) presents high-precision observations and timing analyses of 68 millisecond pulsars, extending baselines to nearly 16 yr and adding 21 new pulsars. Observations were conducted with Arecibo, the Green Bank Telescope, and the VLA over 327 MHz–3 GHz, using upgraded UPPI backends and processed with the reproducible PINT pipeline (Luo et al. 2021). The release includes narrowband and wideband time-of-arrival data sets, calibrated pulse profiles, and refined timing models incorporating new astrometry, binary solu-tions, frequency-dependent delays, and red-noise measurements for 23 pulsars.

Publicly archived alongside the data are Jupyter notebook workflows and configuration files that enable full regeneration of results. The NANOGrav 15 yr Data Set underpins the companion analyses reporting evidence for the GWB (Agazie et al. 2023a,b) and provides the foundation for all targeted searches carried out here.

## 3. CONTINUOUS WAVE SIGNAL MODEL

CWs produce long-lived, nearly monochromatic signals that persist across the full observational span of PTAs. While evidence for the existence of a nHz GWB is sound, no individual CW source has yet been identified. A PTA is sensitive to CWs through the timing residuals they induce in an array of millisecond pulsars. A vector of timing residuals per pulsar, $\delta t$, fit with the GW signal, $s$, is modeled as:

$$\delta t = M\epsilon + n_{\rm white} + n_{\rm red} + s\,. \tag{1}$$

Here, $M$ is the design matrix that describes the linearized timing model and $\epsilon$ denotes the offsets from the nominal timing parameters. $n_{\rm white}$ and $n_{\rm red}$ respectively denote white and red noise. We model the stochastic background as a Hellings–Downs (HD; Hellings & Downs 1983) cross-correlated process following Agazie et al. (2023c), marginalizing over the timing model parameters. To model HD correlations efficiently, we first sample using a Common Uncorrelated Red Noise (CURN) model and then apply importance reweighting to obtain samples under the HD-correlated model (Hourihane et al. 2023). Unless otherwise specified, all results reported in this work use varied CURN parameters that are subsequently reweighted to the HD model to reduce the risk of degeneracy between the CW and GWB models.

### 3.1. *Continuous Wave Model*

The timing residual induced by a point-source CW is:

$$s(t) = F_+(\hat\Omega)\Delta s_+(t) + F_\times(\hat\Omega)\Delta s_\times(t), \tag{2}$$

where $F_{+,\times}(\hat\Omega)$ are the antenna pattern functions and $\Delta s_{+,\times}(t) = s_{+,\times}(t) - s_{+,\times}(t_p)$ are the Earth–minus–pulsar terms. The retarded pulsar time is $t_p = t - L(1+\hat\Omega\cdot\hat p)$. Visual representations of this coordiante system can be found in e.g. Chamberlin & Siemens (2012); Mingarelli & Sidery (2014).

For a SMBHB emitting at frequency $f_{\rm gw}$, the signal appears as a sinusoid with spatially dependent phase and amplitude determined by $\mathcal{M}_c$, $D_L$, $\iota$, $\psi$, and $\hat\Omega$.

For a circular, slowly evolving binary, the plus and cross polarizations are:

$$s_+(t) = \frac{\mathcal{M}_c^{5/3}}{D_L\omega(t)^{1/3}}\big[-\sin 2\Phi(t)(1+\cos^2\iota)\cos 2\psi - 2\cos 2\Phi(t)\cos\iota\sin 2\psi\big], \tag{3}$$

$$s_\times(t) = \frac{\mathcal{M}_c^{5/3}}{D_L\omega(t)^{1/3}}\big[-\sin 2\Phi(t)(1+\cos^2\iota)\sin 2\psi + 2\cos 2\Phi(t)\cos\iota\cos 2\psi\big]. \tag{4}$$

The orbital phase evolves as

$$\Phi(t) = \Phi_0 + \frac{1}{32\mathcal{M}_c^{5/3}}\left(\omega_0^{-5/3} - \omega(t)^{-5/3}\right), \tag{5}$$

$$\omega(t) = \omega_0\left(1 - \frac{256}{5}\mathcal{M}_c^{5/3}\omega_0^{8/3}(t-t_0)\right)^{-3/8}, \tag{6}$$

where $\omega_0$ is the orbital frequency at reference epoch $t_0$, which we set to the end of the data span (Corbin & Cornish 2010).

The strain amplitude is

$$h_0 = 1.3\times10^{-14}\left(\frac{\mathcal{M}_c}{10^9 M_\odot}\right)^{5/3}\left(\frac{f_{gw}}{10\text{ nHz}}\right)^{2/3}\left(\frac{100\text{ Mpc}}{D_L}\right). \tag{7}$$

Frequency evolution follows

$$\frac{df_{gw}}{dt} = 8.5\times10^{-5}\text{ nHz yr}^{-1}\left(\frac{\mathcal{M}_c}{10^9 M_\odot}\right)^{5/3}\left(\frac{f_{gw}}{10\text{ nHz}}\right)^{11/3}, \tag{8}$$

and the coalescence time is

$$t_c = 4.4\times10^4\text{ yr}\left(\frac{10^9 M_\odot}{\mathcal{M}_c}\right)^{5/3}\left(\frac{10\text{ nHz}}{f_{gw}}\right)^{8/3}. \tag{9}$$

In this work we fix $\hat\Omega$, $D_L$, and $f_{gw}$ using electromagnetic information, replacing uninformative priors with delta functions to enhance sensitivity to $h_0$. Based on Equation 8, a typical target evolves by only $\sim 10^{-3}$ nHz over the 15-year baseline ($T_{obs}$), which is negligible compared to the array's frequency resolution of $1/T_{obs} \approx 2$ nHz. Consequently, the choice of reference epoch $t_0$ does not impact the analysis.

In this work we fix $\hat\Omega$, $D_L$, and $f_{\rm gw}$ using electromagnetic information, replacing uninformative priors with delta functions to enhance sensitivity to $h_0$. This choice specifies hypothesis tests rather than asserting negligible uncertainty, and is motivated by results such as Arzoumanian et al. (2020), who showed that broad frequency priors produce similar $h_0$ limits. The choice of reference epoch $t_0$ may also impact analysis results if the target experiences significant frequency evolution. A broader treatment of $f_{\rm gw}$ will be explored separately.

## 4. METHODS FOR TARGETED SEARCHES

In this section, we describe the statistical framework and robustness tests used to assess the presence of a CW signal in our target list. These tests are critical for targeted searches because the source parameters (specifically sky location and frequency) are fixed by electromagnetic priors. Consequently, the relative amplitude and phase of the CW response in each pulsar are fully determined by the array geometry. This rigidity creates a risk that noise fluctuations in specific pulsars, combined with a particular array configuration, could mimic a GW signal. To distinguish genuine astrophysical signals from such spurious artifacts, we employ three distinct validation strategies: signal coherence tests (Section 4.1), dropout analyses (Section 4.2), and GWB anisotropy constraints (Section 4.3).

### 4.1. *Signal Coherence Tests*

A defining prediction of a genuine continuous wave (CW) source is phase coherence across the pulsar timing array. Noise artifacts, by contrast, typically appear incoherent from pulsar to pulsar. Following Bécsy et al. (2025), we evaluate coherence using two complementary approaches: a three-model Bayesian comparison and a scrambling-based null distribution. The primary drawback of these coherence tests is that they are conducted assuming fixed noise and the presence of CURN, but these assumptions are unlikely to spuriously favor CW coherence (Bécsy et al. 2025).

#### 4.1.1. *Three-model coherence framework*

We compare three hypotheses:

(i) NOISE: intrinsic pulsar and interstellar noise only,

(ii) CW: a coherent CW added to noise, with a single phase evolution shared across all pulsars,

(iii) INCOH: an incoherent model in which each pulsar receives its own independent sinusoid at the targeted frequency.

In the incoherent model,

$$s_a^{\rm INCOH}(t) = A_a \cos[2\pi f_{\rm gw} t + \phi_a]\,, \tag{10}$$

where $A_a$ and $\phi_a$ are independent for each pulsar. This model absorbs any single-pulsar or few-pulsar noise feature that mimics a periodicity at the targeted $f_{\rm gw}$ but lacks cross-pulsar coherence.

The Bayes factors satisfy the identity

$$\mathcal{B}^{\rm CW}_{\rm NOISE} = \mathcal{B}^{\rm CW}_{\rm INCOH}\,\mathcal{B}^{\rm INCOH}_{\rm NOISE}\,. \tag{11}$$

Here $\mathcal{B}^{\rm INCOH}_{\rm NOISE}$ measures evidence for *any* sinusoid (coherent or not), while $\mathcal{B}^{\rm CW}_{\rm INCOH}$ measures the incremental support for phase coherence across the array. A genuine astrophysical signal requires both terms to exceed unity, with $\mathcal{B}^{\rm CW}_{\rm INCOH}$ being the key discriminant.

#### 4.1.2. *Model-scrambling coherence tests*

To obtain an empirical null distribution of Bayes factors expected from noise alone, we also perform model-scrambling tests that deliberately destroy coherence while preserving noise properties. This follows the original strategy used in PTA CW searches.

The model scrambling approach runs a large collection of analyses in which the CW model is randomized in ways that erase cross-pulsar correlations. The resulting ensemble provides a baseline null distribution of Bayes factors. The Bayes factor from the unscrambled dataset can then be compared to this distribution to test for statistically significant coherence.

We employ two scrambling procedures:

- *Sky shuffling*: randomly exchanging the sky positions of pulsars in the antenna pattern, which destroys the geometric correlation structure expected from a CW.
- *Phase shifting*: adding an independent random phase shift to the CW signal in each pulsar, which removes phase coherence while maintaining the same frequency content.

Both procedures remove the inter-pulsar coherence that characterizes a true CW signal. For a significant detection, the unscrambled Bayes factor should lie as an outlier relative to the scrambled ensemble. If it does not, the apparent signal is likely attributable to noise fluctuations or pulsar-specific systematics.

### 4.2. *Dropout Analysis*

We employ the Dropout Analysis of Aggarwal et al. (2019) as a robustness test to evaluate the contribution of individual pulsars to a candidate CW signal. In this framework the astrophysical CW parameters are held fixed, and each pulsar is assigned a binary switch parameter $\kappa_a \in \{0, 1\}$ that controls whether the CW response is included in that pulsar's residuals:

$$s_a^{\rm drop}(t) = \kappa_a\, s_a(t). \tag{12}$$

The posterior distribution of the $\kappa_a$ parameters then directly quantifies which pulsars support the presence of a signal. A genuine CW should be coherently supported by many pulsars across the array, whereas a noise artifact or pulsar-specific systematic typically manifests as support concentrated in only one or a few pulsars. This test is especially valuable in targeted searches, where the geometry of the CW response is fully determined by the assumed host galaxy coordinates and any apparent signal must therefore be checked for its consistency across the PTA.

### 4.3. *GWB Anisotropy from a CW*

Here we compute the angular power spectrum, $C_\ell$, for a GWB signal dominated by a single CW in an effort to understand its properties. The result will help us to understand if a binary candidate is a true SMBHB system, since there should be some GWB contribution at each of their GW frequencies.

We model the sky map of the characteristic strain squared, $h_c^2(\hat{\Omega})$, as the sum of a perfectly uniform isotropic component and a single point source. This idealized model is valid in the limit where one bright, deterministic source stands out against a backdrop of a very large number of faint, unresolved sources that average out to a smooth continuum.

Omitting the explicit dependence from the frequency, $f$, for brevity, the total field is given by:

$$h_c^2(\hat{\Omega}) = h_{c,\mathrm{iso}}^2 + h_{c,\mathrm{CW}}^2 \cdot \delta(\hat{\Omega}, \hat{\Omega}_0)\,, \tag{13}$$

where $h_{c,\mathrm{iso}}^2$ is the constant value of the isotropic background, $h_{c,\mathrm{CW}}^2$ is the flux of the CW source, and $\delta(\hat{\Omega}, \hat{\Omega}_0)$ is the Dirac delta function localizing the source to a direction $\hat{\Omega}_0$ on the sky. The characteristic strain $h_c^2$ is directly related to the GW power spectral density $S_h(f) = h_c^2(f)/(12\pi^2 f^3)$ at a given frequency $f$. Equation (13) can therefore be interpreted as a physical decomposition of the total GW power on the sky. It separates the signal into two distinct contributions: the uniform power from a stochastic, isotropic GWB, and the highly localized power from a single, monochromatic CW source.

To compute the angular power spectrum, we first expand this field in the basis of real spherical harmonics,

$$h_c^2(\hat{\Omega}) = \sum_{\ell=0}^{\infty} \sum_{m=-\ell}^{\ell} a_{\ell m} Y_{\ell m}(\hat{\Omega})\,, \tag{14}$$

following the notation of Mingarelli et al. (2013). The coefficients $a_{\ell m}$ are found by projecting the field onto the basis functions:

$$\begin{aligned} a_{\ell m} = {} & h_{c,\mathrm{iso}}^2 \int_{S^2} Y_{\ell m}(\hat{\Omega})\, d\hat{\Omega} \\ & + h_{c,\mathrm{CW}}^2 \int_{S^2} \delta(\hat{\Omega}, \hat{\Omega}_0) Y_{\ell m}(\hat{\Omega})\, d\hat{\Omega}\,. \end{aligned} \tag{15}$$

The first integral is non-zero only for the monopole ($\ell = 0, m = 0$), where $Y_{00} = 1/\sqrt{4\pi}$. The second integral is evaluated using the sifting property of the delta function. This yields:

$$a_{\ell m} = h_{c,\mathrm{iso}}^2 \sqrt{4\pi}\, \delta_{\ell 0}\delta_{m0} \; + \; h_{c,\mathrm{CW}}^2 \cdot Y_{\ell m}(\hat{\Omega}_0)\,. \tag{16}$$

The presence of the CW source breaks the statistical isotropy of the sky. We compute the ensemble-averaged angular power spectrum

$$C_\ell^{\mathrm{tot}} \equiv \left\langle \hat{C}_\ell \right\rangle = \frac{1}{2\ell+1} \sum_{m=-\ell}^{\ell} \left\langle |a_{\ell m}|^2 \right\rangle\,. \tag{17}$$

For the monopole ($\ell = 0$), the power is:

$$\begin{aligned} C_0 = a_{00}^2 &= \left( h_{c,\mathrm{iso}}^2 \sqrt{4\pi} + \frac{h_{c,\mathrm{CW}}^2}{\sqrt{4\pi}} \right)^2 \\ &= 4\pi \left( h_{c,\mathrm{iso}}^2 + \frac{h_{c,\mathrm{CW}}^2}{4\pi} \right)^2. \end{aligned} \tag{18}$$

For all higher multipoles ($\ell > 0$), the isotropic contribution vanishes. Using the spherical harmonic addition theorem,

$$\sum_{m=-\ell}^{\ell} \left[ Y_{\ell m}(\hat{\Omega}_0) \right]^2 = \frac{2\ell+1}{4\pi},$$

we obtain the power spectrum:

$$C_\ell = \frac{h_{c,\mathrm{CW}}^4}{4\pi}, \qquad \ell > 0. \tag{19}$$

Thus, a single bright source produces a "white" power spectrum—equal power at all $\ell > 0$—a hallmark of pointlike anisotropy. The ratio of anisotropic power to monopole power is:

$$\frac{C_\ell}{C_0} = \frac{h_{c,\mathrm{CW}}^4}{16\pi^2 \left( h_{c,\mathrm{iso}}^2 + \frac{h_{c,\mathrm{CW}}^2}{4\pi} \right)^2}\,, \tag{20}$$

which is constant for all $\ell > 0$. See Figure 2 of Mingarelli et al. (2017) or Figure 5 of Burke-Spolaor et al. (2019) for visualizations. Appendix A derives the corresponding expression for a discrete GWB rather than a perfectly isotropic background, which is more realistic for PTA frequency bands.

### 4.4. *Software*

Throughout this work we use `enterprise` (Ellis et al. 2020) with `PTMCMCSampler` (Ellis & van Haasteren 2017) as our primary Bayesian pipeline. `enterprise` provides modular noise models, analytic marginalization over timing-model parameters, and support for both CURN and HD-correlated stochastic backgrounds. We sample in the full CURN hyperparameter space and obtain HD posteriors via importance reweighting following Hourihane et al. (2023).

**Table 1.** Source parameters and Bayes factors for eight targets with a Bayes factor greater than one under the CURN model. The top two candidates are J1536+0441 and J0729+4008. Including HD correlations leaves only these two with Bayes factors greater than one. Digits in parentheses represent uncertainties in the least significant digit. The electromagnetic period has no reported error; thus the uncertainty on $f_{\rm GW}$ is unknown.

| Name | RA | Dec | $z$ | $P_{\rm EM}$ [day] | $f_{\rm GW}$ [nHz] | $\mathcal{B}^{\rm CURN+CW}_{\rm CURN}$ | $\mathcal{B}^{\rm HD+CW}_{\rm HD}$ |
|---|---|---|---|---|---|---|---|
| SDSS J153636.22+044127.0 | 15h 36m 36.20s | +04d 41m 26.9s | 0.38894(4) | 1110 | 20.84 | 3.37(5) | 1.91(4) |
| SDSS J072908.71+400836.6 | 07h 29m 08.60s | +40d 08m 37.0s | 0.07404(2) | 1612 | 14.36 | 2.44(3) | 3.7(1) |
| HS 0926+3608 | 09h 29m 52.10s | +35d 54m 49.6s | 2.15 | 1561 | 2.15 | 1.26(1) | 0.97(2) |
| SDSS J113050.21+261211.4 | 11h 30m 50.20s | +26d 12m 11.8s | 1.01(2) | 2173 | 10.65 | 1.11(2) | 0.71(4) |
| SDSS J140704.43+273556.6 | 14h 07m 04.50s | +27d 35m 56.3s | 2.2158(2) | 1561 | 14.82 | 1.08(5) | 0.89(5) |
| SDSS J081133.43+065558.1 | 08h 11m 33.40s | +06d 55m 58.3s | 1.266(2) | 1586 | 14.59 | 1.10(2) | 1.00(7) |
| SDSS J115141.81+142156.6 | 11h 51m 41.80s | +14d 21m 57.0s | 1.002(1) | 1492 | 15.51 | 1.04(3) | 0.97(3) |
| SDSS J082926.01+180020.7 | 08h 29m 26.00s | +18d 00m 20.7s | 0.810(1) | 1449 | 15.98 | 1.01(7) | 0.96(8) |

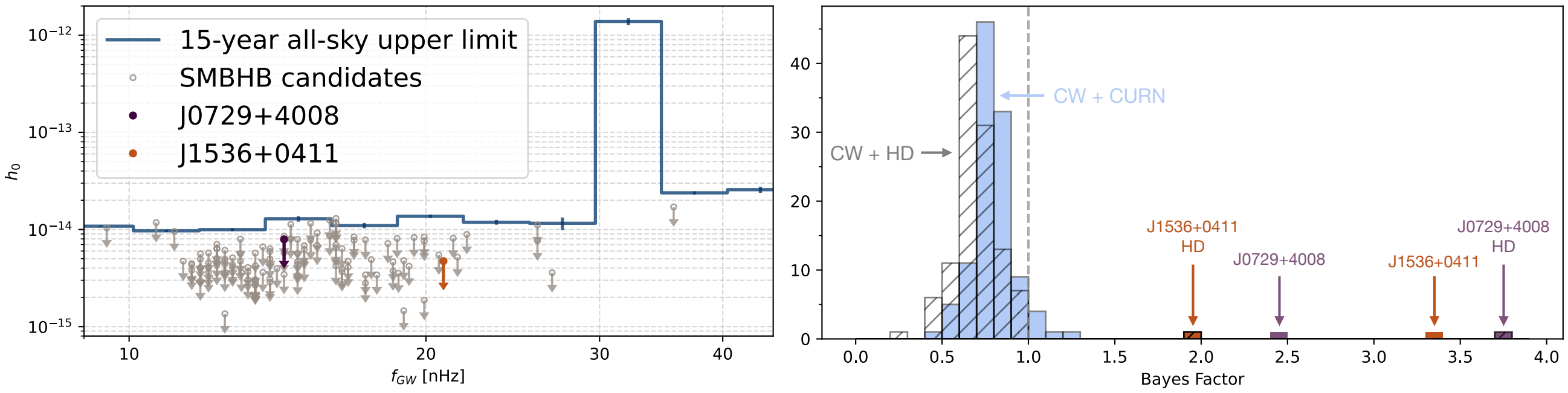

**Figure 1.** Left: 95% upper limits on GW strain from targeted CW+CURN searches, compared to the sky-averaged limits of Agazie et al. (2023c). Right: Bayes factor distributions for all 114 SMBHB candidates under both CW+CURN and CW+HD models. Mean Bayes factors are $0.83 \pm 0.31$ (median 0.78) for CURN and $0.73 \pm 0.32$ (median 0.69) for HD. Two sources stand out: J1536+0441 and J0729+4008. For CW+CURN vs CURN-only the Bayes factors are 3.15(2) and 2.21(2); under HD correlations they are 1.91(4) and 3.7(1), respectively. Both cases are discussed in Section 6.

Signal–coherence tests and dropout analyses are performed with `FurgeHullam` (Bécsy 2024), which implements the incoherent-signal model of Bécsy et al. (2025), constructs the CW/INCOH/NOISE model hierarchy, and provides sky-shuffling and phase-shifting tools for building scrambled null distributions. `FurgeHullam` assumes fixed red noise and a CURN to carry out these analyses rapidly. Dropout probabilities follow the method of Aggarwal et al. (2019).

All analyses use Python workflows consistent with NANOGrav's reproducibility standards, and all scripts for targeted searches, coherence tests, scrambled ensembles, and dropout analyses are available on the Data Repository linked at the end of this manuscript.

#### 4.4.1. *Comparing `QuickCW` and `enterprise`*

For validation and efficiency, we also apply `QuickCW` (Bécsy et al. 2022a), a reformulated CW likelihood built on `enterprise` that separates the CW parameters into slowly varying "shape" parameters and fast "projection" parameters, yielding a speedup of approximately a factor of four.

`QuickCW` has been adapted to operate in a targeted search mode to be described by an upcoming publication (Agarwal et al. in prep), and we briefly describe the adaptation here. To use `QuickCW` in the targeted search setting, we impose narrow priors on sky position and GW frequency, matching the delta-function constraints used in `enterprise`. Because `QuickCW` samples in strain amplitude rather than in luminosity distance, we recover $d_L$ through post-processing: each posterior sample is mapped to an implied $d_L(\mathcal{M}_c, h_0, f_{\rm gw})$ and samples inconsistent with the host-galaxy distance are removed and renormalized. This produces a posterior equivalent to the targeted search likelihood. We also ensure the sampled $\mathcal{M}_c$ range is compatible with the minimum allowed strain amplitude.

`QuickCW` supports sampling in the CURN hyperparameters, enabling the same importance-reweighting step used in `enterprise` to obtain HD-correlated posteriors. With these adaptations, the Bayes factors, upper limits, and posteriors from `QuickCW` agree closely with those from `enterprise`, demonstrating that it is a reliable and efficient framework for large-scale targeted CW searches.

## 5. RESULTS

### 5.1. *Search Overview and Bayes Factor Distribution*

We conducted a targeted CW source search on 114 candidate sources, comprising the CRTS candidates, the two OVRO candidates, and 3C 66B. For all targets, we fixed RA, DEC, $D_L$, and $f_{\rm gw}$ from electromagnetic data, assuming the observed periodicity originates from and traces the binary orbital period when determining $f_{\rm gw}$. At the $\sim$ 10–30 nHz frequencies probed, the GWB is expected to have an amplitude of $A_{\rm GWB} \approx 2.4 \times 10^{-15}$ (Agazie et al. 2023a), though the background is predicted to deviate from a power law and become less prevalent above $\sim$ 26 nHz due to source discreteness (Sesana et al. 2008; Agazie et al. 2025). As such, for each target we compute a Bayes factor comparing a CW + GWB model to GWB alone. We consider the two options of modeling the GWB: either as a CURN or with HD cross correlations, where the HD-correlated model is more physically accurate (Agazie et al. 2023a).

Taking the GWB to be a CURN, we find that there are only eight targets for which the Bayes factor $\mathcal{B}^{\rm CURN+CW}_{\rm CURN} > 1$ (see Table 1). Across the entire target set of 114 sources, the mean Bayes factor is $0.83 \pm 0.31$ with a median of 0.78. Repeating this process for the GWB modeled as an HD-correlated process, we find that the Bayes factor is reduced in 63% of targets, resulting in a mean of $0.73 \pm 0.32$ and a median of 0.32. In this scenario, all targets have Bayes factors consistent with noise; the two with values slightly above unity do not constitute meaningful evidence for a CW signal.

Only J1536+0441 (internally nicknamed "Rohan") and SDSS J0729+4008 ("Gondor") stand out modestly under both scenarios, with Bayes factors given in Table 1. Although these Bayes factors are not statistically significant once the number of targets searched is taken into account, they are the largest outliers in the distribution, and we therefore treat them as case studies. We explore these two candidates in detail in Section 6.

### 5.2. *Chirp Mass and Gravitational Wave Strain Upper Limits for the 114 Targets*

We report upper limits on strain and chirp mass for 114 AGN. Our targeted searches fix $\hat{\Omega}$, $D_L$, and $f_{\rm gw}$, reducing the model dimensionality by four. Full limits are provided in Appendix B, where the upper limits on chirp mass and strain are the 95th quantile with a uniform prior in $\mathcal{M}_c$. These reported upper limits are computed simultaneously with an HD-correlated GWB. Future searches will investigate how the results change, if at all, with different priors.

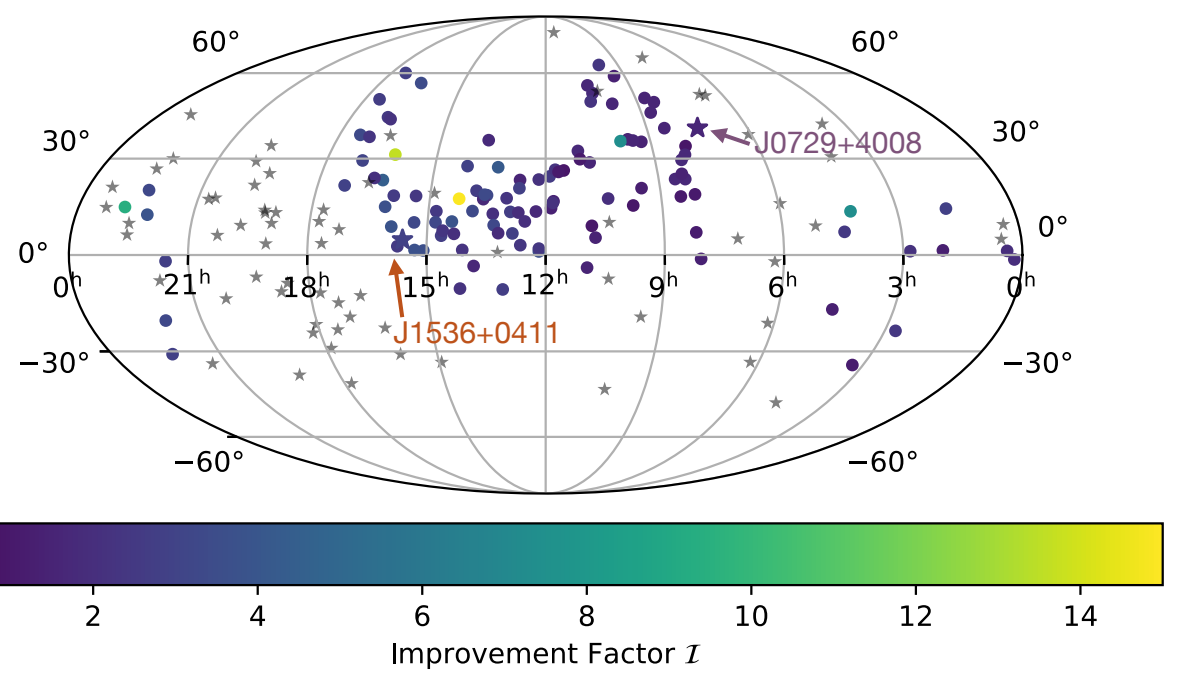

**Figure 2.** Improvement in strain upper limits as a function of sky location for CRTS and OVRO sources examined in our catalog, defined in Equation 21, using a targeted search with a CW + CURN model. The circles are the SMBHB candidates, and the gray stars are the NANOGrav pulsars from Agazie et al. (2023a). The most improved targets are around 15h, but overall there is no clear trend for improvement as a function of sky location.

To quantify how much the targeted searches improve the NANOGrav 15 yr all sky CW strain upper limits, we define an improvement factor between the NANOGrav 15 yr upper limit (NG15) and the targeted search upper limit (TS) as:

$$\mathcal{I} = \rm NG15/TS\,. \tag{21}$$

Across the catalog of 114 candidate SMBHBs, we find an improvement factor, Equation 21, which ranges from $0.82 \leq \mathcal{I} \leq 14.94$, with a median value of 2.24, and a mean of $2.60 \pm 2.01$. Unless otherwise specified, in the following comparisons we use the CURN model of the GWB since the work we comparing to, Agazie et al. (2023c), uses the CURN model.

Overall, we find that the improvements due to the targeted searches are generally larger for closer systems, up to a factor of 15 for $z < 0.5$, but the correlation with $z$ is weak. There does not appear to be an overall improvement in upper limits as a function of sky location, Figure 2, or GW frequency, Figure 4. Within current sensitivity, the improvements are therefore likely driven by fixing host specific parameters rather than by any particular region of the sky or narrow frequency band.

The best overall improvement is for SMBHB candidate SDSS J141425.92+171811.2 at $f_{\rm gw} = 13.0$ nHz and $z = 0.41$. Here the strain upper limit improves from $1.0 \times 10^{-14}$ in the all-sky search to $6.7 \times 10^{-16}$ with the targeted search — a factor of 15×. However, the upper limit increases to $9.77 \times 10^{-16}$ once using an HD-correlated model for the GWB instead of CURN. While this would imply only a 10× improvement over the all-sky results, this is not proper comparison because the all-sky result uses CURN.

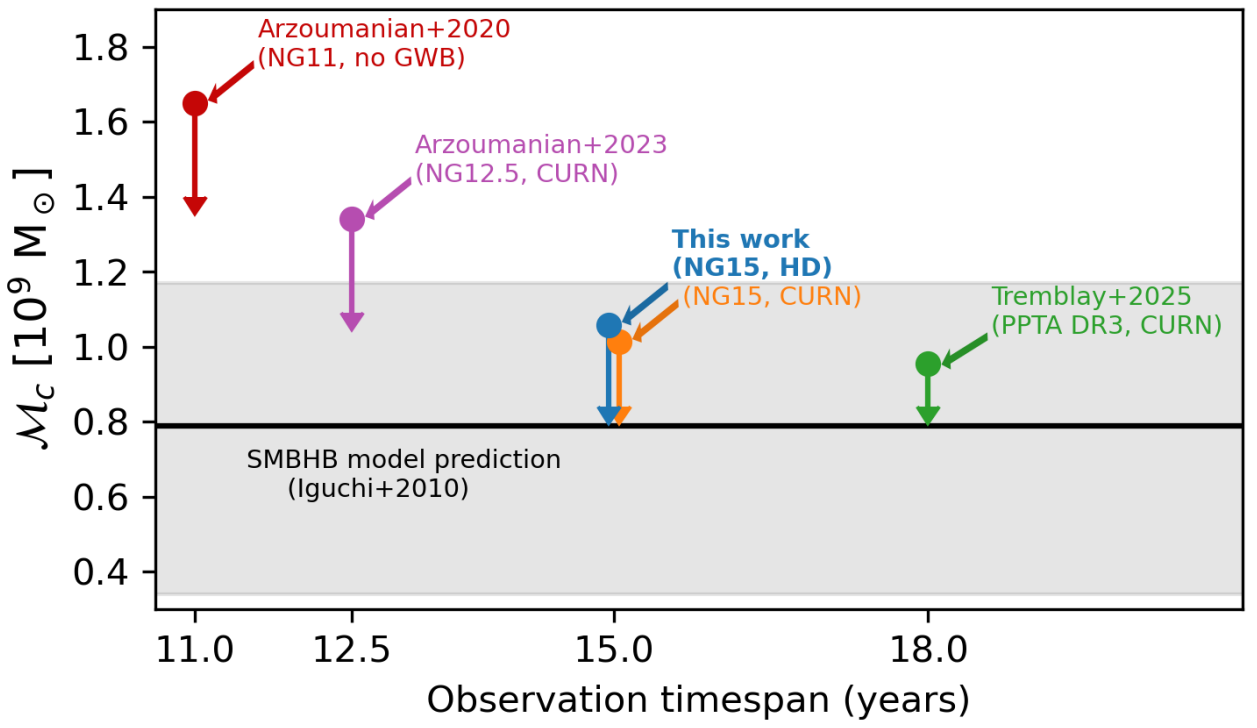

**Figure 3.** Upper limits on the chirp mass of a candidate binary in 3C 66B, assuming a circular binary model. The solid horizontal line marks the best-fit chirp mass prediction from the orbital model of Iguchi et al. (2010), while the surrounding gray shaded band indicates the parameter space allowed by the uncertainties in their astrometric measurements. The 15 yr dataset results in $\mathcal{M}_c^{95\%} = 1.06(3) \times 10^9\ M_\odot$ assuming an HD-correlated model for the GWB. This now constrains the SMBHB model from Iguchi et al. (2010), is a 21% improvement in the chirp mass over the 12.5 yr dataset (Arzoumanian et al. 2023), and is competitive with the limit from PPTA DR3 (Tremblay et al. 2025). Assuming a CURN model for the GWB produces a slightly lower upper limit of $\mathcal{M}_c^{95\%} = 1.01(1) \times 10^9\ M_\odot$, also shown for comparison.

At the other extreme, SDSS J113050.21+261211.4 with $f_{\rm gw} = 10.7,{\rm nHz}$ at $z = 1.01$ shows a degraded upper limit: $9.7 \times 10^{-15}$ for the all-sky search vs. $1.2 \times 10^{-14}$ for the targeted search. This candidate also has the fourth largest Bayes factor in the catalog of $\mathcal{B}^{\rm CURN+CW}_{\rm CURN} = 1.11(2)$ (see Table 1), however after reweighting this Bayes factor is reduced to $\mathcal{B}^{\rm HD+CW}_{\rm HD} = 0.71(4)$ and the upper limit is decreased to $9.0 \times 10^{-15}$ — a minor improvement over the all-sky result.

### 5.3. *Best constraints on 3C 66B*

The new upper limit on the chirp mass of the SMBHB candidate in galaxy 3C 66B, presented here, rules out a region of parameter space previously allowed by electromagnetic observations, demonstrating the power of PTA data to provide independent and complementary constraints on SMBHB masses. The 95% one-sided Bayesian credible intervals on 3C 66B's chirp mass are $\mathcal{M}_c^{95\%} = 1.1(1) \times 10^9\ M_\odot$ or $\mathcal{M}_c^{95\%} = 1.01(1) \times 10^9\ M_\odot$ as found using two independent pipelines based on `QuickCW` and `enterprise` respectively, under the CURN framework. Reweighting the `enterprise` samples to include HD correlations in the GWB model raises the upper limit to $\mathcal{M}_c^{95\%} = 1.06(3) \times 10^9\ M_\odot$. In Figure 3 we see the evolution of this upper limit (assuming a CURN) compared to targeted searches using the NANOGrav 11 yr dataset (Arzoumanian et al. 2020) and 12.5 yr dataset, as well as PPTA DR3 (Tremblay et al. 2025), which has an 18 year timespan. Together with Tremblay et al. (2025), for the first time the CW limits are constraining the chirp mass predicted from the SMBHB model in Iguchi et al. (2010), assuming a circular binary orbit. However, PTA constraints will still be consistent with the SMBHB model if the binary orbit is eccentric, as demonstrated in Agazie et al. (2024); Tian et al. (2025).

## 6. ANALYSIS OF J1536+0441 AND J0729+4008

J1536+0441 and J0729+4008 are the two sources for which the Bayes factors are found to be consistently above unity (see Table 1), and we use them as "test cases" to illustrate how additional electromagnetic information can be brought to bear to assess their reality as GW-emitting SMBH binaries. Here we carry out a series of analyses and tests to assess whether J1536+0441 and J0729+4008 are better explained as genuine SMBHBs or as noise fluctuations and ordinary AGN variability. We combine detailed CW parameter estimation with coherence tests using pulsar scrambling and phase shifts, dropout analyses to diagnose the contribution of individual pulsars and possible noise artifacts, random targeting experiments that quantify the effective trials factor from searching many hosts, consistency checks with GWB amplitude, anisotropy, and population synthesis models, and updated electromagnetic constraints from time domain photometry, spectroscopy, and high–resolution imaging when possible. We examine the following effects: pulsar noise as a potential source of false positives; consistency of these candidates with GWB population synthesis models; and updated electromagnetic data related to optical periodicities for both candidates from CRTS, Zwicky Transient Factory (ZTF, Masci et al. 2019), the Wide-Field Infrared Survey Explorer (WISE, Wright et al. 2010), and The Asteroid Terrestrial-impact Last Alert System (ATLAS, Tonry et al. 2018).

Our tests and their results are summarized in Table 2. We believe that this is a useful roadmap and first step for interpreting low-significance CW outliers in targeted searches, even though the current candidates are fully consistent with noise once the number of targets searched is taken into account. Here we quote the chirp masses of both J1536+0441 and J0729+4008 as the median plus or minus the distance to the 16th and 84th

| Test | J1536+0441 | J0729+4008 | Section | Comments |
|---|---|---|---|---|
| 1. Extended Periodicity | Periodic | Unclear | 6.3.1 | J1536+0441 is still periodic, J0729+4008 is unclear. See Figure 7. |
| 2. Spectral features | Unclear | No | 6.3.2 | No apparent change in J1536+0441's double H$\beta$ lines, but there are changes in the UV end of the slope, Figure 8. J0729+4008's spectrum does not change. |
| 3. CRTS-GWB Consistency | Yes | Yes | 6.4.1 | Between 5 and 8 CRTS sources are expected to be genuine binaries. |
| 4. Population Synthesis | Consistent with GWB | Consistent with GWB | 6.4.2 | J1536+0441 and J0729+4008's total masses are within the 68% credible region at their respective GW frequencies. See Figure 10. |
| 5. GWB Anisotropy | Unclear | Unknown | 6.4.3 | NANOGrav does not search for anisotropy above 10 nHz, but MeerKAT probes up to 21 nHz and finds a small coincident fluctuation at J1536+0441's location. J0729+4008 is not in MeerKAT's field of view. |
| 6. Distinct from GWB | Less distinct | More distinct | 6.4.4 | Less support for J1536+0441 under a GWB model with HD-correlations than a CURN, improved support for J0729+4008. |
| 7. Coherence Tests | $\sim 2.5\sigma$ | $\sim$1.5-1.8$\sigma$ | 6.5.1 | Weak support for coherence in J0729+4008, more support for coherence in J1536+0441. See Figure 11, and Section 6.5.3. |
| 8. Dropout Tests | 48 of 67 pulsars support CWs | 32 of 67 pulsars support CWs | 6.5.2 | J0729+4008's top pulsars are known to have excess noise. See Figure 13. |
| 9. Random Targeting | Indistinguishable from noise | Indistinguishable from noise | 6.5.3 | No support for the two candidates versus random targets. Both candidates are consistent with noise after trials factors corrections. |
| 10. Software Crosscheck | Consistent | Consistent | 6.6 | `enterprise` and `QuickCW` give consistent results. |

**Table 2.** A suite of tests carried out on J1536+0441 and J0729+4008 to try and establish if they are SMBHBs. These tests are divided into electromagnetic, GWB, statistical, and software consistency checks, separated by horizontal lines above. Results are mixed: J1536+0441 remains periodic and shows stronger coherence with cleaner dropout behavior (Tests 1, 7-8), whereas J0729+4008 is better distinguished from an HD-correlated GWB (Test 6). Both are consistent with CRTS–GWB expectations and broadly consistent with population synthesis models (Tests 3–4). Spectral changes are inconclusive (Test 2) and anisotropy constraints are currently uninformative at their frequencies (Test 5), but MeerKAT data at J1536+0441's sky location warrant further investigation. The statistical tests support weak coherence in both candidates (Test 7), but random targeting (Test 9) shows that the results are consistent with noise once accounting for the number of searches conducted.

quantiles. The chirp mass value is therefore the central 68% of the posterior.

### 6.1. *Overview of J1536+0441*

Object J1536+0441 is a $z = 0.38$ AGN and is the furthest from the median Bayes factor of the CRTS population (Figure 1). We find $\mathcal{B}^{\rm HD+CW}_{\rm HD} = 1.91(4)$ at a targeted $f_{\rm gw} = 21$ nHz with a median chirp mass of $\log_{10}\mathcal{M}_c \approx 9.67$ ($4.7 \times 10^9$ $M_\odot$), Figure 5. If J1536+0441 is a genuine SMBHB, it would be slowly evolving with $\dot{f}_{\rm gw} = 16.9^{+8.8}_{-1.2}$ pHz yr$^{-1}$ and $t_c = 470^{+1260}_{-160}$ yr.

This candidate has been studied in more detail electromagnetically, but we had no knowledge of it before carrying out this search. It was originally identified as a binary candidate by Boroson & Lauer (2009) via the $-3,500$ km s$^{-1}$ blueshift between the broad H$\beta$ and narrow [O III] $\lambda$5007 emission lines in its optical spectrum. Such a velocity offset may be a signature of orbital motion under the hypothesis that one black hole in the binary is active and its motion induces a large Doppler shift relative to the rest frame of the host galaxy (Gaskell 1984). Long term spectroscopic monitoring to construct a radial velocity curve (Eracleous et al. 2012; Runnoe et al. 2015, 2017) is consistent with a binary having $M_{\rm tot} > 3.8 \times 10^8$ M$_\odot$ and a separation $a > 0.134$ pc.

This separation is inconsistent with the separation found from the fiducial period (or $f_{\rm GW}$) and mass found here of $a_{\rm rest} \sim 0.017$ pc (Table 3). This tension does not rule out a binary interpretation: the radial veloc-ity technique is primarily sensitive to orbital periods of tens to hundreds of years. A short period binary with sub-pc separation can therefore evade detection in radial velocity monitoring if the broad lines originate from a circumbinary disk. In this scenario, the lack of orbital modulation is expected, while the observed broad-line offsets can be produced by stationary disk asymmetries or spiral arms, consistent with Eracleous et al. (2012) and Runnoe et al. (2015, 2017).

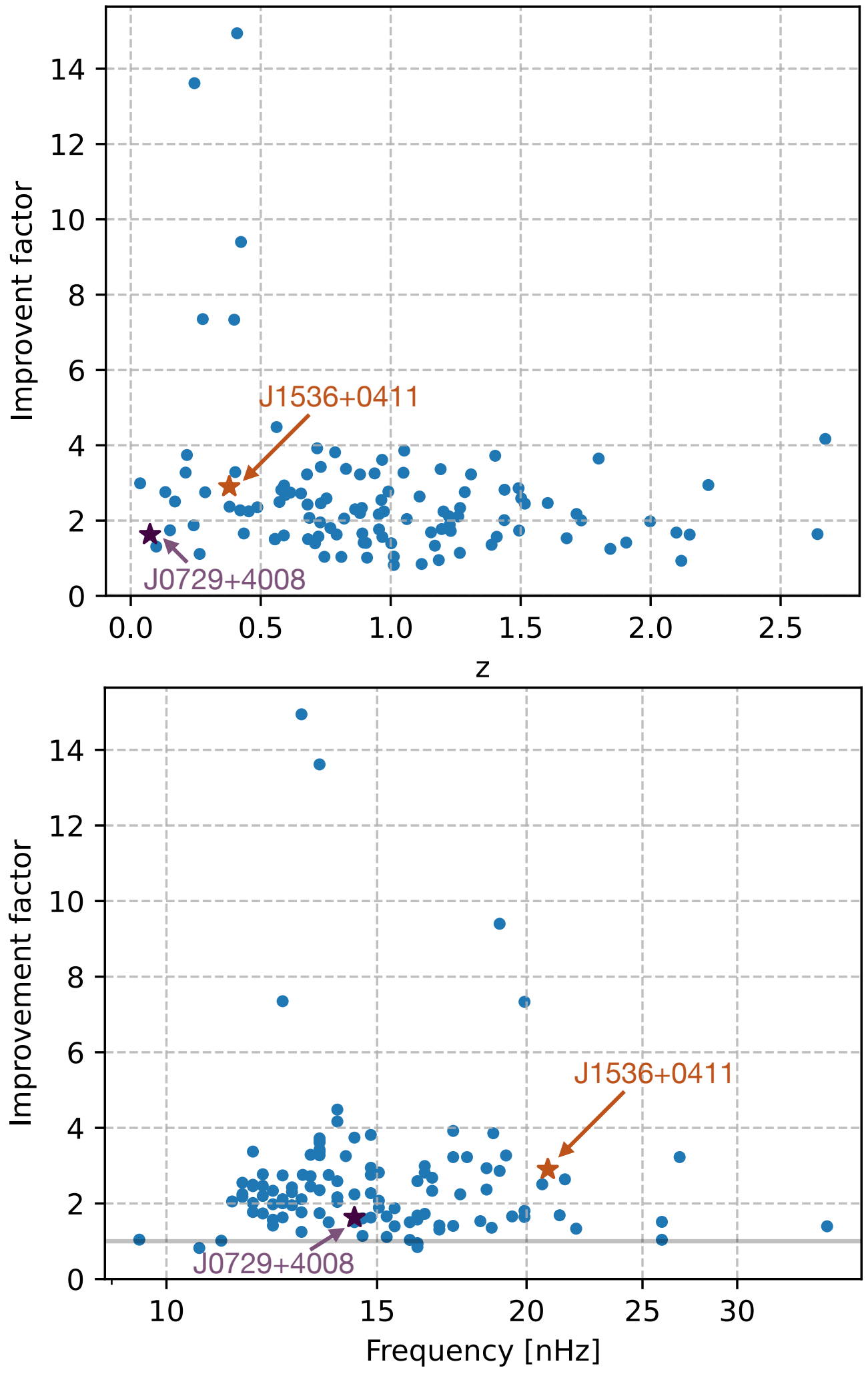

**Figure 4.** Improvements on strain upper limits as a function of $z$ (upper panel) and GW frequency (lower panel) as compared to the all-sky search results, under the CW + CURN model. Above, we see that the targets which are the most improved are at $z < 0.5$. However, below we do not see a clear trend of improvement as a function of frequency, though the most improved sources appear between 13 nHz and 20 nHz.

An ultraviolet spectrum from the Hubble Space Telescope exists for this object, but the broad lines profiles proved too absorbed to address the binary hypothesis (Runnoe et al. 2025).

Furthermore, this candidate appears in high resolution radio and optical imaging. Very Large Array (VLA) observations at 8.5 GHz revealed two radio sources, one consistent with the location of the SDSS source and a secondary, weaker source, with a separation of approximately 5 kpc (Wrobel & Laor 2009). High resolution *Hubble Space Telescope* imaging similarly revealed a companion galaxy at a projected separation of about 5 kpc (Decarli et al. 2009a,b). The broad H$\beta$ emission is potentially associated with a SMBHB within the main galaxy (Lauer & Boroson 2009; Decarli et al. 2009b).

Very long baseline interferometric (VLBI) observations with both the European VLBI Network (EVN, Bondi & Pérez-Torres 2010) and the Very Long Baseline Array (VLBA, Wrobel & Laor 2010; Breiding et al. 2021) showed that each of the two radio sources hosts a compact radio source. The primary radio source is unresolved, indicating that, if it hosts an SMBH binary, the two SMBHs must be closer than 8.5 pc. This constraint is consistent with that implied by the original analysis (separation $\approx 0.1$ pc, Boroson & Lauer 2009), albeit not a strong constraint on the existence of a SMBHB.

The less exotic alternative to the binary scenario for J1536+0441 is that the main galaxy hosts a single radio–quiet (but not radio silent) nucleus that is among the class of well-known disk emitters (Chornock et al. 2010; Gaskell 2010). These objects have complex broad-line profiles consistent with the shape observed for J1536+0441. Recent infrared spectroscopy supports this interpretation, because the He I $\lambda$10830 does not share the profile shape of all the other emission lines (Zhang et al. 2019). This mismatch in line profiles is naturally explained by a single, structured broad-line region rather than by bulk orbital motion of two SMBHs.

### 6.2. *Overview of J0729+4008*

J0729+4008 is a $z = 0.074$ AGN identified by Graham et al. (2015) and was further examined by Charisi et al. (2018), who found that its periodicity was consistent with a relativistic Doppler-boosting model. Doppler boosting can reproduce low amplitude modulations in single SMBH systems as well, so this interpretation is more suggestive than definitive

This AGN shows optical variability with a $\sim 4.5$ yr period. The black hole mass was estimated using the AGN single-epoch virial method by two different teams. Oh et al. (2015) found a mass of $\log(M_{\rm tot}/M_{\odot}) = 7.7$ based on the measured H$\alpha$ line width and using the Greene & Ho (2005) prescription. Later, Liu et al. (2019) found a slightly lower mass combining estimates from both H$\alpha$ (Greene & Ho 2005) and H$\beta$ (Ho & Kim 2015) methods to be $\log(M_{\rm tot}/M_{\odot}) = 7.4$. These virial estimates are known to carry substantial systematic uncertainties particularly when the broad H$\beta$ line is weak, as is the case here (Shen et al. 2011). These caveats make the electromagnetic based mass estimates informative but not decisive. The inferred orbital separation of a binary found from the period and mass found here is $a_{\rm rest} \sim 0.022$ pc. We find no published radio observations analogous to those available for J1536+0441, and the lack of high-resolution radio imaging leaves the host morphology and presence of any jet ambiguous.

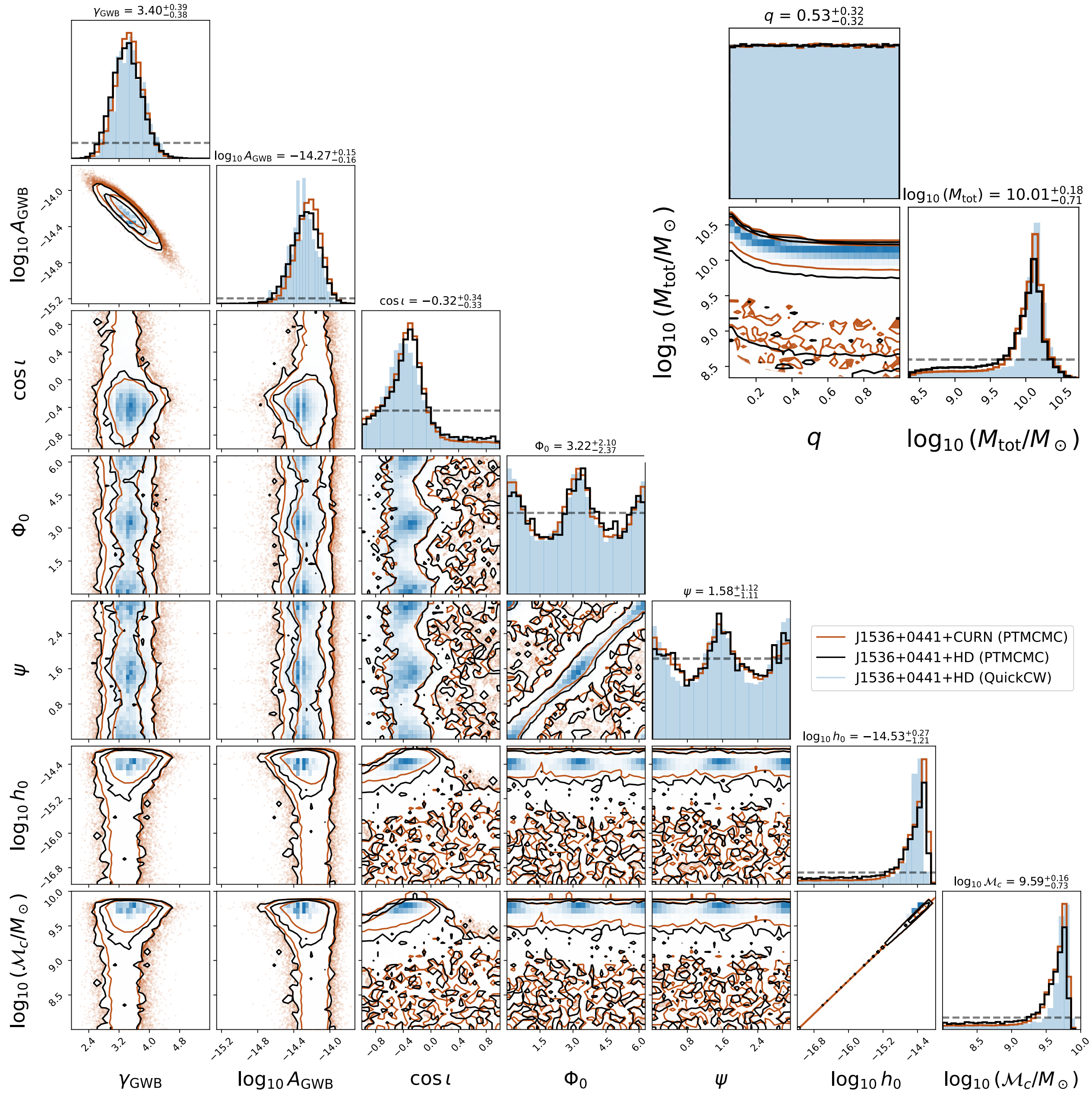

**Figure 5.** SDSS J1536+0441 was identified previously as a sub-pc SMBHB candidate by Lauer & Boroson (2009) with double peaked broad lines, though the existence of this feature did not inform our search. A targeted search was conducted using `PTMCMCSampler` under both an HD-correlated GWB (black) and under a CURN (red), where we report parameter values using HD in the figure. A search was also carried out using `QuickCW` under an HD-correlated GWB (shaded blue) with consistent results. We fix the target RA, DEC, $D_L$, and $f_{\rm GW} = 21$ nHz based on the CRTS periodicity and recover a median chirp mass of $\log_{10}\mathcal{M}_c = 9.59^{+0.16}_{-0.73}$. We find a strain amplitude of $\log_{10} h_0 = -14.53^{+0.27}_{-1.21}$ with modest constraints on the other binary parameters. This candidate shows a slight but statistically insignificant preference for a signal over noise, with a Bayes factor of $\mathcal{B}^{\rm HD+CW}_{\rm HD} = 1.91(4)$. The inset image is a breakdown of the chirp mass into total mass, $M_{\rm tot}$ and mass ratio, $q$, using rejection sampling from Petrov et al. (2024). Here the priors are dashed grey, and one can see that the mass ratio, $q$ is currently unconstrained (prior dominated). Statistics for other targets can be found in Tables 6-8.

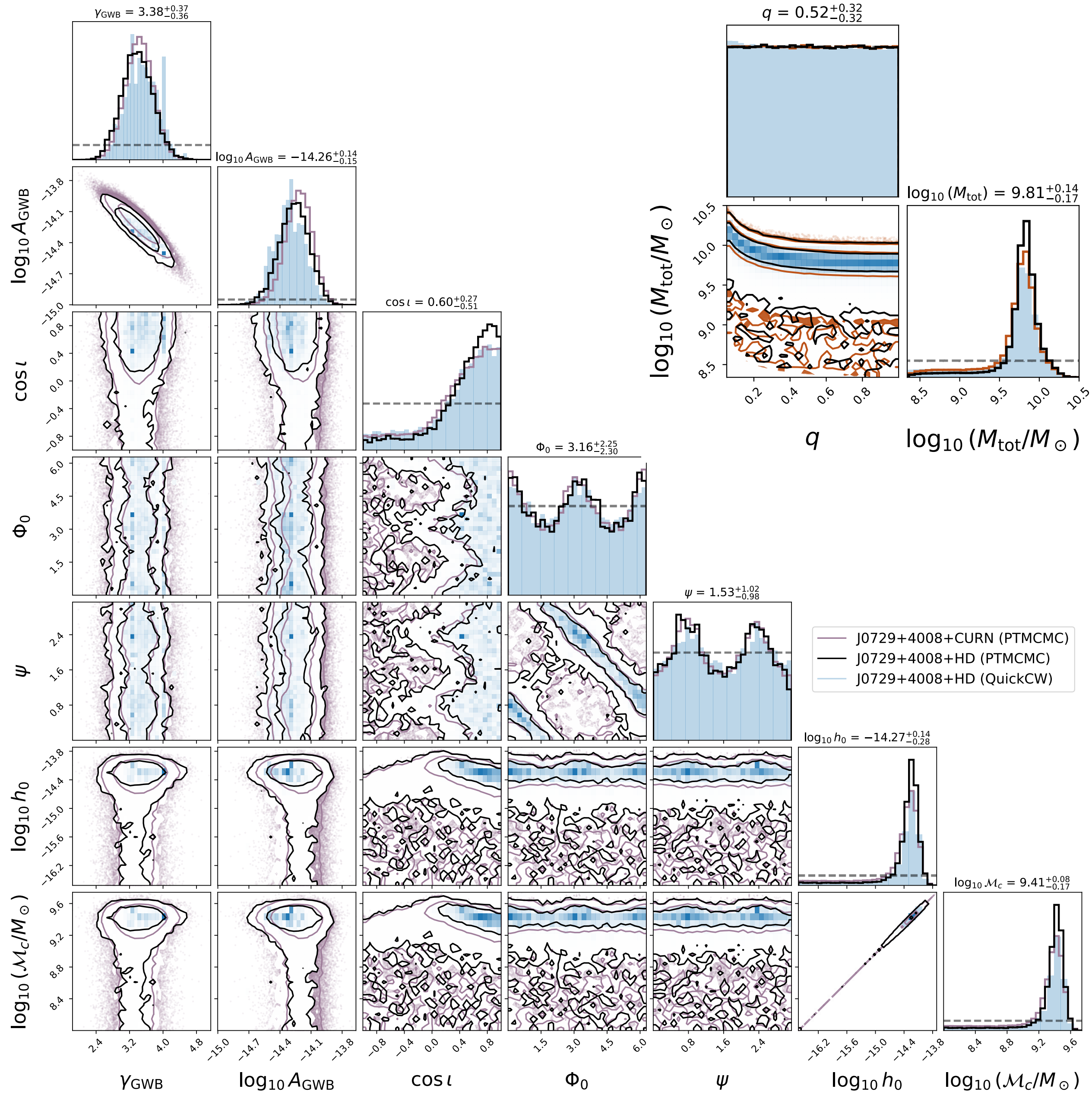

**Figure 6.** J0729+4008 is a type 1.9 quasar, and at $z = 0.074$ it is the closest of the two SMBHB candidates identified in this search. As in Figure 5, we report targeted search results under an HD-correlated GWB (black) and under a CURN (purple), where parameter values in the figure correspond to the HD model. A search was also carried out using `QuickCW` under an HD-correlated GWB (shaded blue) with consistent results. We measure the median value of chirp mass to be $\log_{10}\mathcal{M}_c = 9.41^{+0.08}_{-0.17}\ M_\odot$. The inset shows the decomposition of the chirp mass into total mass, $M_{\rm tot}$, and mass ratio, $q$, using rejection sampling from Petrov et al. (2024). The priors are dashed grey, and we can see that $q$ is unconstrained/dominated by the prior. Interestingly, Charisi et al. (2018) identify J0729+4008 as a weak Doppler-boosted candidate. Furthermore, Guo et al. (2020) pointed out that it has a red SED and a strong broad-line component in H$\alpha$ but a weak broad-line component in H$\beta$. Statistics for other targets can be found in Tables 6-8.

Our search yields $\mathcal{B}^{\rm HD+CW}_{\rm HD} = 3.7(1)$ with a median chirp mass of $\log_{10}\mathcal{M}_c \approx 9.38$ $(2.4 \times 10^9\ M_\odot)$, Figure 6. If J0729+4008 were a genuine binary, the expected GW frequency evolution, Equation 8, would be $\dot{f}_{\rm gw} = 1.28^{+0.67}_{-0.88}$ pHz/yr, and the time to coalescence, Equation 9, would be $t_c = 4120^{+8910}_{-1420}$ yr.

The inferred PTA mass is larger than the electromagnetic estimates. Given the limitations of single-epoch virial masses and the potential for noise leakage or mismodeled pulsar behavior to bias the CW parameters, this discrepancy should not be taken as decisive evidence in either direction. It may reflect intrinsic shortcomings of the virial estimates at low luminosity, residual systematics in the targeted PTA analysis, or simply the fact that the current signal is too weak for stable parameter recovery. If a statistically significant GW signal were identified for this source, extended timing baselines would be required to determine whether the signal arises from astrophysics, noise, or both.

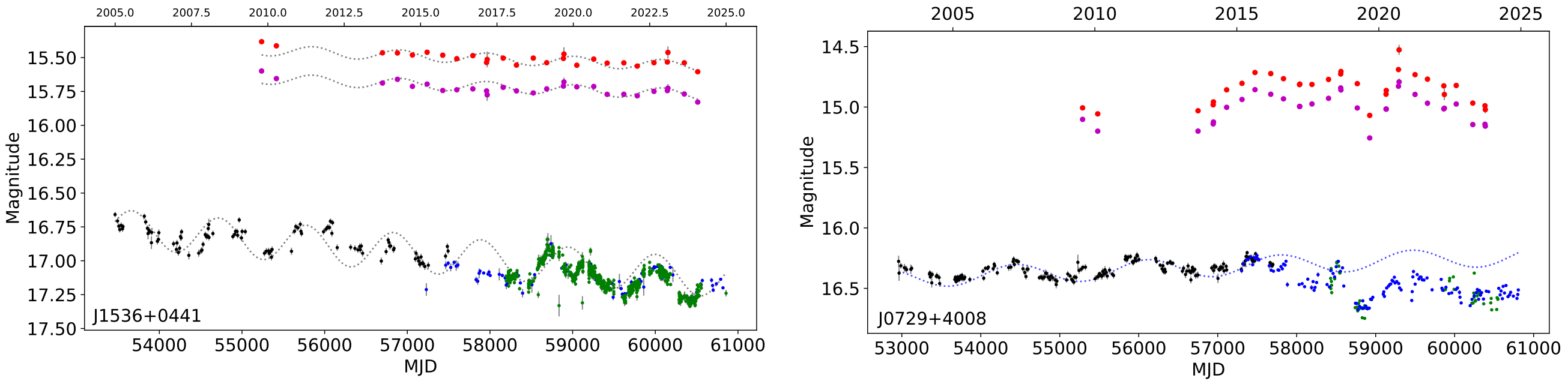

**Figure 7.** Updated light curves for J1536+0441 (left) and J0729+4008 (right), starting with CRTS data (black) and extended by ATLAS (blue) and ZTF (green) data. The blue dashed line shows a sinusoidal signal at the period of the binary candidate. While J1536+0441 is still periodic, J0729+4008 appears to no longer be so. The purple (W2 band) and pink (W1 band) colored dots are WISE data associated with each AGN. Both candidates show WISE modulations proposed by (D'Orazio & Haiman 2017) and could be used to test models for the source continuum variability and the dust geometry.

### 6.3. *Electromagnetic Data*

Electromagnetic data can also test the SMBHB hypothesis for our two candidates by probing whether the photometric and spectroscopic behavior of J1536+0441 and J0729+4008 reflects binary motion or AGN variability. Evaluating the persistence of periodicities and spectral features places the CW results in a broader multimessenger context, and tests the consistency of the binary interpretation.

#### 6.3.1. *Extended Periodicity*

Both J1536+0441 and J0729+4008 exhibit continued modulation over the 10 years since their initial reporting. J1536+0441 has broadly maintained the behavior originally noted to a total now of 6 cycles. Refitting to the extended data suggests a period of 1056 days, which is also shown in its mid-infrared data, albeit with a phase shift of 75 days. It also shows a slow and steady dimming of 0.5 mag over the 20-year baseline, which may reflect long-term variable processes in the disk. Alternatively, this variability could arise from super-orbital cycles driven by overdense "lumps" in a circumbinary disk, as predicted by hydrodynamic simulations that produce $\sim$5-orbit accretion modulations; see e.g. Figure 7 in Duffell et al. 2020). The characteristic timescale of the lump-driven variability can be longer depending on disk viscosity, thickness, and other parameters. However, these interpretations are not unique, and the observed behavior is also compatible with long-term accretion variability unrelated to a SMBHB.

J0729+4008 started to exhibit more short term variability around MJD = 58000, and both the optical and mid-infrared data show flaring activity since then that has overwritten any periodic signal. It remains to be seen if another decade of monitoring data will see the return of modulated behavior. X-ray observations from Chandra and NuSTAR telescopes find nothing to distinguish J0729+4008 from the regular AGN population (Saade et al. 2020, 2024). The disappearance of periodicity does not necessarily rule out a SMBHB, as a low-amplitude periodic component can be intermittently masked by higher-amplitude stochastic variability, and our current statistical tools cannot reliably recover such buried signals. More generally, at least 90% of the CRTS candidates have deviated away from the periodic behavior they exhibited between 2003 and 2014 (Matthew Graham, private communication).

More generally, at least 90% of the CRTS candidates have deviated away from the periodic behavior they exhibited between 2003 and 2014 (Matthew Graham, private communication). This is a pattern seen in other large samples of photometric binary candidates as well. While this certainly reflects a high degree of false positives in these data sets, which we discuss next, we do not have statistical tools able to identify a persistent low amplitude periodic signal occasionally masked by larger amplitude stochastic variability. These objects should continue to be monitored, if only to identify the physical processes, e.g. in the accretion disk, that contribute to short term modulated behavior. This motivates continued electromagnetic monitoring, because long-baseline light curves are required to distinguish be-tween stochastic disk variability and persistent, stable periodic components expected from a SMBHB.

### 6.3.2. *Electromagnetic Spectra*

The H$\beta$ broad emission line forms in gas clouds a few light-weeks to light-months from the accreting black hole. Because the gas is moving at a few thousand km s$^{-1}$, it is a good Doppler tracer of any bulk motion of the SMBH(s) themselves. If a quasar hosts a SMBHB, the H$\beta$ line should either shift periodically, split into two broad moving components, or display asymmetric profiles that evolve coherently with the binary's orbit. Each signature can also be created by a single SMBH, so the litmus test is repeatable, phase-locked behavior across multiple epochs and, ideally, multiple diagnostics (continuum, other lines, VLBI structure).

In Figure 8, we show the broadband spectra for J1536+0441 and J0729+4008, while Figure 9 presents a detailed view of the H$\beta$ and [O III] emission lines. For J1536+0441, while the double-peaked H$\beta$ profile remains stable over the examined range, we observe changes in the UV (blue) continuum slope as well as in the shape of the H$\alpha$ complex and its components. The continuum variations are most naturally explained by accretion-rate or spectral energy distribution changes, whereas the modest H$\alpha$ variations may reflect emissivity or geometry effects rather than orbital motion. The limited diagnostic value of the H$\beta$ line in this object is compounded by its weakness and by well-known systematics in single-epoch virial mass estimators (e.g. Shen & Loeb 2010; Shen et al. 2011). These issues make it difficult to use the current spectroscopy to meaningfully confirm or reject a SMBHB scenario in J1536+0441.

In contrast, J0729+4008 shows no significant changes in continuum slope or in the H$\alpha$ and H$\beta$ line strengths or profiles, consistent with either a steady accretion state or a binary geometry in which velocity signatures are suppressed. Binary configurations with low inclination or small mass ratio can yield absorption dominated or symmetric line profiles without measurable radial velocity evolution, so the lack of line shifts here does not discriminate strongly between a single SMBH and a SMBHB. Overall, the spectral variability remains inconclusive with respect to a SMBHB interpretation. Long-baseline, multi-epoch spectroscopy is required to test for the phase-locked behavior expected in a SMBHB (Eracleous et al. 2012), and will be essential for distinguishing orbital evolution from normal AGN variability in both systems.

### 6.3.3. *Imaging SMBHB Candidates*

To gauge the possibility of imaging structures on the scale of the binary orbit, we compute the characteristic angular separations that the orbits of candidate binaries in J1536+0441 and J0729+4008 subtend on the sky. To do so we calculate the orbital semi-major axis by assuming that the observed electromagnetic variability period is the redshifted binary orbital period and that the total redshifted binary masses inferred in the CW analysis are those quoted by the range in the top-right insets of Figures 5 and 6. Over this range of masses we determine the range of rest-frame semi-major axes. The angular separation on the sky follows from dividing these by the angular diameter distance of the source at the quoted redshifts (1082.2 Mpc for J1536+0441 and 292 Mpc for J0729+4008).

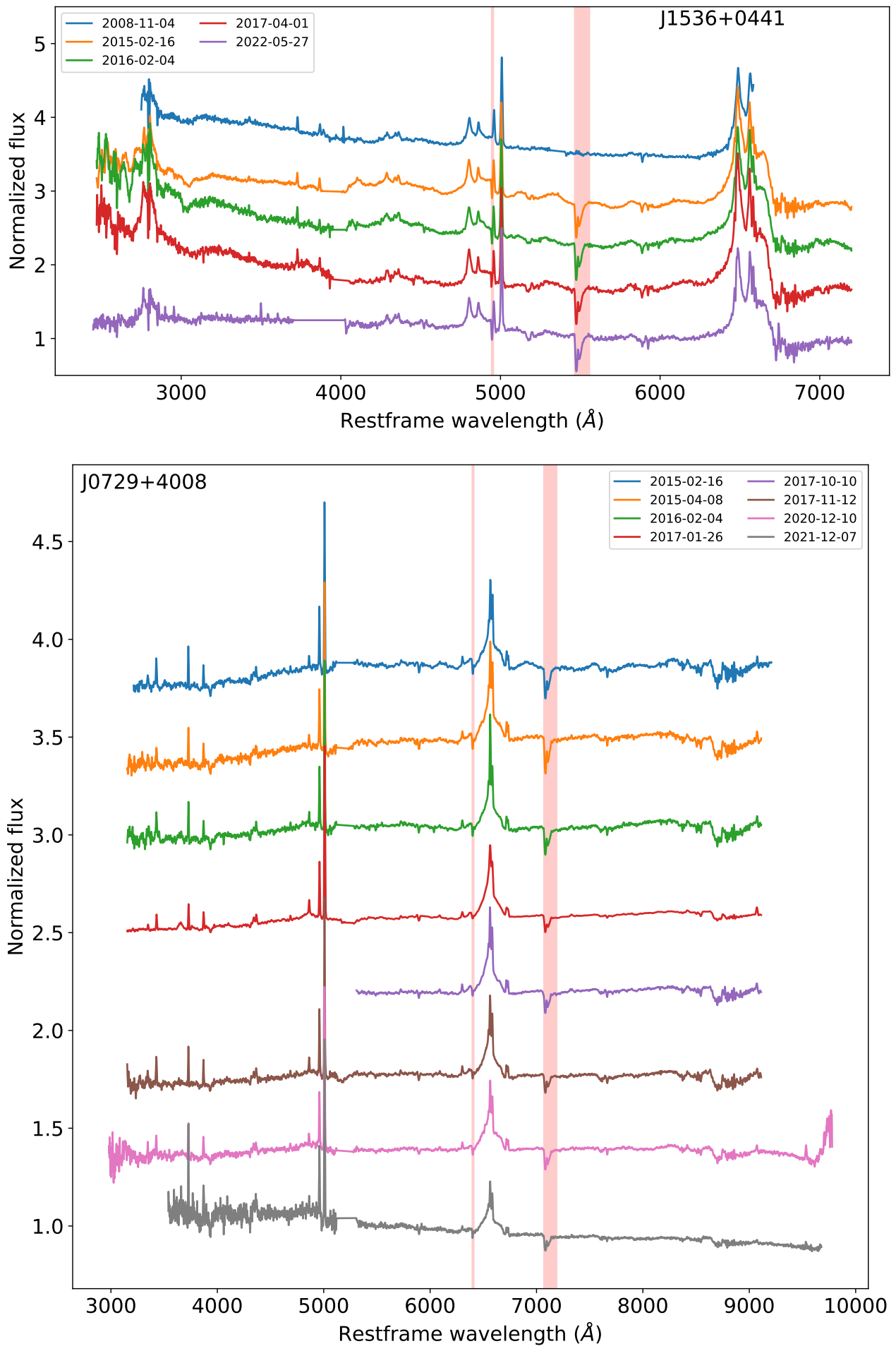

**Figure 8.** Spectra for J1536+0441 (top) and J0729+4008 (bottom). Both sets of spectra were normalized to the [OIII] flux, and the pink bands are the telluric A bands. For J1536+0441, there are no real changes in the double H$\beta$ over the examined range, however, there are changes in the slope of the UV end of the continuum. J0729+4008's spectrum does not seem to be changing in terms of continuum slope or line strengths or profiles for H$\alpha$ or H$\beta$.

| Parameter | J1536+0441 | J0729+4008 | Comments |
|---|---|---|---|
| $z$ | 0.379 | 0.074 | Redshift. Also see Figure 4. |
| Period | 1111 days | 1612 days | Period of optical variability reported in Graham et al. (2015). J1536+0441's best fit period has since decreased slightly and J0729+4008 is no longer periodic, see Section 6.3.1 and Figure 7. |
| $f_{\rm gw}$ | 21 nHz | 14 nHz | Assuming a 1:1 correspondence between optical and orbital periods. see Section 2.1 and Figure 4. |
| EM $M_{\rm tot}$ | $> 6.61 \times 10^8\ M_\odot$ | $2.63 \times 10^7\ M_\odot$ | Total mass estimated from spectra using H$\alpha$ and H$\beta$ line widths by Shen et al. (2011) and Liu et al. (2019), respectively. See Section 6.1. |
| $a_{\rm rest}$ | 0.017 pc | 0.022 pc | Rest frame binary semi-major axis, calculated from rest frame orbital period and total binary mass. See Section 6.3.3. |
| $\theta_a$ | 3.2 $\mu$as | 15.3 $\mu$as | Angular size of binary orbit corresponding to $a_{\rm rest}$. See Section 6.3.3. |
| $\mathcal{B}^{\rm CURN+CW}_{\rm CURN}$ | 3.15(2) | 2.21(2) | Bayes factor for CURN + CW model vs CURN-only model with log-uniform $\mathcal{M}_c$ prior. See Figure 1 and Section 5.1. |
| $\mathcal{B}^{\rm HD+CW}_{\rm HD}$ | 1.91(4) | 3.7(1) | Bayes factor for HD + CW model vs HD-only model with log-uniform $\mathcal{M}_c$ prior. See Figure 1 and Section 6.4.4. |
| $\mathcal{B}^{\rm HD+CW}_{\rm CURN+CW}$ | 110(4) | 300(10) | Bayes factor for HD + CW model vs CURN + CW model. See Section 6.4.4. |
| $\log_{10}\left(\mathcal{M}_c^{\rm HD}/M_\odot\right)$ | $9.59^{+0.16}_{-0.73}$ | $9.41^{+0.08}_{-0.17}$ | Median of $\mathcal{M}_c$ posterior for HD + CW model with log-uniform $\mathcal{M}_c$ prior. See Section 6.1 and Figure 5 for J1536+0441 and Figure 6 for J0729+4008. |
| $\log_{10}\left(M_{\rm tot}^{\rm HD}/M_\odot\right)$ | $10.01^{+0.18}_{-0.71}$ | $9.81^{+0.13}_{-0.17}$ | Total binary mass from decomposing $\mathcal{M}_c^{\rm HD}$ into $M_{\rm tot}$ and mass ratio $q$. See insets in Figure 5 for J1536+0441 and Figure 6 for J0729+4008. |
| $\log_{10}\left(\mathcal{M}_c^{\rm CURN}/M_\odot\right)$ | $9.67^{+0.11}_{-0.34}$ | $9.38^{+0.11}_{-0.30}$ | Median of $\mathcal{M}_c$ posterior for CURN + CW model with log-uniform $\mathcal{M}_c$ prior. See Section 6.1 and Figure 5 for J1536+0441 and Figure 6 for J0729+4008. |
| $\log_{10}\left(M_{\rm tot}^{\rm CURN}/M_\odot\right)$ | $10.05^{+0.16}_{-0.89}$ | $9.75^{+0.17}_{-0.99}$ | Total binary mass from decomposing $\mathcal{M}_c^{\rm CURN}$ into $M_{\rm tot}$ and mass ratio $q$. See insets in Figure 5 for J1536+0441 and Figure 6 for J0729+4008. |

**Table 3.** Parameters of interest for J1536+0441 and J0729+4008. The upper section of the table report measured or inferred electromagnetic data or information, while the bottom of the table report the inferred CW parameters for both binary candidates. The EM period and derived quantities have no reported errors.

Angular sizes of $\sim 15\,\mu$as for J0729+4008's orbit and $\sim 3\,\mu$as for J1536+0441 are at the limit of the astrometric tracking resolution of experiments such as the EHT (Event Horizon Telescope Collaboration et al. 2019; D'Orazio & Loeb 2018) and the GRAVITY+ experiments (e.g., GRAVITY Collaboration et al. 2021; Dexter et al. 2020). These estimates should be regarded as order-of-magnitude indicators rather than imaging predictions, since they rely on an assumed mapping between periodicity and orbital motion and on CW-inferred masses which have large uncertainties. Moreover, whether such experiments could detect or track either source depends sensitively on their fluxes, compactness, and variability at the relevant wavelengths, none of which are yet well constrained. In this sense, the angular scales illustrate that direct astrometric tests of orbital motion are not excluded in principle, but will require future dedicated high resolution observations.

### 6.4. *Consistency with the GWB*

While individual CW signals may have weak evidence, their collective properties should reflect the underlying population responsible for the stochastic GWB. Here we assess whether J1536+0441 and J0729+4008 are consistent with the observed GWB measurements from the NANOGrav 15 yr data. We start with a cross-check of the number of genuine binaries expected in CRTS needed to be consistent with the GWB amplitude, and we then compare J1536+0441 and J0729+4008's mass and frequency to those expected from population synthesis models of the GWB. We also explore if J1536+0441 and J0729+4008 should induce measurable GWB anisotropy, and check how introducing a Hellings-Downs correlated GWB noise model affects the candidates' Bayes factors.

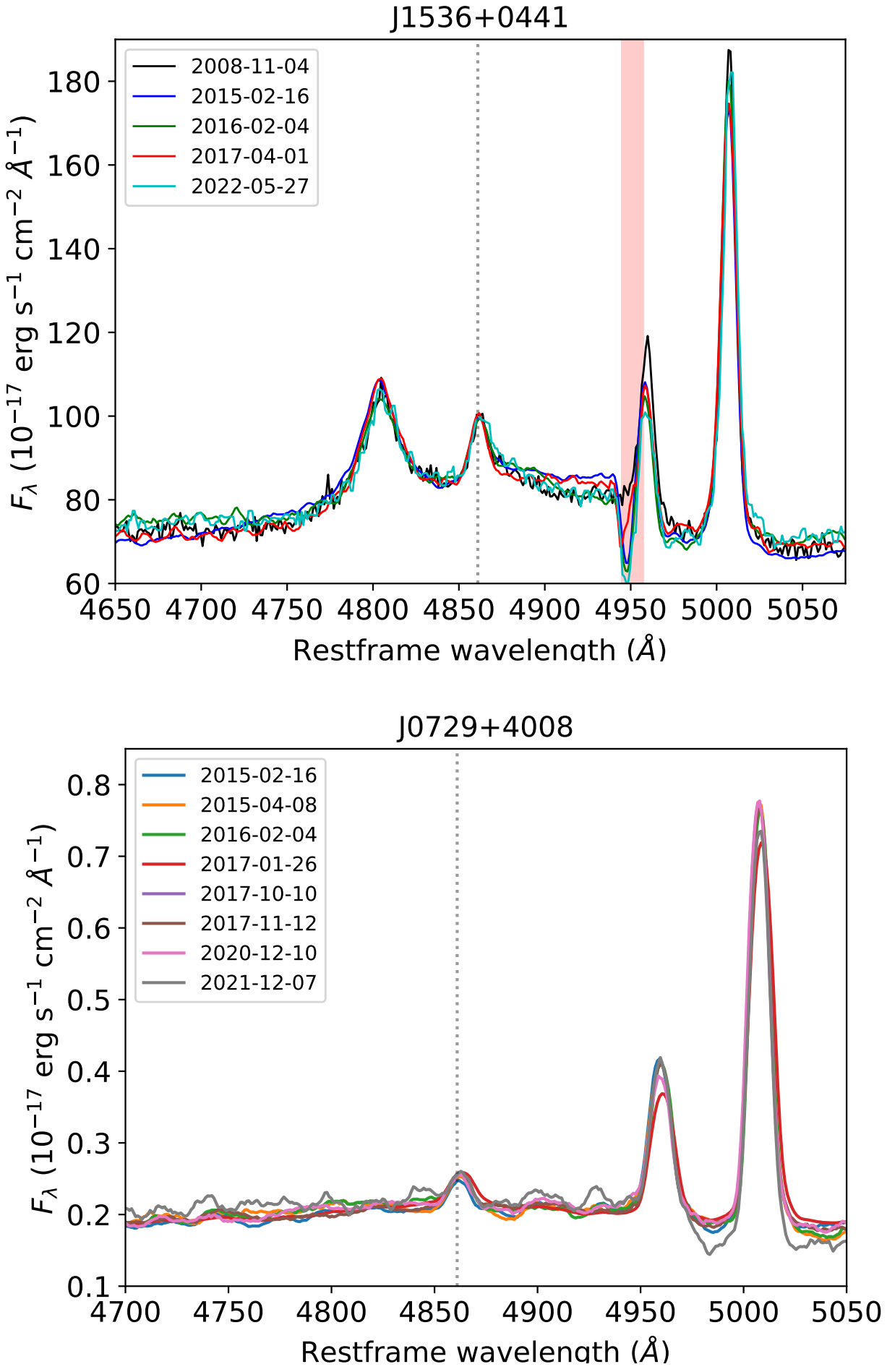

**Figure 9.** Detailed spectral view of the H$\beta$-[O III] region for J1536+0441 (top) and J0729+4008 (bottom). The vertical dotted lines indicate the rest-frame wavelength of H$\beta$, set by the narrow [O III] doublet. For J1536+0441, the double-peaked broad H$\beta$ profile remains stable, without significant velocity shifts over the 14 year baseline of observations (2008–2022). Similarly, J0729+4008 exhibits a stable broad H$\beta$ line profile with no discernible radial velocity evolution between 2015 and 2021.

#### 6.4.1. *CRTS and the GWB*

The primary source of the nHz GWB is likely SMBHBs in their slow inspiral phase. Its amplitude and strain spectrum can therefore give us insights into the cosmic population of SMBHBs (Phinney 2001; Sesana et al. 2008; Casey-Clyde et al. 2022). Any nHz GW emitting SMBHB also contributes to the background at some level, and should therefore have a mass and frequency consistent with GWB population synthesis models. These models should, in turn, yield the correct amplitude of the GWB from e.g. Agazie et al. (2023a).

The SMBHB occupation fraction is the fraction of galaxies that harbor SMBHB systems emitting CWs. Sesana et al. (2018) showed that the CRTS sample was definitively contaminated by false positives, or else it would create a SMBHB population in tension with the GWB upper limits at the time (Arzoumanian et al. 2016). Kelley et al. (2019) further showed that hydrodynamic variability models predict that $5^{+20}_{-1}$ SMBHBs should appear in the CRTS data, compared to only $0.2^{+0.6}_{-0.05}$ expected from Doppler boosting alone, while Casey-Clyde et al. (2025) constrained the maximum number of genuine CRTS binaries to eight. We emphasize that none of these population-level constraints were included as priors in our targeted search; rather, we use them here to provide astrophysical context. Within this framework, our finding of two Bayes factors slightly above unity is fully consistent with expectations from population synthesis models, including the strong contamination of the CRTS sample and the predicted small number of genuine binaries, and the identification of J0729+4008 plausibly explained by Doppler boosting is likewise compatible with these models. These considerations reinforce that our candidates are not statistically or astrophysically anomalous once the expected false-positive rate of the CRTS sample is accounted for.

#### 6.4.2. *Population synthesis models*

In Figure 10, we use the GWB model from Chen et al. (2019) to decompose the GWB amplitude into its constituent number of SMBHBs expected at nHz frequencies as a function of mass. We scale the model to give the GWB amplitude from the NANOGrav 15 yr data (Agazie et al. 2023a), $A = 2.4 \times 10^{-15}$ at a reference frequency of 1/yr, assuming SMBHBs source the GWB with spectral index $\gamma = 13/3$.

While both J1536+0441 and J0729+4008 are consistent with the GWB constraints explored above, J1536+0441's median total mass is rather large compared to the population model with $\log_{10} \mathcal{M}_{\rm tot} = 10.05^{+0.16}_{-0.89}$ at 21 nHz, Figure 10. In fact, only a few CW sources are expected within the lower part of its 68% credible region. If this is the case, future PTA data releases may lead to a lower mass measurement. Electromagnetic observations detailed in Section 6.1 assert that the lower total mass limit for J1536+0441 is $M_{\rm tot} > 3.8 \times 10^8 M_\odot$, also consistent with GWB population synthesis models.

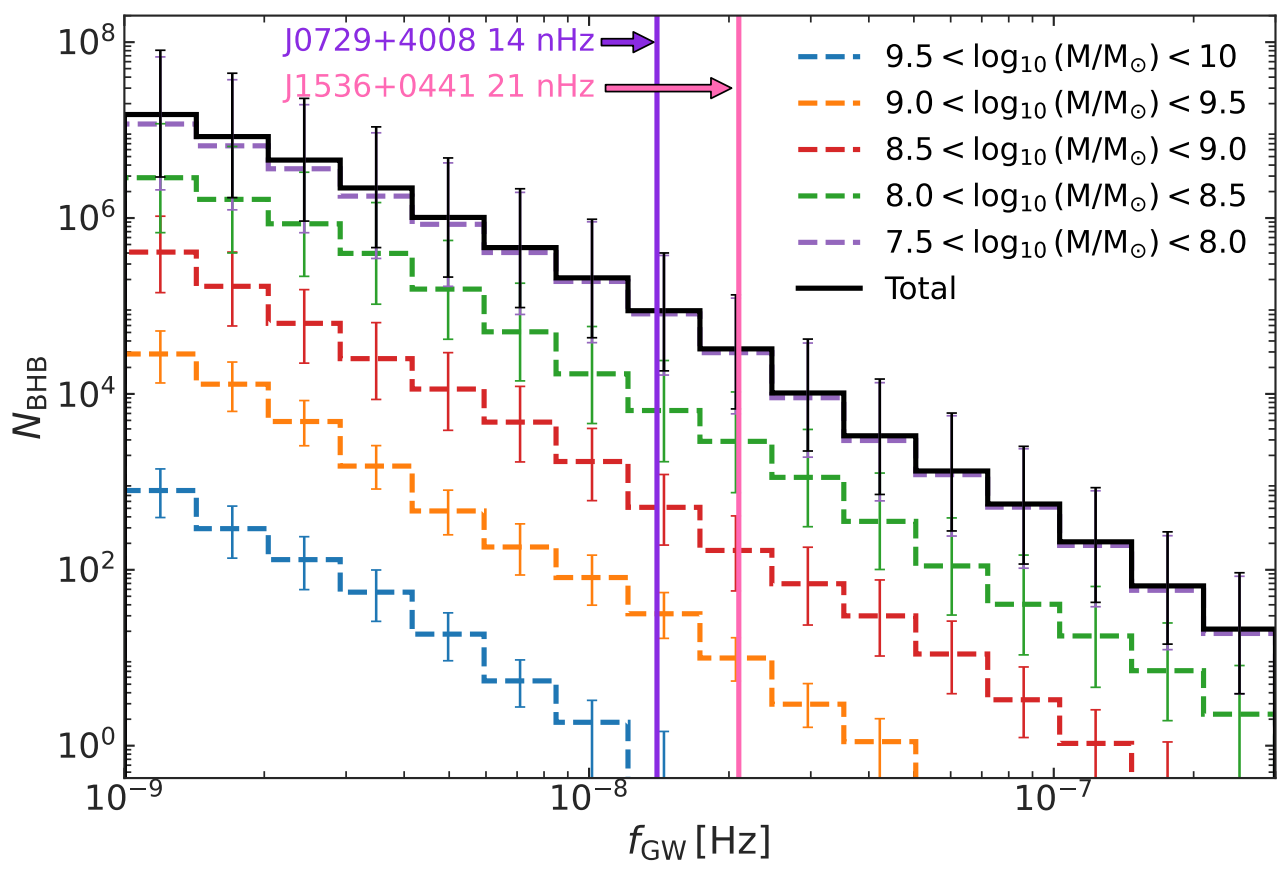

**Figure 10.** Expected SMBHB counts consistent with the NANOGrav 15 yr GWB (Agazie et al. 2023a), from the Chen et al. (2019) population model. The vertical lines indicate the frequencies of our two candidates. J1536+0441 (pink) has $\log_{10} M_{\rm tot} = 10.05^{+0.16}_{-0.89}$; at this mass and frequency (21 nHz), the model predicts $< 1$ source, placing the candidate in the rarefied high-mass tail of the population. J0729+4008 (purple) has $\log_{10} M_{\rm tot} = 9.75^{+0.17}_{-0.99}$ at 14 nHz; the model predicts $\sim 1$ source at the median mass, but the 68% credible region allows for up to $\sim 100$ sources, making it consistent with the bulk population.

J1536+0441's large median total mass at 21 nHz may also indicate that our GWB models should be revised to allow for more massive SMBHBs, also suggested by Sato-Polito et al. (2024). More broadly, both systems fall within the parameter space permitted by population synthesis models once their substantial astrophysical and modeling uncertainties are included, and neither candidate sits in tension with the number or mass distribution of binaries required to reproduce the NANOGrav GWB.

### 6.4.3. *GWB Anisotropy*

If J1536+0441 and J0729+4008 are genuine SMBHBs, their presence may introduce anisotropies in the GWB signal, e.g. Mingarelli et al. (2013); Gardiner et al. (2024). This is especially relevant for massive binaries like J1536+0441, and J0729+4008, which is very nearby (Mingarelli et al. 2017).

Previously, Taylor & Gair (2013); Mingarelli et al. (2017) showed that a single bright CW source should imprint a characteristic $\ell$-independent signature on the angular power spectrum, $C_\ell$. In Section 4.3, Equation 20, we derived a new and simple expression for $C_\ell/C_0$ which we can use to assess if a CW should have been measurable in the GWB's angular power spectrum.

The full observed signal contains both a deterministic component from the CW source and a stochastic component from the unresolved source population, both contributing with power at all angular scales (see Appendix A). Any measurement of the $C_\ell$ spectrum captures the combined effect of both the deterministic CW and the stochastic background. Identifying the CW requires separating its flat spectrum from the stochastic shot noise, which in turn depends on the unknown properties of the source population. This additional stochastic contribution can significantly affect the anisotropic power, potentially masking the characteristic flat spectrum expected from a CW source, see e.g. Figure 2 of Mingarelli et al. (2017) and Figure 5 of Burke-Spolaor et al. (2019). In the most idealized case where the CW source sits atop a perfectly smooth isotropic background, $C_{\ell>0}/C_0 \approx 0.64$ for J1536+0441 and 0.68 for J0729+4008.

A search for such anisotropies in the NANOGrav 15 yr data set found no significant evidence, however the search only explored GW frequencies between 2 nHz and 10 nHz. If they are binaries, neither J1536+0441 nor J0729+4008's signals would have been present since their GW frequencies were not explored, however it is worth reiterating that the posteriors on higher-order multipoles are entirely prior dominated (Agazie et al. 2023), preventing any data-informed inference of the $C_\ell$ spectrum shape. Therefore, NANOGrav $C_\ell$ measurements are currently uninformative for distinguishing between CW and stochastic scenarios. Even future detection of $C_\ell$ would require careful modeling to separate the CW imprint from the stochastic component.

MeerKAT's 4.5 year anisotropy maps (Grunthal et al. 2025), however, do probe the $f_{\rm GW} = 21\,{\rm nHz}$ and $f_{\rm GW} = 14\,{\rm nHz}$ bins relevant for J1536+0441 and J0729+4008. J0729+4008's sky position lies outside MeerKAT's field of view, whereas J1536+0441 is within it. In the MeerKAT anisotropy maps at $f_{\rm GW} \simeq 21\,{\rm nHz}$, J1536+0441's position coincides with an S/N$\sim$2 fluctuation which is fully consistent with isotropic GWB fluctuations. This coincidence at 21 nHz will be monitored in upcoming data releases.

### 6.4.4. *GWB Confusion*

Our analysis framework first searches for CWs in the presence of a CURN, which only models the autocorrelated component of a GWB, and then switches to the HD-correlated GWB model using likelihood reweighting (Hourihane et al. 2023). Comparing these two stages helps diagnose how cross-correlations influence the inferred CW signal and whether any apparent CW support may instead be tied to the treatment of the GWB. The HD model is physically preferable because the GWB induces cross-correlations between pulsars following the Hellings–Downs curve, and in the NANOGrav 15 yr dataset these cross-correlations are strongly favored over a CURN, with a Bayes factor of $\sim$ 220 to $\sim$ 1000 depending on the Fourier basis used (Agazie et al. 2023a). Agazie et al. (2023c) also found that the Bayes factor for a candidate CW at 4 nHz decreases when a CURN analysis is reweighted to include HD correlations. Ferranti et al. (2025) further observe this effect in simulations without any CW present, arguing that a mismodelled GWB may leak power into the CW model. While this leakage has been observed most clearly at the lowest PTA frequencies, it could plausibly occur up to the "knee" frequency of $\sim$ 26 nHz for an astrophysical GWB (Agazie et al. 2025). At the same time, individual CWs contribute to the GWB and are expected on average to generate HD-like correlations (Cornish & Sesana 2013; Bécsy et al. 2022b), making it non-trivial to determine when an emerging CW becomes distinguishable from the background. It is therefore natural to consider how J1536+0441 and J0729+4008 behave under both CURN and HD assumptions.

For the J1536+0441 search, reweighting from CURN to HD decreases the CW Bayes factor from $\mathcal{B}^{\rm CURN+CW}_{\rm CURN} = 3.15(2)$ to $\mathcal{B}^{\rm HD+CW}_{\rm HD} = 1.91(4)$. This reduction suggests a degeneracy between the GWB and the CW at the sky location and frequency of J1536+0441. It is possible that the higher Bayes factor under the CURN model reflects the CW model absorbing unmodelled cross-correlations in the GWB (Agazie et al. 2023c; Ferranti et al. 2025). However, the opposite scenario cannot be excluded: in the low signal regime, a real but weak CW may be partially absorbed by the cross-correlated GWB model, reducing its apparent significance. The reweighted Bayes factor for the HD-correlated GWB itself is also lower than the corresponding value with no CW present, $\mathcal{B}^{\rm HD+CW}_{\rm CURN+CW} = 110(4)$, reinforcing that J1536+0441 and the GWB model are not cleanly separable. Reweighting to the HD model also decreases the upper limit on J1536+0441's chirp mass to $\mathcal{M}_c^{95\%} = 6.5(1) \times 10^9\, M_\odot$. This behavior underscores that modest CW support at this frequency is sensitive to the treatment of the GWB and should be interpreted cautiously.

For the J0729+4008 search, reweighting from CURN to HD increases the CW Bayes factor from $\mathcal{B}^{\rm CURN+CW}_{\rm CURN} = 2.21(2)$ to $\mathcal{B}^{\rm HD+CW}_{\rm HD} = 3.7(1)$, with no appreciable change to the chirp mass upper limit. This suggests that J0729+4008's apparent signal is unlikely to originate from the GWB, and that improved modelling of the cross-correlated background helps separate the two contributions. The reweighted Bayes factor for the HD-correlated GWB model increases to $\mathcal{B}^{\rm HD+CW}_{\rm CURN+CW} = 300(10)$, indicating that an unmodelled CW component was introducing additional common noise in the CURN-only analysis. Whether this behavior is expected from a genuine CW or could arise from mismodelled pulsar-specific noise remains unclear; this will be revisited with future data and simulations.

**Table 4.** Bayes factors between various models considered for the coherence test: CWs, an incoherent CW, and noise (which is the CURN), see Section 6.5.1. Both candidates show mild support for coherence.

| Object | $\mathcal{B}^{\rm CW}_{\rm NOISE}$ | $\mathcal{B}^{\rm INCOH}_{\rm NOISE}$ | $\mathcal{B}^{\rm CW}_{\rm INCOH}$ |
|---|---|---|---|
| J1536+0441 | $3.58^{+0.33}_{-0.30}$ | $0.69^{+0.03}_{-0.03}$ | $5.22^{+0.71}_{-0.63}$ |
| J0729+4008 | $0.77^{+0.05}_{-0.04}$ | $0.50^{+0.02}_{-0.02}$ | $1.53^{+0.17}_{-0.15}$ |

### 6.5. *Statistical Tests*

To assess whether the modest Bayes factors for J1536+0441 and J0729+4008 reflect genuine phase coherent signals or ordinary noise fluctuations, we turn to a suite of statistical tests that probe coherence, informativeness of the targets, and pulsar-by-pulsar support for a CW. These tests target different failure modes. Coherence tests evaluate whether the putative signal shares a common phase evolution across the PTA. Dropout analyses reveal whether the apparent signal is driven by a small subset of pulsars. Random targeting experiments quantify how often similar features arise by chance when many hosts are examined. This last test innately addresses the multiple comparisons problem in which multiple analyses are carried out on the same dataset. These tools provide a sharper view of whether a given candidate behaves as a real SMBHB would or whether the features are more naturally explained by noise.

#### 6.5.1. *Results of Coherence Tests*

We carry out a test of signal coherence via model scrambling, see Section 4.1. Figure 11 shows the null distribution of Bayes factors for both candidates produced via 1000 phase–scrambling runs (blue dashed) and 300 sky–shuffling runs (green dotted). Also shown are the unscrambled foreground results (vertical black lines).

We see that the unscrambled values lie in the tail of the null distribution for both candidates, though the strength of support differs. J0729+4008 shows weak coherence with $p$–values of 0.07 ($\sim 1.5\sigma$) via shuffling and 0.04 ($\sim 1.8\sigma$) via shifting. J1536+0441 shows somewhat stronger coherence with $p = 0.007$ ($\sim 2.5\sigma$) for both scrambling approaches. The apparent mismatch between modest CW Bayes factors and small coherence $p$–values; this is expected because the coherence test isolates the phase pattern across pulsars rather than the overall strength of the evidence for a CW. Even a low-BF candidate can yield a low $p$–value if its phase structure is unlikely to arise from noise.

The contrast with Bécsy et al. (2025)—where $p \sim 0.01$ occurred only when the CW Bayes factor exceeded 100—likely reflects two differences: (i) their analysis used only 10–15 pulsars, whereas the NANOGrav 15 yr dataset contains many more; and (ii) the targeted nature of our search. In the all-sky case the search can re-optimize the sky location in each scramble, while in a targeted search the sky location is fixed. This fixed-location condition naturally sharpens the distinction between scrambled and unscrambled runs. A full exploration of this effect will be discussed in future work.

Finally, the quoted $p$–values do not incorporate the fact that we examined 114 targets. Assuming independence, a conservative Bonferroni factor would inflate all $p$–values above unity. The proper treatment of correlation among targets will be pursued in future simulations, so these $p$–values should be regarded as local rather than global significances.

The second coherence test is a model comparison between the coherent CW model, the noise-only model (NOISE), and an incoherent sinusoid model (INCOH). Table 4 shows the corresponding Bayes factors. The CW–NOISE Bayes factors are lower than in Table 1 because `FurgeHullam` fixes the noise parameters during this test. The final column shows a weak preference for coherence in both candidates, with J1536+0441 somewhat stronger. Importantly, neither candidate prefers the incoherent model, slightly strengthening the cross-pulsar interpretation noted above.

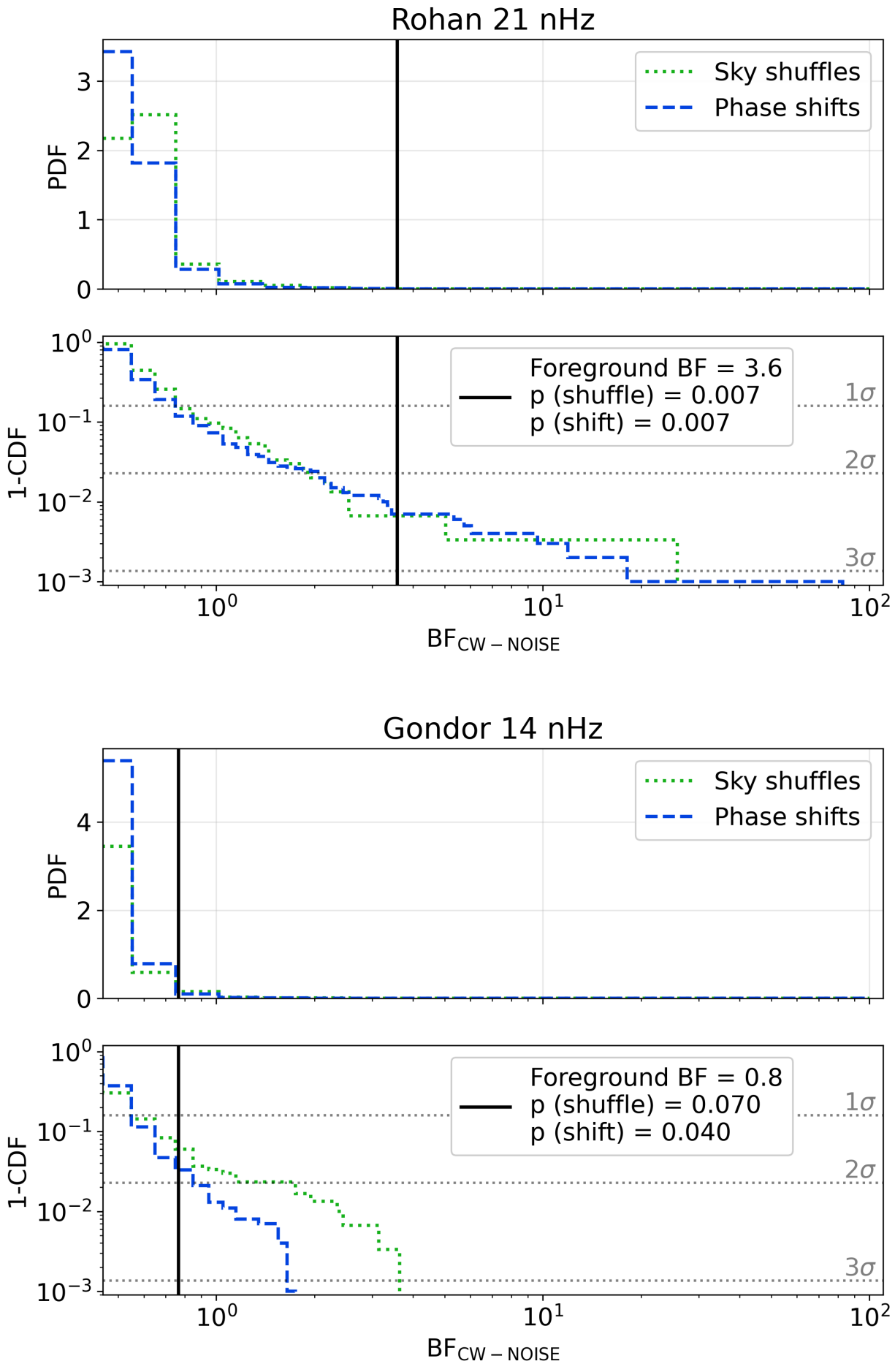

**Figure 11.** Coherent and Incoherent search results for J1536+0441 (upper panel) and J0729+4008 (lower panel). Shown are the null distributions of Bayes factors via sky shuffling (green) and phase shifting (blue) along with the Bayes factors found without any model scrambling (black). We find weak support for J0729+4008, with a $p$-value of $p = 0.04$ for the phase shift analysis and $p = 0.07$ for the sky shuffles. J1536+0441 is the stronger candidate of the two, with $p = 0.007$ with both sky shuffles and phase shifts.

### 6.5.2. *Dropout Analysis and Pulsar Noise*

To assess robustness against pulsar-specific noise, we performed a dropout analysis following Aggarwal et al. (2019).

For J1536+0441, 48 of the 67 pulsars favor a CW signal with $\mathcal{B}_{\rm dropout} > 1$, and six pulsars contribute most strongly, with PSRs J1911+1347 and J1741+1351 being the top contributors. Among the top six pulsars, PSR J1600−3053 appears and is known to exhibit challenging scattering-related noise (Alam et al. 2021; Chalumeau et al. 2022; Larsen et al. 2024).

For J0729+4008, 32 pulsars favor a CW signal. The highest-ranked contributors include PSRs J0613−0200 and J1713+0747. J0613−0200 produced a spurious ∼15 nHz feature in the NANOGrav 11 yr CW search (Aggarwal et al. 2019), and its significance is reduced with improved chromatic modelling (Larsen et al. 2024). PSR J1713+0747 has long been recognized as sensitive to high-frequency noise (Figure 14). Their prominence in J0729+4008 suggests that pulsar specific noise is likely contributing, motivating custom noise model development for the pulsars.

These dropout Bayes factors are modest. Larger values were found in earlier NANOGrav data releases, e.g. $\mathcal{B}_{\rm dropout} = 23$ for PSR J0613−0200 in the 9 yr search at 15 nHz and $\sim 100$ for PSR J1713+0747 in the 11 yr search at 109 nHz. This history underscores the need for refined pulsar-noise models before any low-significance CW feature can be interpreted. A systematic exploration of custom noise models for all pulsars is planned for future work.

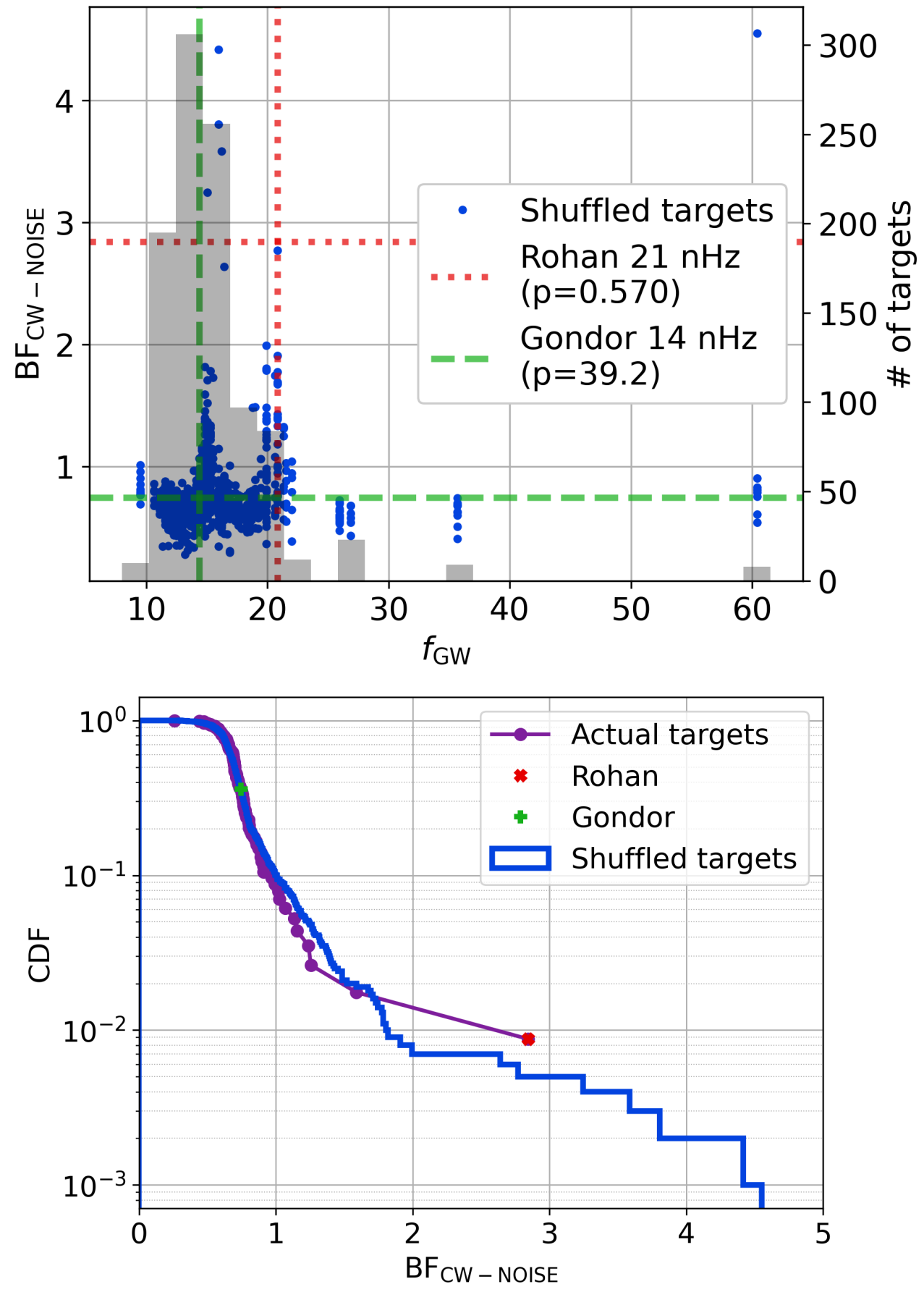

**Figure 12.** Results of random targeting test: (top) Bayes factors and frequencies of 1,000 randomly shuffled targets, along with J1536+0441 (Rohan) and J0729+4008 (Gondor). Note the excess of large Bayes factors around 14 nHz and 21 nHz. (bottom) Cumulative distribution function (CDF) of 1,000 random shuffles targets (shown in the background histogram) and our 114 actual targets. The fact that these distributions closely follow each other indicate that our targets are consistent with an uninformed random targeting.

#### 6.5.3. *Random Targeting Test*

To quantify the effect of examining 114 targets, we performed a random-targeting experiment in which the frequencies, sky positions, and distances of the real targets were randomly shuffled. This preserves the overall parameter distribution while removing any coherence among individual triplets. This test answers if any of the signals present in the dataset can be unambiguously isolated to one of the candidate galaxies we are searching for, or if they can be produced just as frequently by a random galaxy.

We generated 1,000 randomized target lists and computed Bayes factors for each. The top panel of Figure 12 shows their distribution as a function of frequency, along with those of J1536+0441 and J0729+4008. These candidates lie near frequencies where random realizations already tend to show higher Bayes factors. When compared to the empirical background, the corresponding $p$–values are $p = 0.57$ (J1536+0441) and $p = 0.39$ (J0729+4008). These values already incorporate the relevant multiple trial correction because they are evaluated directly against the empirical distribution of the 114 targets.

The bottom panel shows the cumulative distribution functions of the randomized and actual targets. The two distributions are statistically indistinguishable, with only a small upward fluctuation at J1536+0441. The random-targeting test naturally yields several notions of $p$–value (raw, Bonferroni-scaled, and global), which differ for all but the loudest target. J0729+4008's global $p$–value is $p = 0.96$, reflecting the near-perfect agreement between the real and randomized CDFs.

Although this fixed-noise analysis differs from the fully marginalized analyses elsewhere in the paper, the qualitative behavior is instructive. The second-loudest target in this test, HS 0926+3608, lies near 14 nHz with BF $\sim$ 1.6, echoing the behavior seen for J0729+4008. This repeatability across methodologies supports the conclusion that neither candidate rises above the background of an all-sky search and that none of the 114 targets exhibit statistically significant evidence for a CW.

### 6.6. *Software Crosscheck: Independent pipelines*

To verify that our results are not artifacts of a particular inference framework, we performed crosschecks between `enterprise` and `QuickCW`. For a representative subset of 15 targets, we compare strain and chirp mass upper limits as well as Bayes factors under the CURN hypothesis; the results are listed in Table 5. These tests were carried out during early analyses in which the CURN hyperparameters were fixed to their median values from Agazie et al. (2023a). Results elsewhere in this paper (e.g., Tables 1, 6–8) use varied CURN parameters with importance reweighting to the HD model.

As shown in Table 5, the two pipelines produce consistent 95% upper limits and broadly similar Bayes factors across the sample. Modest differences in Bayes factors arise from the slightly different implementations of prior bounds on $\log_{10}\mathcal{M}_c$ in the two codes, but do not affect any scientific conclusion. In particular, the mild CW support for J1536+0441 and J0729+4008 is recovered by both pipelines, indicating that these features are not sampling artifacts.

For Table 5, the 95% upper limits on $h_0$ and $\mathcal{M}_c$ use one-sided Bayesian credible intervals, and Bayes factors are computed with the Savage–Dickey method under a log-uniform prior on $\mathcal{M}_c$. Upper limits are obtained with uniform priors in $\mathcal{M}_c$, and uncertainties follow the procedures of Arzoumanian et al. (2018, 2020).

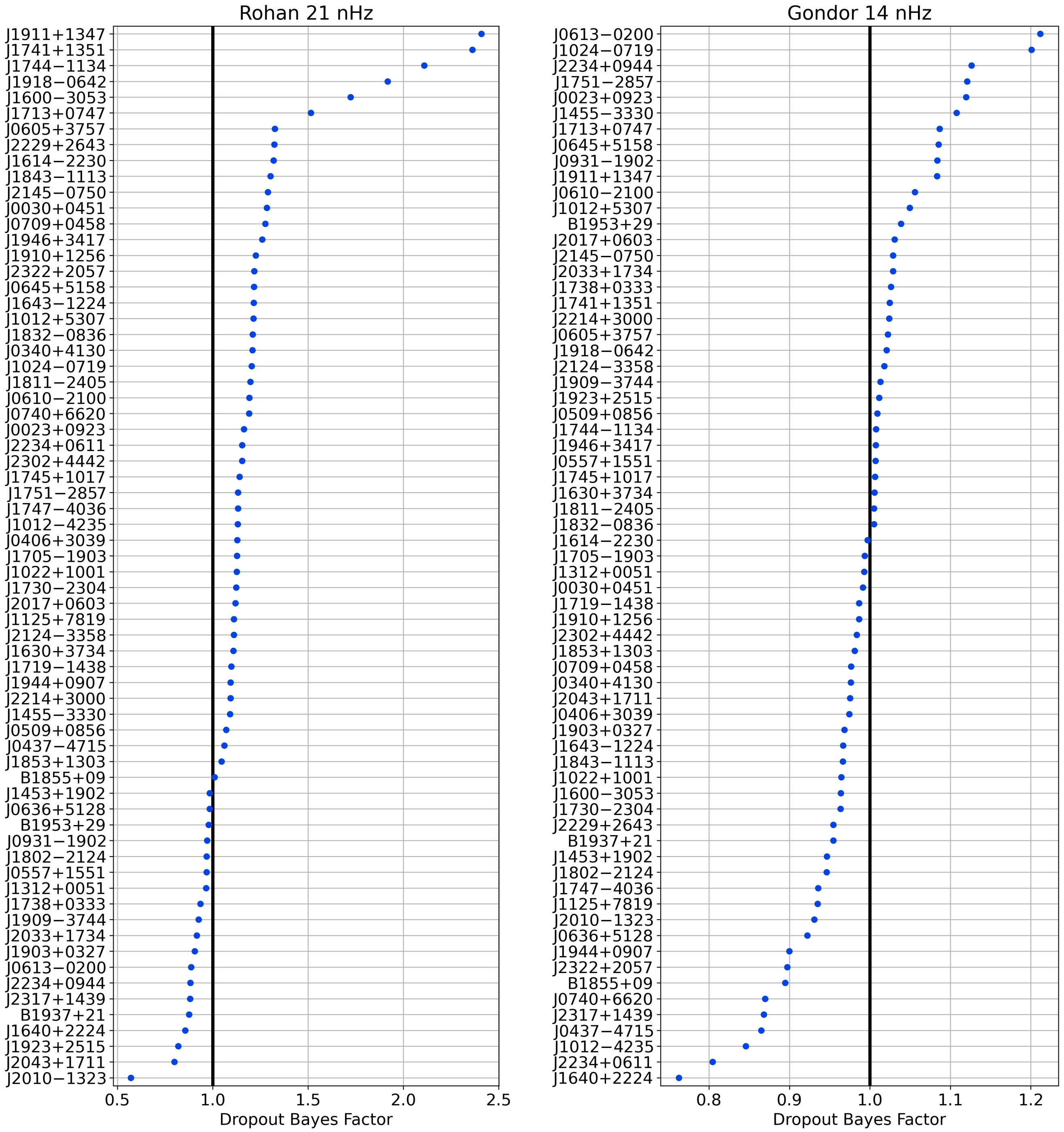

**Figure 13.** Dropout analysis for J1536+0441 and J0729+4008. J1536+0441 prefers a CW signal in 48 of the 67 pulsars, while only 32 pulsars prefer a CW signal for J0729+4008. Several pulsars are of particular interest: PSR J0613−0200 is the top pulsar for J0729+4008 but has previously been identified as problematic in Aggarwal et al. (2019), and benefiting from custom noise models in Larsen et al. (2024). The presence of PSR J1713+0747 as a top pulsar for J0729+4008 may be concerning for binary hypothesis, given its known high frequency noise characteristics, Figure 14, however, its impact appears to be negligible in all-sky searches near the target frequencies 14 nHz and 21 nHz.

| | QuickCW | | | enterprise | | |
|---|---|---|---|---|---|---|
| Object Name | $\log_{10}\mathcal{M}_c^{95\%}$ | $\log_{10} h_0^{95\%}$ | $\mathcal{B}^{\rm CURN+CW}_{\rm CURN}$ | $\log_{10}\mathcal{M}_c^{95\%}$ | $\log_{10} h_0^{95\%}$ | $\mathcal{B}^{\rm CURN+CW}_{\rm CURN}$ |
| 3C 66B | 9.07(2) | −13.85(4) | 0.62(8) | 8.980(3) | −13.883(5) | 0.293(3) |
| HS 0926+3608 | 10.462(5) | −14.25(1) | 0.71(2) | 10.552(3) | −13.932(5) | 1.28(1) |
| HS 1630+2355 | 10.154(8) | −14.32(1) | 0.48(3) | 10.153(4) | −14.158(7) | 0.817(5) |
| PKS 2131−021 | 10.245(8) | −14.37(1) | 0.64(3) | 10.203(3) | −14.275(5) | 0.775(6) |
| PKS J0805−0111 | 10.391(6) | −14.06(1) | 0.61(2) | 10.373(3) | −13.928(5) | 0.98(3) |
| SDSS J072908.71+400836.6 (Gondor) | 9.573(5) | −14.002(8) | 2.20(8) | 9.567(3) | −14.007(4) | 2.44(3) |
| SDSS J092911.35+203708.5 | 10.679(7) | −13.84(1) | 0.55(2) | 10.528(3) | −13.929(5) | 0.88(2) |
| SDSS J114857.33+160023.1 | 10.21(2) | −14.41(3) | 0.45(1) | 10.259(6) | −14.17(1) | 0.716(9) |
| SDSS J131706.19+271416.7 | 10.579(3) | −14.185(4) | 0.58(3) | 10.384(3) | −14.345(4) | 0.78(2) |
| SDSS J133516.17+183341.4 | 10.20(2) | −14.40(3) | 0.502(7) | 10.121(3) | −14.363(5) | 0.76(1) |
| SDSS J134855.27−032141.4 | 10.315(9) | −14.45(2) | 0.44(2) | 10.474(3) | −14.022(6) | 0.810(7) |
| SDSS J140704.43+273556.6 | 10.407(4) | −14.355(6) | 0.76(9) | 10.406(3) | −14.191(4) | 1.06(4) |
| SDSS J153636.22+044127.0 (Rohan) | 9.860(2) | −14.229(3) | 3.46(6) | 9.829(2) | −14.140(3) | 3.37(5) |
| SDSS J160730.33+144904.3 | 10.25(2) | −14.53(4) | 0.46(2) | 10.233(3) | −14.398(5) | 0.69(1) |
| SDSS J164452.71+430752.2 | 10.33(1) | −14.41(2) | 0.43(1) | 10.367(5) | −14.190(8) | 0.782(6) |
| SNU J13120+0641 | 9.790(7) | −14.21(1) | 0.64(5) | 9.814(3) | −13.999(6) | 0.811(6) |

**Table 5.** Comparing upper limits and Bayes factors for a selection of the top targets from Xin et al. (2021) using independent `QuickCW` and `enterprise` search pipelines. For the purpose of the software comparison, these searches fix CURN parameters at the median values from Agazie et al. (2023a). For upper limits under a varied, HD-correlated GWB model see Tables 6-8 and Section 6.6. Numbers in parentheses are uncertainties to the least significant digit of the reported values.

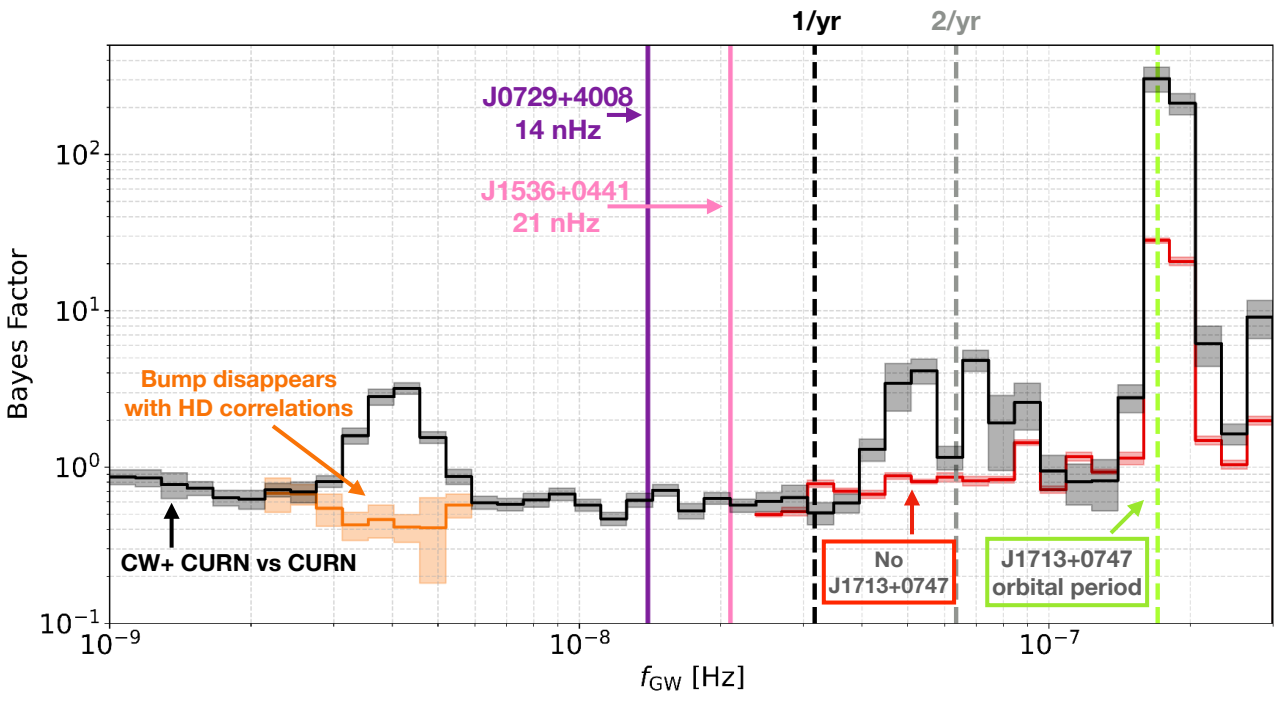

**Figure 14.** Bayes factors as a function of GW frequency, adapted from the original figure by Agazie et al. (2023c). In the most recent NANOGrav all-sky CW search, Agazie et al. (2023c) found a Bayes factor peak between 3 and 5 nHz which disappeared when Hellings and Downs cross-correlations were applied (orange line). Here we can see clearly that there was no Bayes factor greater than unity at 14 nHz (purple line; J0729+4008) nor was there one at 21 nHz (pink line; J1536+0441).

## 7. DISCUSSION AND CONCLUSIONS

Using the NANOGrav 15 yr data set, we carried out the first catalog-based, targeted search for CWs from 114 SMBHB candidates corresponding to AGN with periodic light curves. By fixing sky position (see Figure 2), redshift, and GW frequency to values inferred from long-term optical or radio variability, we improved our sensitivity to CWs: the median 95 % strain upper limit improved by a factor of $0.82 \leq \mathcal{I} \leq 14.94$, with a median value of 2.24 and mean of $2.60 \pm 2.01$ relative to all-sky analyses at the same frequencies (Agazie et al. 2023c). We placed strain and chirp-mass upper limits for all targets (Tables 6–8), and for the first time excluded previously allowed mass ranges for a putative SMBHB in 3C 66B (Iguchi et al. 2010).

Bayesian model comparison against a model that includes only an HD-correlated GWB shows no significant evidence for a CW in any target (Table 1). The two targets with Bayes factors slightly above unity are fully consistent with noise after accounting for the number of searches. J1536+0441 at $f_{\mathrm{gw}} = 21\,\mathrm{nHz}$ yields $\mathcal{B}^{\mathrm{HD+CW}}_{\mathrm{HD}} = 1.91(4)$ and $\log_{10}\mathcal{M}_c = 9.59^{+0.16}_{-0.73}$, with a coherence-test $p$-value of 0.007 ($2.5\sigma$). J0729+4008 at $f_{\mathrm{gw}} = 14\,\mathrm{nHz}$ yields $\mathcal{B}^{\mathrm{HD+CW}}_{\mathrm{HD}} = 3.7(1)$ and $\log_{10}\mathcal{M}_c = 9.41^{+0.08}_{-0.17}$, with $p = 0.04$–$0.07$ ($1.5$–$1.8\sigma$). While these values indicate weak support, applying a conservative trials factor such as the Bonferroni correction drives all $p$-values to $p \sim 1$, implying full consistency with noise fluctuations.

To further evaluate these candidates, we performed a battery of tests summarized in Table 2. J1536+0441's periodicity persists in extended CRTS, ATLAS, ZTF, and WISE light curves, whereas J0729+4008's periodicity has faded following a mid-IR flare. Population-level expectations provide a consistent astrophysical frame: hydrodynamic variability models predict $5^{+20}_{-1}$ true binaries in the CRTS sample (Kelley et al. 2019), and GWB-consistency arguments limit the number to at most eight (Casey-Clyde et al. 2025). Having only two targets with $\mathcal{B} > 1$ fits comfortably within this envelope.

J0729+4008 lies in a region of parameter space compatible with many low-mass binaries, whereas J1536+0441 occupies a sparse, very high-mass region. This could indicate either residual noise contamination or that population models may require a heavier tail, as suggested by Sato-Polito et al. (2024). If either source dominated its GW frequency bin, it would imprint anisotropy at the $\sim 65\,\%$ level. NANOGrav's anisotropy search does not probe 14–21 nHz (Agazie et al. 2023), but MeerKAT does (Grunthal et al. 2025). J0729+4008 lies outside MeerKAT's field of view; J1536+0441 lies within it and coincides with an S/N $\sim$ 2 fluctuation. This is fully consistent with isotropic GWB structure, but the region will be monitored in future datasets.

Pulsar dropout tests illustrate the limitations of current evidence. In earlier data sets, some pulsars produced large dropout Bayes factors (e.g., J0613−0200 at 15 nHz in the 9 yr and J1713+0747 at 109 nHz in the 11 yr). Our results are far smaller. Here J0613−0200 contributes most to J0729+4008 with only $\mathcal{B}_{\mathrm{dropout}} = 1.3$, and J1713+0747 shows no signs of contaminating the relevant 14–21 nHz range (Figure 14). For J1536+0441, evidence is distributed across one third of the array with a maximum $\mathcal{B}_{\mathrm{dropout}} = 1.8$; for J0729+4008, the signal rests disproportionately on a few noisy pulsars. These patterns reinforce that low-level excesses must be interpreted cautiously and that targeted noise modeling will be essential going forward.

We are now planning simulations to inject and recover J1536+0441- and J0729+4008-like signals to study detection thresholds, false positives, and the expected number of pulsars that should favor a CW for a given sky location. Because searches across many targets are not independent, we will also investigate improved methods for handling multiple comparisons beyond simple Bonferroni corrections.

High-cadence, low-RMS pulsar timing from future IPTA data combinations will clarify the interpretation of both candidates. The discovery and timing of additional millisecond pulsars—including those found in Fermi-LAT unassociated sources—could improve sensi-tivity along the relevant lines of sight within a few years (Burt et al. 2011; Mingarelli et al. 2017; Xin et al. 2021). Gamma-ray pulsar timing (Kerr & Parthasarathy 2022) will eventually provide an independent probe unaffected by chromatic noise (Agazie et al. 2023; Larsen et al. 2024; Hazboun et al. 2025), offering a clean cross-check for CW candidates.

On the electromagnetic side, spectroscopic monitoring, continued ZTF and WISE follow-up, and broadband variability studies may confirm or refute binary interpretations. Future work could also incorporate information about $\cos\iota$ for Doppler-boost candidates and impose priors on mass ratio or eccentricity (D'Orazio & Charisi 2023). Both candidates show noteworthy long-term variability: J0729+4008 flared in the infrared before its optical dimming, and J1536+0441 shows a slow decline in optical brightness. WISE data for J1536+0441 exhibit modulations expected from periodically varying continua, which may constrain dust geometry and the nature of the variability (D'Orazio & Haiman 2017).

In summary, a one target pilot study has evolved into a 114-object survey, improving strain and chirp-mass upper limits beyond the best all-sky constraints. This work establishes a roadmap for a systematic SMBHB detection framework. In the coming years, improved pulsar timing, electromagnetic follow-up, and a refined detection protocol will clarify whether any of these SMBHB candidates could be genuine. As long-lived, monochromatic CW sources emitting in both gravitational and electromagnetic radiation, such systems would offer exceptional opportunities for multimessenger studies of binary evolution and may serve as critical anchors for LISA in the milliHertz band.

# APPENDIX

## A. ANISOTROPY FROM A SINGLE BRIGHT SOURCE IN A DISCRETE GWB

In a realistic setting, the isotropic background is not a smooth field but arises from the superposition of a finite, discrete number of unresolved sources. This intrinsic discreteness introduces shot noise, which arises from the Poisson sampling of the underlying source population. On the other hand, the bright source considered here is a deterministic signal. The shot noise contributes with power across all angular scales (Taylor & Gair 2013; Mingarelli et al. 2017). This additional power from the discreteness of the background (see e.g. Agazie et al. 2025 can affect the detectability of a bright CW source imprint on the angular statistics.

The total GW intensity is the sum of stochastic and deterministic components:

$$h^2_{c,\mathrm{tot}}(f,\hat{\Omega}) = h^2_{c,\mathrm{stoch}}(f,\hat{\Omega}) + h^2_{c,\mathrm{CW}}(f)\delta^{(2)}(\hat{\Omega}-\hat{\Omega}_0)\,, \tag{A1}$$

where $h^2_{c,\mathrm{stoch}}(f,\hat{\Omega})$ is the random field from the GWB, $h^2_{c,\mathrm{CW}}(f)$ is the CW source characteristic strain, and $\hat{\Omega}_0$ is its sky position (Rosado et al. 2015; Agazie et al. 2025). For a discrete population, $h^2_{c,\mathrm{stoch}}(f,\hat{\Omega}) = \sum_k h_k^2 f_k/\Delta f_{\mathrm{GW}}$, where $\Delta f_{\mathrm{GW}} = 1/T_{\mathrm{obs}}$ is the frequency bin width (Babak et al. 2016).

Due to the linearity of spherical harmonic transforms, the total coefficients decompose as:

$$a^{\mathrm{tot}}_{\ell m} = a^{\mathrm{stoch}}_{\ell m} + a^{\mathrm{CW}}_{\ell m} = a^{\mathrm{stoch}}_{\ell m} + h^2_{c,\mathrm{CW}} Y^*_{\ell m}(\hat{\Omega}_0)\,. \tag{A2}$$

Since the CW source breaks isotropy, we compute the angle-averaged power spectrum. Expanding $\langle |a^{\mathrm{tot}}_{\ell m}|^2\rangle$ yields four terms: the stochastic power $C^{\mathrm{stoch}}_\ell$, the CW power $(h^2_{c,\mathrm{CW}})^2 |Y_{\ell m}(\hat{\Omega}_s)|^2$, and two cross-terms. The cross-terms vanish except for the monopole ($\ell = 0$), where the stochastic field has a non-zero mean (Allen & Ottewill 1997; Sato-Polito & Kamionkowski 2024):

$$\bar{I}(f) = \frac{f}{4\pi\Delta f}\int d\vec{\theta}\,\frac{dN_{\Delta f}}{d\vec{\theta}} h^2(f,\vec{\theta})\,, \tag{A3}$$

where $\vec{\theta}$ denotes the source parameters (masses, redshift, orbital parameters, etc.), $dN_{\Delta f}/d\vec{\theta}$ is the differential source distribution, and $h(f,\vec{\theta})$ is the strain amplitude from a source with parameters $\vec{\theta}$. Neglecting Large Scale Structure clustering, the stochastic background power spectrum from the discrete sources is (Sato-Polito & Kamionkowski 2024)

$$C^{\mathrm{stoch}}_\ell(f) = \delta_{\ell 0} 4\pi \left(\frac{f}{4\pi\Delta f}\int d\vec{\theta}\,\frac{dN_{\Delta f}}{d\vec{\theta}} h^2(f,\vec{\theta})\right)^2 + \left(\frac{f}{4\pi\Delta f}\right)^2 \int d\vec{\theta}\,\frac{dN_{\Delta f}}{d\vec{\theta}} h^4(f,\vec{\theta})\,. \tag{A4}$$

The first term is the monopole from the squared mean field, while the second is the shot noise from source discreteness. Using the spherical harmonic addition theorem, the final angle-averaged power spectrum becomes:

$$C_\ell^{\rm tot}(f) = C_\ell^{\rm stoch}(f) + \frac{(h^2_{c,\rm CW}(f))^2}{4\pi} + 2\delta_{\ell 0}\bar{I}(f)h^2_{c,\rm CW}(f)\,. \tag{A5}$$

This final expression reveals the critical role of shot noise in this scenario. For $\ell > 0$, the angular power receives contributions from both the shot noise of the stochastic part and the white-noise spectrum of the CW source, while the monopole ($\ell = 0$) contains additional contributions from the mean field and the cross-term. This makes it clear that the shot noise fundamentally alters the ratio $C_\ell/C_0$ compared to the idealized case of a smooth isotropic background plus CW source. Without properly accounting for the shot noise from source discreteness, one would significantly underestimate the anisotropic power at $\ell > 0$, potentially misinterpreting or failing to detect the signature of a bright CW source against the stochastic background.

Finally, a definitive identification of a CW source requires detecting its characteristic $\ell$-independent contribution to the anisotropic power spectrum. Current PTA datasets, however, yield posteriors on higher-order multipoles that are often dominated by priors, making it difficult to robustly measure the shape of the $C_\ell$ spectrum. Additional motion-induced effects may introduce degeneracies with the anisotropy imprint from a continuous wave, arising from the peculiar velocities of both the observer and the nearby GWB sources. In addition, our motion relative to the cosmic rest frame induces a kinematic dipole at $\ell = 1$, which however is smaller than the expected anisotropies induced by SMBHBs (Tasinato 2023).

## B. UPPER LIMITS ON CRTS CANDIDATES

| Name | RA | Dec | z | Period [days] | Freq [nHz] | $M_{\rm tot}$ [$M_\odot$] | $\mathcal{M}_c^{95\%}$ [$M_\odot$] | $h_0^{95\%}$ |
|---|---|---|---|---|---|---|---|---|
| MCG 5−40−026[a] | 17h 01m 07.80s | +29d 24m 24.6s | 0.04 | 1407 | 16.4 | $7.24\times10^{6}$ | $1.29\times10^{9}$ | $5.38\times10^{-15}$ |
| SDSS J072908.71+400836.6 | 07h 29m 08.60s | +40d 08m 37.0s | 0.07 | 1612 | 14.4 | $2.63\times10^{7}$ | $3.38\times10^{9}$ | $1.16\times10^{-14}$ |
| 2MASS J04352649−1643460 | 04h 35m 26.50s | −16d 43m 45.7s | 0.10 | 1369 | 16.9 | $6.03\times10^{7}$ | $3.92\times10^{9}$ | $1.23\times10^{-14}$ |
| PGC 3096192 | 12h 50m 29.00s | +06d 36m 11.1s | 0.13 | 1561 | 14.8 | $1.15\times10^{7}$ | $3.55\times10^{9}$ | $6.85\times10^{-15}$ |
| SDSS J110554.78+322953.7 | 11h 05m 54.80s | +32d 29m 54.1s | 0.15 | 1724 | 13.4 | $1.74\times10^{8}$ | $4.52\times10^{9}$ | $8.38\times10^{-15}$ |
| HS 0423+0658 | 04h 26m 30.20s | +07d 05m 30.3s | 0.17 | 1123 | 20.6 | – | $4.01\times10^{9}$ | $8.00\times10^{-15}$ |
| HS 2219+1944 | 22h 22m 21.10s | +19d 59m 48.1s | 0.21 | 1724 | 13.4 | – | $3.87\times10^{9}$ | $4.46\times10^{-15}$ |
| SDSS J121018.34+015405.9 | 12h 10m 18.30s | +01d 5m 06.2s | 0.22 | 1612 | 14.4 | $3.47\times10^{8}$ | $4.12\times10^{9}$ | $5.05\times10^{-15}$ |
| SNU J13120+0641 | 13h 12m 04.70s | +06d 41m 07.6s | 0.24 | 1492 | 15.5 | $1.38\times10^{9}$ | $6.54\times10^{9}$ | $1.01\times10^{-14}$ |
| SDSS J161013.67+311756.4 | 16h 10m 13.70s | +31d 17m 56.s | 0.25 | 1724 | 13.4 | $8.23\times10^{7}$ | $1.82\times10^{9}$ | $1.07\times10^{-15}$ |
| CT 638 | 03h 18m 06.50s | −34d 26m 37.4s | 0.27 | 1514 | 15.3 | – | $9.57\times10^{9}$ | $1.70\times10^{-14}$ |
| RXS J04117+1324 | 04h 11m 46.90s | +13d 24m 16.5s | 0.28 | 1850 | 12.5 | $1.45\times10^{8}$ | $2.95\times10^{9}$ | $1.99\times10^{-15}$ |
| BZQJ1305−1033 | 13h 05m 33.00s | −10d 33m 19.1s | 0.29 | 1693 | 13.7 | $3.16\times10^{8}$ | $5.25\times10^{9}$ | $5.30\times10^{-15}$ |
| SDSS J123821.84+030024.2 | 12h 38m 21.80s | +03d 00m 24.6s | 0.38 | 1249 | 18.5 | $8.32\times10^{8}$ | $6.58\times10^{9}$ | $6.79\times10^{-15}$ |
| SDSS J153636.22+044127.0 | 15h 36m 36.20s | +04d 41m 26.9s | 0.38 | 1110 | 20.8 | $6.61\times10^{8}$ | $6.33\times10^{9}$ | $6.92\times10^{-15}$ |
| SDSS J135225.80+132853.2 | 13h 52m 25.80s | +13d 28m 53.3s | 0.40 | 1754 | 13.2 | $5.62\times10^{8}$ | $6.07\times10^{9}$ | $4.44\times10^{-15}$ |
| KUV 09484+3557 | 09h 51m 23.90s | +35d 42m 49.2s | 0.40 | 1162 | 19.9 | $2.04\times10^{8}$ | $3.83\times10^{9}$ | $2.74\times10^{-15}$ |
| SDSS J141425.92+171811.2 | 14h 14m 25.90s | +17d 18m 11.6s | 0.41 | 1785 | 13.0 | $3.89\times10^{8}$ | $2.50\times10^{9}$ | $9.77\times10^{-16}$ |
| SDSS J133631.45+175613.8 | 13h 36m 31.40s | +17d 56m 14.1s | 0.42 | 1561 | 14.8 | $1.07\times10^{9}$ | $8.73\times10^{9}$ | $8.31\times10^{-15}$ |
| SDSS J224829.47+144418.0 | 22h 48m 29.40s | +14d 44m 18.4s | 0.42 | 1218 | 19.0 | $7.24\times10^{8}$ | $3.52\times10^{9}$ | $2.14\times10^{-15}$ |
| RXS J10304+5516 | 10h 30m 25.0s | +55d 16m 23.4s | 0.44 | 1514 | 15.3 | $2.69\times10^{8}$ | $1.07\times10^{10}$ | $1.14\times10^{-14}$ |
| SDSS J140600.26+013252.2 | 14h 06m 00.30s | +01d 32m 52.4s | 0.45 | 1999 | 11.6 | $2.57\times10^{8}$ | $8.63\times10^{9}$ | $6.32\times10^{-15}$ |
| SDSS J094715.56+631716.4 | 09h 47m 15.60s | +63d 17m 17.3s | 0.49 | 1724 | 13.4 | $1.66\times10^{9}$ | $8.46\times10^{9}$ | $6.20\times10^{-15}$ |
| SDSS J114749.70+163106.7 | 11h 47m 49.70s | +16d 31m 06.8s | 0.55 | 1449 | 16.0 | $1.66\times10^{8}$ | $1.32\times10^{10}$ | $1.25\times10^{-14}$ |
| SDSS J084146.19+503601.1 | 08h 41m 46.30s | +50d 36m 00.5s | 0.56 | 1693 | 13.7 | $2.75\times10^{7}$ | $1.21\times10^{10}$ | $9.71\times10^{-15}$ |
| MS 10548−0335 | 10h 57m 22.30s | −03d 51m 31.3s | 0.56 | 892 | 26.0 | – | $1.02\times10^{10}$ | $1.12\times10^{-14}$ |
| SDSS J161854.64+230859.1 | 16h 18m 54.60s | +23d 08m 59.3s | 0.56 | 1666 | 13.9 | $2.34\times10^{8}$ | $7.33\times10^{9}$ | $4.21\times10^{-15}$ |
| SDSS J144754.62+132610.0 | 14h 47m 54.60s | +13d 26m 10.4s | 0.57 | 1960 | 11.8 | $3.8\times10^{8}$ | $9.66\times10^{9}$ | $5.85\times10^{-15}$ |
| 6QZ J221925.1−305408 | 22h 19m 25.20s | −30d 54m 08.1s | 0.58 | 1407 | 16.4 | – | $8.42\times10^{9}$ | $5.71\times10^{-15}$ |
| HS 0946+4845 | 09h 50m 00.70s | +48d 31m 29.9s | 0.59 | 1586 | 14.6 | $3.89\times10^{8}$ | $1.38\times10^{10}$ | $1.18\times10^{-14}$ |
| HS 1715+2131 | 17h 17m 20.10s | +21d 28m 15.0s | 0.59 | 1249 | 18.5 | – | $7.94\times10^{9}$ | $5.48\times10^{-15}$ |
| SDSS J134820.42+194831.5 | 13h 48m 20.40s | +19d 48m 31.9s | 0.59 | 1388 | 16.7 | $4.27\times10^{7}$ | $8.78\times10^{9}$ | $6.00\times10^{-15}$ |
| SDSS J143820.60+055447.9 | 14h 38m 20.60s | +05d 54m 48.1s | 0.61 | 1850 | 12.5 | $1.55\times10^{8}$ | $9.39\times10^{9}$ | $5.32\times10^{-15}$ |
| SDSS J125414.23+131348.1 | 12h 54m 14.20s | +13d 13m 48.4s | 0.66 | 1754 | 13.2 | $8.71\times10^{8}$ | $9.69\times10^{9}$ | $5.36\times10^{-15}$ |
| SDSS J144755.57+100040.0 | 14h 47m 55.60s | +10d 00m 40.4s | 0.68 | 861 | 26.9 | $8.51\times10^{8}$ | $7.39\times10^{9}$ | $5.26\times10^{-15}$ |
| RX J024252.3−232633 | 02h 42m 51.90s | −23d 26m 34.0s | 0.68 | 1817 | 12.7 | – | $1.08\times10^{10}$ | $6.02\times10^{-15}$ |
| SDSS J082716.85+490534.0 | 08h 27m 16.90s | +49d 05m 34.9s | 0.68 | 1612 | 14.4 | $9.12\times10^{8}$ | $1.61\times10^{10}$ | $1.25\times10^{-14}$ |
| SDSS J132103.41+123748.2 | 13h 21m 03.40s | +12d 37m 48.1s | 0.69 | 1538 | 15.1 | $8.13\times10^{8}$ | $1.31\times10^{10}$ | $9.11\times10^{-15}$ |
| SBS 0920+590 | 09h 23m 58.70 | +58d 49m 06.3s | 0.71 | 648 | 35.7 | $5.75\times10^{8}$ | $1.74\times10^{10}$ | $2.50\times10^{-14}$ |
| SDSS J221016.97+122213.9 | 22h 10m 17.00s | +12d 22m 14.0s | 0.72 | 1333 | 17.4 | $1.00\times10^{9}$ | $7.90\times10^{9}$ | $4.10\times10^{-15}$ |
| SDSS J124044.49+231045.8 | 12h 40m 44.50s | +23d 10m 46.1s | 0.72 | 1428 | 16.2 | $8.71\times10^{8}$ | $1.41\times10^{10}$ | $1.02\times10^{-14}$ |
| UM 234 | 00h 23m 03.20s | +01d 15m 33.9s | 0.73 | 1817 | 12.7 | $1.55\times10^{9}$ | $1.30\times10^{10}$ | $7.47\times10^{-15}$ |
| SDSS J152157.02+181018.6 | 15h 21m 57.00s | +18d 10m 19.2s | 0.73 | 1960 | 11.8 | $8.91\times10^{7}$ | $1.17\times10^{10}$ | $5.94\times10^{-15}$ |
| 163107.34+560905.3 | 16h 31m 07.40s | +56d 09m 05.1s | 0.73 | 1724 | 13.4 | $2.40\times10^{8}$ | $9.08\times10^{9}$ | $4.26\times10^{-15}$ |

[a]EM mass from Pal et al. (2024)

**Table 6.** Electromagnetic and GW-constrained upper limits on chirp mass and strain amplitude for candidate SMBHBs. EM-derived total masses are reported when available in Graham et al. (2015), except when indicated otherwise.

| Name | RA | Dec | z | Period [days] | Freq [nHz] | $M_{\rm tot}$ [$M_\odot$] | $\mathcal{M}_c^{95\%}$ [$M_\odot$] | $h_0^{95\%}$ |
|---|---|---|---|---|---|---|---|---|
| SDSS J080648.65+184037.0 | 08h 06m 48.60 | +18d 40m 37.3s | 0.75 | 892 | 26.0 | $9.77 \times 10^{7}$ | $1.59 \times 10^{10}$ | $1.63 \times 10^{-14}$ |
| SDSS J115346.39+241829.4 | 11h 53m 46.40s | +24d 18m 29.8s | 0.75 | 1666 | 13.9 | $9.33 \times 10^{8}$ | $1.26 \times 10^{10}$ | $7.29 \times 10^{-15}$ |
| SDSS J081617.73+293639.6 | 08h 16m 17.80s | +29d 36m 40.7s | 0.77 | 1162 | 19.9 | $5.89 \times 10^{9}$ | $1.43 \times 10^{10}$ | $1.11 \times 10^{-14}$ |
| SDSS J155449.11+084204.8 | 15h 54m 49.10s | +08d 42m 05.4s | 0.79 | 1561 | 14.8 | $7.08 \times 10^{8}$ | $1.01 \times 10^{10}$ | $4.95 \times 10^{-15}$ |
| SDSS J104758.34+284555.8 | 10h 47m 58.30s | +28d 45m 56.2s | 0.79 | 1850 | 12.5 | $5.25 \times 10^{8}$ | $1.55 \times 10^{10}$ | $8.95 \times 10^{-15}$ |
| SDSS J082926.01+180020.7 | 08h 29m 26.00s | +18d 00m 20.7s | 0.81 | 1449 | 16.0 | $2.63 \times 10^{8}$ | $2.19 \times 10^{10}$ | $1.82 \times 10^{-14}$ |
| HS 1630+2355 | 16h 33m 02.70s | +23d 49m 28.8s | 0.82 | 2040 | 11.3 | $7.24 \times 10^{9}$ | $1.42 \times 10^{10}$ | $6.91 \times 10^{-15}$ |
| FBQS J17239+3748 | 17h 23m 54.30s | +37d 48m 41.7s | 0.83 | 1960 | 11.8 | $2.40 \times 10^{9}$ | $1.06 \times 10^{10}$ | $4.33 \times 10^{-15}$ |
| SDSS J130040.62+172758.4 | 13h 00m 40.60s | +17d 27m 58.5s | 0.86 | 1817 | 12.7 | $7.59 \times 10^{8}$ | $1.33 \times 10^{10}$ | $6.35 \times 10^{-15}$ |
| SDSS J131909.08+090814.7 | 13h 19m 09.10s | +09d 08m 15.1s | 0.88 | 1298 | 17.8 | $4.68 \times 10^{8}$ | $1.02 \times 10^{10}$ | $4.99 \times 10^{-15}$ |
| HE 1408−1003 | 14h 10m 40.30s | −10d 17m 29.7s | 0.88 | 1923 | 12.0 | – | $1.42 \times 10^{10}$ | $6.64 \times 10^{-15}$ |
| SDSS J143621.29+072720.8 | 14h 36m 21.30s | +07d 27m 21.1s | 0.89 | 1886 | 12.3 | $1.07 \times 10^{9}$ | $1.37 \times 10^{10}$ | $6.25 \times 10^{-15}$ |
| FBQS J081740.1+232731 | 08h 17m 40.20s | +23d 27m 32.0s | 0.89 | 1190 | 19.5 | $3.55 \times 10^{9}$ | $1.70 \times 10^{10}$ | $1.21 \times 10^{-14}$ |
| SDSS J091554.50+352949.6 | 09h 15m 54.50s | +35d 29m 49.9s | 0.90 | 1369 | 16.9 | $1.12 \times 10^{9}$ | $1.73 \times 10^{10}$ | $1.14 \times 10^{-14}$ |
| SDSS J104430.25+051857.2 | 10h 44m 30.30s | +05d 18m 56.8s | 0.91 | 1333 | 17.4 | $1.74 \times 10^{9}$ | $1.74 \times 10^{10}$ | $1.14 \times 10^{-14}$ |
| CSO 67 | 11h 03m 27.50s | +29d 48m 11.2s | 0.91 | 2082 | 11.1 | $2.24 \times 10^{9}$ | $2.35 \times 10^{10}$ | $1.40 \times 10^{-14}$ |
| SDSS J133127.31+182416.9 | 13h 31m 27.30s | +18d 24m 17.1s | 0.94 | 1638 | 14.1 | $2.45 \times 10^{9}$ | $1.29 \times 10^{10}$ | $5.80 \times 10^{-15}$ |
| US 3204 | 02h 49m 28.90s | +01d 09m 25.0s | 0.95 | 1666 | 13.9 | $8.91 \times 10^{8}$ | $1.68 \times 10^{10}$ | $8.72 \times 10^{-15}$ |
| SDSS J102349.38+522151.2 | 10h 23m 49.50s | +52d 21m 51.8s | 0.96 | 1785 | 13.0 | $3.89 \times 10^{9}$ | $1.67 \times 10^{10}$ | $8.26 \times 10^{-15}$ |
| SDSS J154409.61+024040.0 | 15h 44m 09.60s | +02d 40m 39.8s | 0.96 | 1999 | 11.6 | $5.75 \times 10^{8}$ | $1.39 \times 10^{10}$ | $5.57 \times 10^{-15}$ |
| SDSS J150450.16+012215.5 | 15h 04m 50.20s | +01d 22m 15.8s | 0.97 | 1724 | 13.4 | $1.58 \times 10^{9}$ | $1.08 \times 10^{10}$ | $4.04 \times 10^{-15}$ |
| SDSS J082827.84+400333.9 | 08h 28m 27.80s | +40d 03m 34.1s | 0.97 | 1886 | 12.3 | $7.41 \times 10^{8}$ | $1.86 \times 10^{10}$ | $9.33 \times 10^{-15}$ |
| SDSS J114438.34+262609.4 | 11h 44m 38.30s | +26d 26m 10.1s | 0.97 | 1314 | 17.6 | $2.4 \times 10^{9}$ | $1.38 \times 10^{10}$ | $7.17 \times 10^{-15}$ |
| SDSS J172656.96+600348.5 | 17h 26m 56.90s | +60d 03m 49.1s | 0.99 | 1923 | 12.0 | $1.41 \times 10^{9}$ | $1.35 \times 10^{10}$ | $5.26 \times 10^{-15}$ |
| SDSS J115141.81+142156.6 | 11h 51m 41.80s | +14d 21m 57.0s | 1.00 | 1492 | 15.5 | $1.29 \times 10^{9}$ | $2.17 \times 10^{10}$ | $1.35 \times 10^{-14}$ |
| SDSS J113050.21+261211.4 | 11h 30m 50.20s | +26d 12m 11.8s | 1.01 | 2173 | 10.7 | $2.09 \times 10^{9}$ | $2.94 \times 10^{10}$ | $1.73 \times 10^{-14}$ |
| SDSS J113916.47+254412.6 | 11h 39m 16.40s | +25d 44m 13.0s | 1.01 | 2439 | 9.5 | $1.45 \times 10^{9}$ | $2.85 \times 10^{10}$ | $1.52 \times 10^{-14}$ |
| SDSS J152035.23+095925.2 | 15h 20m 35.20s | +09d 59m 25.7s | 1.05 | 1203 | 19.2 | $1.32 \times 10^{9}$ | $1.28 \times 10^{10}$ | $6.14 \times 10^{-15}$ |
| SDSS J142301.96+101500.1 | 14h 23m 02.00s | +10d 15m 00.0s | 1.05 | 1233 | 18.8 | $2.88 \times 10^{9}$ | $1.17 \times 10^{10}$ | $5.20 \times 10^{-15}$ |
| SDSS J102255.21+172155.7 | 10h 22m 55.20s | +17d 21m 56.0s | 1.06 | 1666 | 13.9 | $4.90 \times 10^{8}$ | $1.88 \times 10^{10}$ | $9.25 \times 10^{-15}$ |
| 4C 50.43[a] | 17h 31m 03.70s | +50d 07m 35.7s | 1.11 | 1075 | 21.5 | $1.51 \times 10^{8}$ | $1.45 \times 10^{10}$ | $7.61 \times 10^{-15}$ |
| SDSS J080237.60+340446.3 | 08h 02m 37.60s | +34d 04m 46.6s | 1.12 | 1428 | 16.2 | $9.12 \times 10^{8}$ | $2.84 \times 10^{10}$ | $1.90 \times 10^{-14}$ |
| SDSS J083349.55+232809.0 | 08h 33m 49.60s | +23d 28m 09.2s | 1.16 | 1086 | 21.3 | $2.51 \times 10^{9}$ | $1.96 \times 10^{10}$ | $1.19 \times 10^{-14}$ |
| PKS 0157+011[b] | 02h 00m 03.90s | +01d 25m 12.6s | 1.17 | 1051 | 22.0 | $7.94 \times 10^{8}$ | $2.07 \times 10^{10}$ | $1.30 \times 10^{-14}$ |
| SDSS J104941.01+085548.4 | 10h 49m 41.00s | +08d 55m 48.5s | 1.19 | 1428 | 16.2 | $2.34 \times 10^{9}$ | $2.76 \times 10^{10}$ | $1.69 \times 10^{-14}$ |
| SDSS J133516.17+183341.4 | 13h 35m 16.10s | +18d 33m 41.8s | 1.19 | 1724 | 13.4 | $5.75 \times 10^{9}$ | $1.32 \times 10^{10}$ | $4.34 \times 10^{-15}$ |
| SDSS J133807.69+360220.3 | 13h 38m 07.70s | +36d 02m 20.3s | 1.20 | 1960 | 11.8 | $1.38 \times 10^{9}$ | $2.05 \times 10^{10}$ | $8.23 \times 10^{-15}$ |
| SDSS J103111.52+491926.5 | 10h 31m 11.50s | +49d 19m 27.2s | 1.20 | 1612 | 14.4 | $1.10 \times 10^{9}$ | $1.93 \times 10^{10}$ | $8.42 \times 10^{-15}$ |
| SDSS J114857.33+160023.1 | 11h 48m 57.40s | +16d 00m 22.7s | 1.22 | 1850 | 12.5 | $7.94 \times 10^{9}$ | $1.83 \times 10^{10}$ | $6.91 \times 10^{-15}$ |
| SDSS J124157.90+130104.1 | 12h 41m 57.90s | +13d 01m 04.7s | 1.23 | 1538 | 15.1 | $8.91 \times 10^{8}$ | $2.13 \times 10^{10}$ | $9.99 \times 10^{-15}$ |
| SDSS J133654.44+171040.3 | 13h 36m 54.40s | +17d 10m 40.8s | 1.23 | 1407 | 16.4 | $1.74 \times 10^{9}$ | $1.97 \times 10^{10}$ | $9.32 \times 10^{-15}$ |
| SDSS J121056.83+231912.5 | 12h 10m 56.80s | +23d 19m 13.0s | 1.26 | 1785 | 13.0 | $6.03 \times 10^{8}$ | $1.85 \times 10^{10}$ | $6.91 \times 10^{-15}$ |
| SDSS J081133.43+065558.1 | 08h 11m 33.40s | +06d 55m 58.3s | 1.27 | 1586 | 14.6 | $2.45 \times 10^{9}$ | $2.98 \times 10^{10}$ | $1.65 \times 10^{-14}$ |
| SDSS J170616.24+370927.0 | 17h 06m 16.20s | +37d 09m 27.0s | 1.27 | 1388 | 16.7 | $1.26 \times 10^{9}$ | $1.67 \times 10^{10}$ | $6.89 \times 10^{-15}$ |

[a]EM mass from Saade et al. (2020)
[b]EM mass from Foustoul et al. (2025)

**Table 7.** Electromagnetic and GW-constrained upper limits on chirp mass and strain amplitude for candidate SMBHBs. Continuation of Table 6

| Name | RA | Dec | z | Period [days] | Freq [nHz] | $M_{\rm tot}$ $[M_\odot]$ | $\mathcal{M}_c^{95\%}$ $[M_\odot]$ | $h_0^{95\%}$ |
|---|---|---|---|---|---|---|---|---|
| PKS 2131−021 | 21h 34m 10.309s | −01d 53m 17.238s | 1.29 | 1780 | 13.0 | – | $1.59 \times 10^{10}$ | $5.29 \times 10^{-15}$ |
| BZQ J2156−2012 | 21h 56m 33.70s | −20d 12m 30.2s | 1.31 | 1333 | 17.4 | – | $1.39 \times 10^{10}$ | $4.98 \times 10^{-15}$ |
| PKS J0805−0111 | 08h 05m 12.888s | −01d 11m 13.795s | 1.39 | 1237 | 18.7 | – | $2.37 \times 10^{10}$ | $1.18 \times 10^{-14}$ |
| QNZ3:54 | 15h 18m 06.60s | +01d 31m 34.9s | 1.40 | 1724 | 13.4 | $1.86 \times 10^9$ | $1.40 \times 10^{10}$ | $3.92 \times 10^{-15}$ |
| BZQ J0842+4525 | 08h 42m 15.30s | +45d 25m 45.0s | 1.41 | 1886 | 12.3 | $3.02 \times 10^9$ | $2.45 \times 10^{10}$ | $9.30 \times 10^{-15}$ |
| 3C 298.0 | 14h 19m 08.20s | +06d 28m 35.1s | 1.44 | 1960 | 11.8 | $3.72 \times 10^9$ | $2.18 \times 10^{10}$ | $7.25 \times 10^{-15}$ |
| SDSS J014350.13+141453.0 | 01h 43m 50.00s | +14d 14m 54.9s | 1.44 | 1538 | 15.1 | $1.62 \times 10^9$ | $1.88 \times 10^{10}$ | $6.70 \times 10^{-15}$ |
| SDSS J124119.04+203452.7 | 12h 41m 19.00s | +20d 34m 53.4s | 1.49 | 1218 | 19.0 | $2.51 \times 10^9$ | $1.81 \times 10^{10}$ | $7.02 \times 10^{-15}$ |
| SDSS J121457.39+132024.3 | 12h 14m 57.40s | +13d 20m 24.5s | 1.49 | 1923 | 12.0 | $2.88 \times 10^9$ | $2.43 \times 10^{10}$ | $8.41 \times 10^{-15}$ |
| SDSS J155647.78+181531.5 | 15h 56m 47.80s | +18d 15m 32.1s | 1.50 | 1428 | 16.2 | $3.24 \times 10^9$ | $1.80 \times 10^{10}$ | $6.21 \times 10^{-15}$ |
| SDSS J121018.66+185726.0 | 12h 10m 18.70s | +1d 57m 27.0s | 1.52 | 1754 | 13.2 | $3.39 \times 10^9$ | $1.92 \times 10^{10}$ | $5.96 \times 10^{-15}$ |
| SDSS J165136.76+434741.3 | 16h 51m 36.80s | +43d 47m 41.9s | 1.60 | 1923 | 12.0 | $2.19 \times 10^9$ | $2.07 \times 10^{10}$ | $5.92 \times 10^{-15}$ |
| SDSS J093819.25+361858.7 | 09h 38m 19.30s | +36d 18m 58.9s | 1.68 | 1265 | 18.3 | $2.09 \times 10^9$ | $2.56 \times 10^{10}$ | $1.05 \times 10^{-14}$ |
| SDSS J164452.71+430752.2 | 16h 44m 52.70s | +43d 07m 52.9s | 1.72 | 1999 | 11.6 | $1.41 \times 10^10$ | $2.35 \times 10^{10}$ | $6.52 \times 10^{-15}$ |
| SDSS J123147.27+101705.3 | 12h 31m 47.30s | +10d 17m 05.4s | 1.73 | 1850 | 12.5 | $1.58 \times 10^9$ | $2.45 \times 10^{10}$ | $7.28 \times 10^{-15}$ |
| SDSS J160730.33+144904.3 | 16h 07m 30.30s | +14d 49m 04.2s | 1.80 | 1724 | 13.4 | $6.61 \times 10^9$ | $1.71 \times 10^{10}$ | $4.00 \times 10^{-15}$ |
| SDSS J092911.35+203708.5 | 09h 29m 11.30s | +20d 37m 09.2s | 1.85 | 1785 | 13.0 | $8.32 \times 10^9$ | $3.36 \times 10^{10}$ | $1.17 \times 10^{-14}$ |
| SDSS J082121.88+250817.5 | 08h 21m 22.00s | +25d 08m 16.2s | 1.91 | 1886 | 12.3 | $3.39 \times 10^9$ | $3.26 \times 10^{10}$ | $1.03 \times 10^{-14}$ |
| UM 211 | 00h 12m 10.90s | −01d 22m 07.6s | 2.00 | 1886 | 12.3 | – | $2.76 \times 10^{10}$ | $7.38 \times 10^{-15}$ |
| SDSS J134855.27−0321 | 13h 48m 55.30s | −03d 21m 41.4s | 2.10 | 1428 | 16.2 | $7.76 \times 10^9$ | $3.00 \times 10^{10}$ | $9.56 \times 10^{-15}$ |
| SDSS J094450.76+151236.9 | 09h 44m 50.70s | +15d 12m 37.5s | 2.12 | 1428 | 16.2 | $4.07 \times 10^9$ | $4.30 \times 10^{10}$ | $1.73 \times 10^{-14}$ |
| HS 0926+3608 | 09h 29m 52.10s | +35d 54m 49.6s | 2.15 | 1561 | 14.8 | $8.91 \times 10^9$ | $3.55 \times 10^{10}$ | $1.16 \times 10^{-14}$ |
| SDSS J140704.43+273556.6 | 14h 07m 04.50s | +27d 35m 56.3s | 2.22 | 1561 | 14.8 | $8.71 \times 10^9$ | $2.55 \times 10^{10}$ | $6.41 \times 10^{-15}$ |
| SDSS J080809.56+311519.1 | 08h 08m 09.50s | +31d 15m 18.9s | 2.64 | 1162 | 19.9 | $2.29 \times 10^8$ | $3.78 \times 10^{10}$ | $1.22 \times 10^{-14}$ |
| SDSS J131706.19+271416.7 | 13h 17m 06.20s | +27d 14m 16.7s | 2.67 | 1666 | 13.9 | $8.32 \times 10^9$ | $2.43 \times 10^{10}$ | $4.53 \times 10^{-15}$ |

**Table 8.** Electromagnetic and GW-constrained upper limits on chirp mass and strain amplitude for candidate SMBHBs. Continuation of Table 7

## ACKNOWLEDGEMENTS

C.M.F.M. thanks F. van den Bosch, B. Farr, J. Greene, K. Grunthal, D. Holz, N. Navon, V. Ozoliņš, T. A. Prince, C. M. Urry, A. Vecchio. The work contained herein has been carried out by the NANOGrav collaboration, which receives support from the National Science Foundation (NSF) Physics Frontier Center award numbers 1430284 and 2020265, the Gordon and Betty Moore Foundation, NSF AccelNet award number 2114721, an NSERC Discovery Grant, and CIFAR. The Arecibo Observatory is a facility of the NSF operated under cooperative agreement (AST-1744119) by the University of Central Florida (UCF) in alliance with Universidad Ana G. Mendez (UAGM) and Yang Enterprises (YEI), Inc. The Green Bank Observatory is a facility of the NSF operated under cooperative agreement by Associated Universities, Inc. L.B. acknowledges support from the National Science Foundation under award AST-2307171. the National Aeronautics and Space Administration under award 80NSSC22K0808, the Research Corporation for Science Advancement under Cottrell Scholar Award No. 27553. P.R.B. is supported by the Science and Technology Facilities Council, grant number ST/W000946/1. P.N. acknowledges support from the Gordon and Betty Moore Foundation and the John Templeton Foundation that fund the Black Hole Initiative (BHI) at Harvard University where she serves as one of the PIs. S.B.S. gratefully acknowledges the support of a Sloan Fellowship, and the support of NSF under award #2408649. The work of R.B., N.La., X.S., J.T., and D.W. is partly supported by the George and Hannah Bolinger Memorial Fund in the College of Science at Oregon State University. M.C., P.P., and S.R.T. acknowledge support from NSF AST-2007993. M.C. was supported by the Vanderbilt Initiative in Data Intensive Astrophysics (VIDA) Fellowship. M.C. acknowledges support by the European Union (ERC, MMMonsters, 101117624). Support for this work was provided by the NSF through the Grote Reber Fellowship Program administered by Associated Universities,

Inc./National Radio Astronomy Observatory. Pulsar research at UBC is supported by an NSERC Discovery Grant and by CIFAR. K.C. is supported by a UBC Four Year Fellowship (6456). M.E.D. acknowledges support from the Naval Research Laboratory by NASA under contract S-15633Y. T.D. and M.T.L. received support by an NSF Astronomy and Astrophysics Grant (AAG) award number 2009468 during this work. E.C.F. is supported by NASA under award number 80GSFC24M0006. G.E.F., S.C.S., and S.J.V. are supported by NSF award PHY-2011772. K.A.G. and S.R.T. acknowledge support from an NSF CAREER award #2146016. M.J.G. is supported by NSF #2108402. A.D.J. and M.V. acknowledge support from the Caltech and Jet Propulsion Laboratory President's and Director's Research and Development Fund. A.D.J. acknowledges support from the Sloan Foundation. N.La. acknowledges the support from Larry W. Martin and Joyce B. O'Neill Endowed Fellowship in the College of Science at Oregon State University. Part of this research was carried out at the Jet Propulsion Laboratory, California Institute of Technology, under a contract with the National Aeronautics and Space Administration (80NM0018D0004). D.R.L. and M.A.M. are supported by NSF #1458952. M.A.M. is supported by NSF #2009425. C.M.F.M. was supported in part by the National Science Foundation under Grants No. NSF PHY-1748958, AST-2106552, and NASA LPS 80NSSC24K0440. C.M.F.M.'s research was supported in part by grant NSF PHY-2309135 to the Kavli Institute for Theoretical Physics (KITP). A.Mi. is supported by the Deutsche Forschungsgemeinschaft under Germany's Excellence Strategy - EXC 2121 Quantum Universe - 390833306. The Dunlap Institute is funded by an endowment established by the David Dunlap family and the University of Toronto. K.D.O. was supported in part by NSF Grant No. 2207267. T.T.P. acknowledges support from the Extragalactic Astrophysics Research Group at Eötvös Loránd University, funded by the Eötvös Loránd Research Network (ELKH), which was used during the development of this research. H.A.R. is supported by NSF Partnerships for Research and Education in Physics (PREP) award No. 2216793. S.M.R. and I.H.S. are CIFAR Fellows. Portions of this work performed at NRL were supported by ONR 6.1 basic research funding. J.D.R. also acknowledges support from start-up funds from Texas Tech University. J.S. is supported by an NSF Astronomy and Astrophysics Postdoctoral Fellowship under award AST-2202388, and acknowledges previous support by the NSF under award 1847938. C.A.W. acknowledges support from CIERA, the Adler Planetarium, and the Brinson Foundation through a CIERA-Adler postdoctoral fellowship. O.Y. is supported by the National Science Foundation Graduate Research Fellowship under Grant No. DGE-2139292. E.E. was funded in part by a NASA-CT Space Grant, PTE Federal Award No: 80NSSC25M7127. Y.C. was supported by a STARS II Fellowship and a NASA-CT Space Grant, PTE Federal Award No: 80NSSC20M0129.

## AUTHOR CONTRIBUTIONS

The authors are listed alphabetically to acknowledge that a long-term effort like NANOGrav, spanning over a decade, is fundamentally a collective enterprise. All contributors participated in collaborative activities that enabled the results presented in this work, and each reviewed the manuscript, including its text and figures, before submission. Additional specific contributions to this paper are as follows.

C.M.F.M. conceived of the project, led the searches, and coordinated the writing of this paper. The analysis was performed by F.H., B.L., N.A., L.W., J.A.C.C., R.S., C.M.F.M., and S.B.S. who produced the figures and tables. F.H. and B.L. developed new and independent targeted search pipelines to run with `enterprise`, with the help of R.S. C.A.W. originally developed targeted search methodologies and enterprise-based code adaptations for CW analyses under the supervision of S.B.S., S.T., and S.V., and has provided training and documentation on these techniques for the collaboration. Those developments and trainings have informed the broader CW efforts within NANOGrav, including the work presented here. Additional specific contributions to the original `enterprise`-based targeted search code for the NANOGrav 11-year Data Set can be found in Arzoumanian et al. (2020). N.A. and L.W. carried out the `QuickCW` searches with help from B.B., S.B.S., and L.D. The paper was written by C.M.F.M, F.H., B.L, F.S. and B.B. with contributions from N.A., S.B.S., and Q.Z.. The analysis of the GWB anisotropy was derived by F.S.. The dropout searches, incoherent and coherent searches were carried out by B.B.. Posterior reweighting analyses to account for HD correlations were carried out by B.L. for `enterprise` searches and N.A for `QuickCW` searches. The extension of the CRTS light curves for SDSS J1536+0441 and J0729+4008 was carried out by M.G. and P.C.. The discussion about electromagnetic counterparts for SDSS J1536+0441 and J0729+4008 was written by C.M.F.M, F.H., T.J.L., M.G., P.C., S.B.S., J.R., P.N., M.C., D.D., T.B., and K.G.. Additional comments during review were contributed by J.R., D.D., T.J.L., S.C., L.B., C.A.W., S.V., S.R.T..

## DATA AVAILABILITY

A repository to reproduce results of the `enterprise` searches can be found on Zenodo in the dataset "The NANOGrav 15 yr Data Set: Targeted Searches for Supermassive Black Hole Binaries": doi:10.5281/zenodo.17289574. This repository contains posterior samples from MCMC chains, Hellings-Downs weight factors, electromagnetic target priors, as well as Jupyter Notebooks and Python scripts for carrying out analyses.

## REFERENCES

Agarwal, N., Burke-Spolaor, S., Larsen, B., et al. in prep

Agazie, G., Anumarlapudi, A., Archibald, A. M., et al. 2023a, ApJL, 951, L8, doi: 10.3847/2041-8213/acdac6

Agazie, G., Alam, M. F., Anumarlapudi, A., et al. 2023b, ApJL, 951, L9, doi: 10.3847/2041-8213/acda9a

Agazie, G., Anumarlapudi, A., Archibald, A. M., et al. 2023c, ApJL, 951, L50, doi: 10.3847/2041-8213/ace18a

Agazie, G., Anumarlapudi, A., Archibald, A. M., et al. 2023, The Astrophysical Journal Letters, 956, L3, doi: 10.3847/2041-8213/acf4fd

Agazie, G., Anumarlapudi, A., Archibald, A. M., et al. 2023, ApJL, 951, L10, doi: 10.3847/2041-8213/acda88

Agazie, G., Arzoumanian, Z., Baker, P. T., et al. 2024, ApJ, 963, 144, doi: 10.3847/1538-4357/ad1f61

Agazie, G., Anumarlapudi, A., Archibald, A. M., et al. 2025, ApJ, 978, 31, doi: 10.3847/1538-4357/ad93d5

Aggarwal, K., Arzoumanian, Z., Baker, P. T., et al. 2019, ApJ, 880, 116, doi: 10.3847/1538-4357/ab2236

Alam, M. F., Arzoumanian, Z., Baker, P. T., et al. 2021, ApJS, 252, 5, doi: 10.3847/1538-4365/abc6a1

Allen, B., & Ottewill, A. C. 1997, PhRvD, 56, 545, doi: 10.1103/PhysRevD.56.545

Amaro-Seoane, P., Andrews, J., Arca Sedda, M., et al. 2023, Living Reviews in Relativity, 26, 2, doi: 10.1007/s41114-022-00041-y

Arzoumanian, Z., Brazier, A., Burke-Spolaor, S., et al. 2016, ApJ, 821, 13, doi: 10.3847/0004-637X/821/1/13

Arzoumanian, Z., Baker, P. T., Brazier, A., et al. 2018, ApJ, 859, 47, doi: 10.3847/1538-4357/aabd3b

—. 2020, ApJ, 900, 102, doi: 10.3847/1538-4357/ababa1

Arzoumanian, Z., Baker, P. T., Blecha, L., et al. 2023, ApJL, 951, L28, doi: 10.3847/2041-8213/acdbc7

Babak, S., Petiteau, A., Sesana, A., et al. 2016, MNRAS, 455, 1665, doi: 10.1093/mnras/stv2092

Barausse, E., Dey, K., Crisostomi, M., et al. 2023, PhRvD, 108, 103034, doi: 10.1103/PhysRevD.108.103034

Bardati, J., Ruan, J. J., Haggard, D., & Tremmel, M. 2024a, ApJ, 961, 34, doi: 10.3847/1538-4357/ad055a

Bardati, J., Ruan, J. J., Haggard, D., Tremmel, M., & Horlaville, P. 2024b, ApJ, 977, 265, doi: 10.3847/1538-4357/ad9471

Bécsy, B. 2024, Classical and Quantum Gravity, 41, 225017, doi: 10.1088/1361-6382/ad84b0

Bécsy, B., Cornish, N. J., & Digman, M. C. 2022a, PhRvD, 105, 122003, doi: 10.1103/PhysRevD.105.122003

Bécsy, B., Cornish, N. J., & Kelley, L. Z. 2022b, ApJ, 941, 119, doi: 10.3847/1538-4357/aca1b2

Bécsy, B., Cornish, N. J., Petrov, P., et al. 2025, Classical and Quantum Gravity, 42, 175016, doi: 10.1088/1361-6382/adfd36

Begelman, M. C., Blandford, R. D., & Rees, M. J. 1980, Nature, 287, 307, doi: 10.1038/287307a0

Bogdanović, T., Miller, M. C., & Blecha, L. 2022, Living Reviews in Relativity, 25, 3, doi: 10.1007/s41114-022-00037-8

Bondi, M., & Pérez-Torres, M. A. 2010, ApJL, 714, L271, doi: 10.1088/2041-8205/714/2/L271

Boroson, T. A., & Lauer, T. R. 2009, Nature, 458, 53, doi: 10.1038/nature07779

Breiding, P., Burke-Spolaor, S., Eracleous, M., et al. 2021, ApJ, 914, 37, doi: 10.3847/1538-4357/abfa9a

Britzen, S., Zajaček, M., Gopal-Krishna, et al. 2023, ApJ, 951, 106, doi: 10.3847/1538-4357/accbbc

Burke-Spolaor, S., Taylor, S. R., Charisi, M., et al. 2019, A&A Rv, 27, 5, doi: 10.1007/s00159-019-0115-7

Burt, B. J., Lommen, A. N., & Finn, L. S. 2011, ApJ, 730, 17, doi: 10.1088/0004-637X/730/1/17

Casey-Clyde, J. A., Mingarelli, C. M. F., Greene, J. E., et al. 2025, The Astrophysical Journal, 987, 106, doi: 10.3847/1538-4357/adce05

Casey-Clyde, J. A., Mingarelli, C. M. F., Greene, J. E., et al. 2022, ApJ, 924, 93, doi: 10.3847/1538-4357/ac32de

Chalumeau, A., Babak, S., Petiteau, A., et al. 2022, MNRAS, 509, 5538, doi: 10.1093/mnras/stab3283

Chamberlin, S. J., & Siemens, X. 2012, PhRvD, 85, 082001, doi: 10.1103/PhysRevD.85.082001

Charisi, M., Haiman, Z., Schiminovich, D., & D'Orazio, D. J. 2018, MNRAS, 476, 4617, doi: 10.1093/mnras/sty516

Charisi, M., Taylor, S. R., Runnoe, J., Bogdanovic, T., & Trump, J. R. 2022, MNRAS, 510, 5929, doi: 10.1093/mnras/stab3713

Chen, S., Sesana, A., & Conselice, C. J. 2019, MNRAS, 488, 401, doi: 10.1093/mnras/stz1722

Chornock, R., Bloom, J. S., Cenko, S. B., et al. 2010, ApJL, 709, L39, doi: 10.1088/2041-8205/709/1/L39

Corbin, V., & Cornish, N. J. 2010, arXiv e-prints, arXiv:1008.1782, doi: 10.48550/arXiv.1008.1782

Cornish, N. J., & Sesana, A. 2013, Classical and Quantum Gravity, 30, 224005, doi: 10.1088/0264-9381/30/22/224005

Davis, M. C., Grace, K. E., Trump, J. R., et al. 2024, ApJ, 965, 34, doi: 10.3847/1538-4357/ad276e

de la Parra, P. V., Kiehlmann, S., Mróz, P., et al. 2025, ApJ, 987, 191, doi: 10.3847/1538-4357/addc60

Decarli, R., Dotti, M., Falomo, R., et al. 2009b, ApJL, 703, L76, doi: 10.1088/0004-637X/703/1/L76

Decarli, R., Treves, A., Falomo, R., et al. 2009a, SDSS J1536+0441: A quasar pair, not a binary black hole, Tech. Rep. 2061, The Astronomer's Telegram

Dexter, J., Lutz, D., Shimizu, T. T., et al. 2020, ApJ, 905, 33, doi: 10.3847/1538-4357/abc24f

Dittmann, A. J., & Ryan, G. 2024, ApJ, 967, 12, doi: 10.3847/1538-4357/ad2f1e

D'Orazio, D. J., & Charisi, M. 2023, arXiv e-prints, arXiv:2310.16896, doi: 10.48550/arXiv.2310.16896

D'Orazio, D. J., & Di Stefano, R. 2018, MNRAS, 474, 2975, doi: 10.1093/mnras/stx2936

D'Orazio, D. J., Duffell, P. C., & Tiede, C. 2024, ApJ, 977, 244, doi: 10.3847/1538-4357/ad938b

D'Orazio, D. J., & Haiman, Z. 2017, MNRAS, 470, 1198, doi: 10.1093/mnras/stx1269

D'Orazio, D. J., Haiman, Z., Duffell, P., MacFadyen, A., & Farris, B. 2016, MNRAS, 459, 2379, doi: 10.1093/mnras/stw792

D'Orazio, D. J., Haiman, Z., & Schiminovich, D. 2015, Nature, 525, 351, doi: 10.1038/nature15262

D'Orazio, D. J., & Loeb, A. 2018, ApJ, 863, 185, doi: 10.3847/1538-4357/aad413

Duffell, P. C., D'Orazio, D., Derdzinski, A., et al. 2020, ApJ, 901, 25, doi: 10.3847/1538-4357/abab95

El-Badry, K., Hogg, D. W., & Rix, H.-W. 2025, arXiv e-prints, arXiv:2509.10601, doi: 10.48550/arXiv.2509.10601

Ellis, J., & van Haasteren, R. 2017, jellis18/PTMCMCSampler: Official Release, doi: 10.5281/zenodo.1037579

Ellis, J. A., Vallisneri, M., Taylor, S. R., & Baker, P. T. 2020, ENTERPRISE: Enhanced Numerical Toolbox Enabling a Robust PulsaR Inference SuitE, Zenodo, doi: 10.5281/zenodo.4059815

EPTA Collaboration, InPTA Collaboration, Antoniadis, J., et al. 2023, A&A, 678, A50, doi: 10.1051/0004-6361/202346844

Eracleous, M., Boroson, T. A., Halpern, J. P., & Liu, J. 2012, ApJS, 201, 23, doi: 10.1088/0067-0049/201/2/23

Event Horizon Telescope Collaboration, Akiyama, K., Alberdi, A., et al. 2019, ApJL, 875, L1, doi: 10.3847/2041-8213/ab0ec7

Farris, B. D., Duffell, P., MacFadyen, A. I., & Haiman, Z. 2014, ApJ, 783, 134, doi: 10.1088/0004-637X/783/2/134

Ferranti, I., Shaifullah, G., Chalumeau, A., & Sesana, A. 2025, A&A, 694, A194, doi: 10.1051/0004-6361/202451809

Foustoul, V., Webb, N. A., Mignon-Risse, R., et al. 2025, A&A, 699, A55, doi: 10.1051/0004-6361/202452818

Gardiner, E. C., Kelley, L. Z., Lemke, A.-M., & Mitridate, A. 2024, ApJ, 965, 164, doi: 10.3847/1538-4357/ad2be8

Gaskell, C. M. 1984, Annals of the New York Academy of Sciences, 422, 349, doi: 10.1111/j.1749-6632.1984.tb23380.x

—. 2010, Nature, 463, E1, doi: 10.1038/nature08665

Goldstein, J. M., Sesana, A., Holgado, A. M., & Veitch, J. 2019, MNRAS, 485, 248, doi: 10.1093/mnras/stz420

Graham, M. J., Djorgovski, S. G., Stern, D., et al. 2015, MNRAS, 453, 1562, doi: 10.1093/mnras/stv1726

GRAVITY Collaboration, Abuter, R., Amorim, A., et al. 2021, A&A, 647, A59, doi: 10.1051/0004-6361/202040208

Greene, J. E., & Ho, L. C. 2005, ApJ, 630, 122, doi: 10.1086/431897

Grunthal, K., Nathan, R. S., Thrane, E., et al. 2025, MNRAS, 536, 1501, doi: 10.1093/mnras/stae2573

Guo, H., Liu, X., Zafar, T., & Liao, W.-T. 2020, MNRAS, 492, 2910, doi: 10.1093/mnras/stz3566

Hazboun, J. S., Simon, J., Baier, J., et al. 2025, arXiv e-prints, arXiv:2511.22597, doi: 10.48550/arXiv.2511.22597

Heckman, T. M., Miley, G. K., van Breugel, W. J. M., & Butcher, H. R. 1981, ApJ, 247, 403, doi: 10.1086/159050

Hellings, R. W., & Downs, G. S. 1983, ApJL, 265, L39, doi: 10.1086/183954

Hincks, A. D., Ma, X., Naess, S. K., et al. 2025, arXiv e-prints, arXiv:2504.04278, doi: 10.48550/arXiv.2504.04278

Ho, L. C., & Kim, M. 2015, ApJ, 809, 123, doi: 10.1088/0004-637X/809/2/123

Horlaville, P., Ruan, J. J., Eracleous, M., et al. 2025, arXiv e-prints, arXiv:2504.21145, doi: 10.48550/arXiv.2504.21145

Hourihane, S., Meyers, P., Johnson, A., Chatziioannou, K., & Vallisneri, M. 2023, PhRvD, 107, 084045, doi: 10.1103/PhysRevD.107.084045
Iguchi, S., Okuda, T., & Sudou, H. 2010, ApJL, 724, L166, doi: 10.1088/2041-8205/724/2/L166
Jenet, F. A., Lommen, A., Larson, S. L., & Wen, L. 2004, ApJ, 606, 799, doi: 10.1086/383020
Katz, J. I., Anderson, S. F., Margon, B., & Grandi, S. A. 1982, ApJ, 260, 780, doi: 10.1086/160297
Kelley, L. Z., Haiman, Z., Sesana, A., & Hernquist, L. 2019, MNRAS, 485, 1579, doi: 10.1093/mnras/stz150
Kerr, M., & Parthasarathy, A. 2022, Science, 376, doi: 10.1126/science.abm3231
Khusid, N. M., Mingarelli, C. M. F., Natarajan, P., Casey-Clyde, J. A., & Barnacka, A. 2023, ApJ, 955, 25, doi: 10.3847/1538-4357/ace16f
Kiehlmann, S., de la Parra, P. V., Sullivan, A. G., et al. 2025, ApJ, 985, 59, doi: 10.3847/1538-4357/adc567
Larsen, B., Mingarelli, C. M. F., Hazboun, J. S., et al. 2024, ApJ, 972, 49, doi: 10.3847/1538-4357/ad5291
Lauer, T. R., & Boroson, T. A. 2009, ApJ, 703, 930, doi: 10.1088/0004-637X/703/1/930
Lee, K. J., Wex, N., Kramer, M., et al. 2011, MNRAS, 414, 3251, doi: 10.1111/j.1365-2966.2011.18622.x
Liu, H.-Y., Liu, W.-J., Dong, X.-B., et al. 2019, ApJS, 243, 21, doi: 10.3847/1538-4365/ab298b
Liu, T., & Vigeland, S. J. 2021, ApJ, 921, 178, doi: 10.3847/1538-4357/ac1da9
Luo, J., Ransom, S., Demorest, P., et al. 2021, ApJ, 911, 45, doi: 10.3847/1538-4357/abe62f
Masci, F. J., Laher, R. R., Rusholme, B., et al. 2019, PASP, 131, 018003, doi: 10.1088/1538-3873/aae8ac
Merritt, D., & Ekers, R. D. 2002, Science, 297, 1310, doi: 10.1126/science.1074688
Miles, M. T., Shannon, R. M., Reardon, D. J., et al. 2025, MNRAS, 536, 1489, doi: 10.1093/mnras/stae2571
Milosavljević, M., & Merritt, D. 2003, in American Institute of Physics Conference Series, Vol. 686, The Astrophysics of Gravitational Wave Sources, ed. J. M. Centrella (AIP), 201–210, doi: 10.1063/1.1629432
Mingarelli, C. M. F., Grover, K., Sidery, T., Smith, R. J. E., & Vecchio, A. 2012, PhRvL, 109, 081104, doi: 10.1103/PhysRevLett.109.081104
Mingarelli, C. M. F., & Sidery, T. 2014, PhRvD, 90, 062011, doi: 10.1103/PhysRevD.90.062011
Mingarelli, C. M. F., Sidery, T., Mandel, I., & Vecchio, A. 2013, PhRvD, 88, 062005, doi: 10.1103/PhysRevD.88.062005
Mingarelli, C. M. F., Lazio, T. J. W., Sesana, A., et al. 2017, Nature Astronomy, 1, 886, doi: 10.1038/s41550-017-0299-6
Mingarelli, C. M. F., Blecha, L., Bogdanović, T., et al. 2025, Nature Astronomy, 9, 183, doi: 10.1038/s41550-025-02482-1
Oh, K., Yi, S. K., Schawinski, K., et al. 2015, ApJS, 219, 1, doi: 10.1088/0067-0049/219/1/1
O'Neill, S., Kiehlmann, S., Readhead, A. C. S., et al. 2022, ApJL, 926, L35, doi: 10.3847/2041-8213/ac504b
Pal, I., Anju, A., Sreehari, H., et al. 2024, ApJ, 976, 145, doi: 10.3847/1538-4357/ad8088
Petrov, P., Taylor, S. R., Charisi, M., & Ma, C.-P. 2024, ApJ, 976, 129, doi: 10.3847/1538-4357/ad7b14
Phinney, E. S. 2001, arXiv e-prints, astro, doi: 10.48550/arXiv.astro-ph/0108028
Reardon, D. J., Zic, A., Shannon, R. M., et al. 2023, ApJL, 951, L6, doi: 10.3847/2041-8213/acdd02
Rosado, P. A., & Sesana, A. 2014, MNRAS, 439, 3986, doi: 10.1093/mnras/stu254
Rosado, P. A., Sesana, A., & Gair, J. 2015, Monthly Notices of the Royal Astronomical Society, 451, 2417–2433, doi: 10.1093/mnras/stv1098
Runnoe, J. C., Eracleous, M., Bogdanović, T., Halpern, J. P., & Sigursson, S. 2025, ApJ, 984, 17, doi: 10.3847/1538-4357/adba58
Runnoe, J. C., Eracleous, M., Mathes, G., et al. 2015, ApJS, 221, 7, doi: 10.1088/0067-0049/221/1/7
Runnoe, J. C., Eracleous, M., Pennell, A., et al. 2017, MNRAS, 468, 1683, doi: 10.1093/mnras/stx452
Saade, M. L., Stern, D., Brightman, M., et al. 2020, ApJ, 900, 148, doi: 10.3847/1538-4357/abad31
Saade, M. L., Brightman, M., Stern, D., et al. 2024, ApJ, 966, 104, doi: 10.3847/1538-4357/ad372e
Sato-Polito, G., & Kamionkowski, M. 2024, Phys. Rev. D, 109, 123544, doi: 10.1103/PhysRevD.109.123544
Sato-Polito, G., Zaldarriaga, M., & Quataert, E. 2024, PhRvD, 110, 063020, doi: 10.1103/PhysRevD.110.063020
Schutz, B. F. 1986, Nature, 323, 310, doi: 10.1038/323310a0
Sesana, A., Haiman, Z., Kocsis, B., & Kelley, L. Z. 2018, ApJ, 856, 42, doi: 10.3847/1538-4357/aaad0f
Sesana, A., Vecchio, A., & Colacino, C. N. 2008, MNRAS, 390, 192, doi: 10.1111/j.1365-2966.2008.13682.x
Shen, Y., & Loeb, A. 2010, ApJ, 725, 249, doi: 10.1088/0004-637X/725/1/249
Shen, Y., Richards, G. T., Strauss, M. A., et al. 2011, ApJS, 194, 45, doi: 10.1088/0067-0049/194/2/45
Simon, J., Polin, A., Lommen, A., et al. 2014, ApJ, 784, 60, doi: 10.1088/0004-637X/784/1/60

Steinle, N., Middleton, H., Moore, C. J., et al. 2023, MNRAS, 525, 2851, doi: 10.1093/mnras/stad2408

Sudou, H., Iguchi, S., Murata, Y., & Taniguchi, Y. 2003, Science, 300, 1263, doi: 10.1126/science.1082817

Tasinato, G. 2023, Phys. Rev. D, 108, 103521, doi: 10.1103/PhysRevD.108.103521

Taylor, S. R., & Gair, J. R. 2013, PhRvD, 88, 084001, doi: 10.1103/PhysRevD.88.084001

Tian, L.-W., Bi, Y.-C., Wu, Y.-M., & Huang, Q.-G. 2025, arXiv e-prints, arXiv:2508.14742. https://arxiv.org/abs/2508.14742

Tonry, J. L., Denneau, L., Flewelling, H., et al. 2018, ApJ, 867, 105, doi: 10.3847/1538-4357/aae386

Tremblay, J. C., Goncharov, B., van Haasteren, R., et al. 2025, arXiv e-prints, arXiv:2508.20007, doi: 10.48550/arXiv.2508.20007

Vaughan, S., Uttley, P., Markowitz, A. G., et al. 2016, Monthly Notices of the Royal Astronomical Society, 461, 3145, doi: 10.1093/mnras/stw1412

Wang, L.-F., Shao, Y., Xiao, S.-R., Zhang, J.-F., & Zhang, X. 2025, JCAP, 2025, 095, doi: 10.1088/1475-7516/2025/05/095

Westernacher-Schneider, J. R., Zrake, J., MacFadyen, A., & Haiman, Z. 2022, PhRvD, 106, 103010, doi: 10.1103/PhysRevD.106.103010

Witt, C. A., Charisi, M., Taylor, S. R., & Burke-Spolaor, S. 2022, ApJ, 936, 89, doi: 10.3847/1538-4357/ac8356

Wright, E. L., Eisenhardt, P. R. M., Mainzer, A. K., et al. 2010, AJ, 140, 1868, doi: 10.1088/0004-6256/140/6/1868

Wrobel, J. M., & Laor, A. 2009, ApJL, 699, L22, doi: 10.1088/0004-637X/699/1/L22

—. 2010, ApJL, 714, L295, doi: 10.1088/2041-8205/714/2/L295

Xin, C., Mingarelli, C. M. F., & Hazboun, J. S. 2021, ApJ, 915, 97, doi: 10.3847/1538-4357/ac01c5

Xu, H., Chen, S., Guo, Y., et al. 2023, Research in Astronomy and Astrophysics, 23, 075024, doi: 10.1088/1674-4527/acdfa5

Zhang, S., Zhou, H., Shi, X., et al. 2019, ApJ, 877, 33, doi: 10.3847/1538-4357/ab1aa3

Zhu, X.-J., & Thrane, E. 2020, ApJ, 900, 117, doi: 10.3847/1538-4357/abac5a